\documentclass[journal,twoside]{IEEEtran}
\usepackage{multirow}
\usepackage{booktabs}
\usepackage[font=footnotesize]{subcaption}
\usepackage[font=footnotesize]{caption}
\usepackage{amsmath,amssymb,amsthm}
\usepackage{ragged2e}
\usepackage{float}
\usepackage{tikz}
\usepackage{cite}
\usepackage[T1]{fontenc} 
\usepackage[nolist]{acronym}
\usepackage[linesnumbered,ruled,lined,hanginginout]{algorithm2e}
\usepackage{algpseudocode}
\usepackage{color}

\usetikzlibrary{patterns}
\SetAlgoHangIndent{0em}
\let\endshdefinition\enddefinition
\def\enddefinition{\strut\hfill$\square$\endshdefinition}

\newtheoremstyle{mytheorem}
  {3pt}
  {3pt}
  {\itshape}
  {\parindent}
  {\bfseries\itshape}
  {:}
  {.5em}
  {\thmname{#1}\thmnumber{ #2}\thmnote{ (#3)}}
\makeatother
\theoremstyle{mytheorem}
\newtheorem{definition}{Definition}

\newtheorem*{example*}{Example}
\newtheorem{lemma}{Lemma}
\newtheorem{corollary}{Corollary}
\newtheorem{proposition}{Proposition}

\makeatletter
\g@addto@macro\normalsize{%
  \setlength\abovedisplayskip{3.9pt}
  \setlength\belowdisplayskip{3.9pt}
}

\newcounter{mytempeqncnt}

\begin{document}
\begin{figure*}[t!]
\normalsize
This paper was submitted for publication in IEEE Transactions on Wireless Communications on November 3, 2025. It was finally accepted for publication on August 16, 2026.

\

\textcopyright~2026 IEEE. Personal use of this material is permitted. 
Permission from IEEE must be obtained for all other uses, in any current or future media, including reprinting/republishing this material for advertising or promotional purposes, creating new collective works, for resale or redistribution to servers or lists, or reuse of any copyrighted component of this work in other works.

\

Published in IEEE Transactions on Wireless Communications.

\textbf{Date of Publication:} 31 August 2026. \textbf{DOI:} 10.1109/TWC.2026.3726742.

\vspace{17cm}
\end{figure*}

\bstctlcite{IEEE_nodash:BSTcontrol}
\bstctlcite{IEEEexample:BSTcontrol}

\title{Differential Space-Time Block Coding for Phase-Unsynchronized Cell-Free MIMO Downlink}

\author{Marx M. M. Freitas, \IEEEmembership{Member,~IEEE}, Giovanni Interdonato, \IEEEmembership{Senior Member,~IEEE}, \\ and Stefano Buzzi,~\IEEEmembership{Fellow,~IEEE}\vspace{-5mm}

\thanks{This work was supported by the EU under the Italian National Recovery and Resilience Plan (NRRP) of NextGenerationEU, partnership on ``Telecommunications of the Future'' (PE00000001 - program ``RESTART''); specifically, Structural Project NTN, Cascade Call INFINITE, CUP D93C22000910001, and Structural Project 6GWINET, Cascade Call SPARKS, CUP D43C22003080001. A preliminary and shortened version of the technical content presented here is reported in~\cite{freitas2025PMCellFree}, whereas~\cite{Freitas_UAV_2025} limits its analysis to unmanned aerial vehicle-empowered aerial cell-free massive MIMO systems adopting a differential Alamouti code scheme.}

\thanks{M. M. M. Freitas and S. Buzzi are with the Dept. of Electrical and Information Engineering (DIEI), University of Cassino and Southern Lazio, 03043 Cassino, Italy (e-mail:\{marxmiguelmiranda.defreitas; buzzi\}@unicas.it).

G. Interdonato was with the DIEI, University of Cassino and Southern Lazio, 03043 Cassino, Italy. He is now with Ericsson Research, Ericsson AB, 164 83 Stockholm, Sweden (e-mail: giovanni.interdonato@ericsson.com).

S. Buzzi is also with the Consorzio Nazionale Interuniversitario per le Telecomunicazioni (CNIT), 43124 Parma, Italy, and with the Dipartimento di Elettronica, Informazione e Bioingegneria (DEIB), Politecnico di Milano, 20156 Milan, Italy.}
}

\maketitle

\begin{acronym}
    \acro{5G}{fifth-generation}
	\acro{AP}{access point}
	\acro{AWGN}{additive white Gaussian noise}
	\acro{ASD}{Angular standard deviation}
	\acro{ABC}{Artificial Bee Colony}
	\acro{AoA}{angle of arrival}
    \acro{ADC}{analog-to-digital converter}
	\acro{B5G}{beyond 5G}
	\acro{BA}{Bat Algorithm}
	\acro{BS}{base station}
	\acro{BER}{bit-error rate}
	\acro{BPSK}{binary phase shift keying}
	\acro{CS}{Cuckoo Search}
	\acro{CAPEX}{capital expenditure}
	\acro{CF}{cell-free}
    \acro{CF-mMIMO}{cell-free massive multiple-input multiple-output}
	\acro{CPU}{central processing unit}
	\acro{CC}{computational complexity}
    \acro{CM}{complex multiplications}
	\acro{CTMC}{continuous-time Markov chain}
	\acro{CSI}{channel state information}
	\acro{CSIR}{CSI at the receiver}
	\acro{CSIT}{CSI at the transmitter}
	\acro{CDF}{cumulative distribution function}
	\acro{CHD}{channel hardening}
	\acro{DL}{downlink}
	\acro{DCC}{dynamic cooperation clustering}
	\acro{D-mMIMO}{distributed massive multiple-input multiple-output}
    \acro{DFT}{discrete Fourier transform}
    \acro{DSP}{digital signal processor}
    \acro{DPSK}{differential phase-shift keying}
    \acro{DSTBC}{differential space-time block coding }
    \acro{DQO}{differential quasi-orthogonal}
	\acro{EE}{energy efficiency}
	\acro{FS}{fiber switch}
	\acro{FD}{full duplication}
        \acro{FFT}{fast Fourier transform}
	\acro{FVP}{favorable propagation}
	\acro{FPA}{flower pollination algorithm}
	\acro{FA}{Firefly Algorithm}
	\acro{GA}{Genetic Algorithm}
        \acro{GOPS}{giga operations per second}
	\acro{GWO}{Grey Wolf Algorithm}
 \acro{GPP}{general purpose processor}
	\acro{HCPP}{hard core point process}
	\acro{i.i.d.}{independent and identically distributed}
	\acro{InH-open}{Indoor Hotspot Open Office}
	\acro{IoT}{Internet of Things}
	\acro{IC}{inter-coordinated}
	\acro{LOS}{line-of-sight}
	\acro{LP-MMSE}{local partial MMSE}
	\acro{LSFD}{large‐scale fading decoding }
	\acro{LSFB}{Largest-large‐scale fading}
    \acro{LMMSE}{linear minimum mean square error} 
    \acro{ML}{maximum likelihood}
	\acro{MD}{matched-decision}	
	\acro{MCMC}{markov chain monte carlo}
	\acro{MIMO}{multiple-input multiple-output}
	\acro{MF}{matched filter}
	\acro{m-MIMO}{massive-multiple-input multiple-output}
	\acro{MTBF}{mean time between failures}
	\acro{MR}{maximum ratio}
	\acro{MOFPA}{Multiobjective Flowers Pollination Algorithm}
	\acro{MMSE}{minimum mean square error}
	\acro{NLOS}{non-line-of-sight}
    \acro{NCC}{network-centric}
	\acro{NMSE}{normalized mean square error}
	\acro{NOMA}{non-orthogonal multiple access}
	\acro{NP}{no protection}
	\acro{NS}{non-scalable}
	\acro{NF}{noise figure}
    \acro{NR}{new radio}
	\acro{OMA}{orthogonal multiple access}
	\acro{OFDM}{orthogonal frequency-division multiplexing}
	\acro{OPEX}{operational expenditure}
    \acro{OSTBC}{orthogonal STBC}
	\acro{PRBs}{physical resource blocks}
	\acro{PD}{partial duplication}
	\acro{P-MMSE}{partial MMSE}
	\acro{P-RZF}{partial regularized zero-forcing}
	\acro{PDF}{probability density function}
	\acro{PSO}{particle swarm optimization}
    \acro{PA}{proposed approach}
    \acro{PEP}{pairwise error probability}
    \acro{PSK}{phase-shift keying}
	\acro{QAM}{quadrature amplitude modulation}
	\acro{QoS}{quality-of-service}
	\acro{RF}{radio frequency}
	\acro{RSMA}{rate-splitting multiple access}
	\acro{RMSE}{root-mean-square deviation}
	\acro{RV}{random variable}
	\acro{RS}{radio Stripes}
    \acro{RU}{radio unit}
    \acro{RZF}{regularized zero-forcing}
	\acro{SB}{serial buse}
	\acro{SDMA}{space-division multiple acess}
	\acro{SE}{spectral efficiency}
	\acro{SFP} {small form-factor pluggable}
	\acro{SHR}{self-healing radio}
	\acro{SIC}{successive interference cancellation}
	\acro{SCF}{scalable cell-free}
	\acro{SINR}{signal-to-interference-plus-noise ratio}
	\acro{SNR}{signal-to-noise ratio}
    \acro{STBC}{space-time block coding}
	\acro{TCO}{total cost of ownership}
	\acro{TDD}{time-division duplex}
	\acro{TR}{Technical Report}
  	\acro{TRP}{transmission and reception point}
	\acro{UatF}{use-and-then-forget}
	\acro{UAV}{unmanned aerial vehicle}
	\acro{UE}{user equipment}
	\acro{UL}{uplink}
	\acro{UMi}{Urban Micro}
	\acro{UC}{user-centric}
    \acro{UCC}{User-centric clustering}
    \acro{ULA}{uniform linear array}
    \acro{UPA}{uniform planar array}
    \acro{URLLC}{ultra-reliable low-latency communication}
\end{acronym}

\begin{abstract}

In the downlink of \acf{CF-mMIMO} systems, spectral efficiency gains critically rely on joint coherent transmission, as all \acfp{AP} must align their transmitted signals in phase at the \acf{UE}. Achieving such phase alignment is challenging, as it requires tight synchronization among geographically distributed APs. In this paper, we address this issue by introducing a \acf{DSTBC} approach that bypasses the need for \ac{AP} phase synchronization. We first provide analytic bounds to the achievable spectral efficiency of CF-mMIMO with phase-unsynchronized \acp{AP}. Then, we propose a \ac{DSTBC}-based transmission scheme tailored to \ac{CF-mMIMO}, which operates without CSI and phase synchronization among the APs. We derive a closed-form expression for the resulting \acf{SINR}, enabling quantitative comparisons among different \ac{DSTBC} schemes. Numerical simulations confirm that phase misalignments can significantly impair system performance. In contrast, the proposed \ac{DSTBC} scheme can mitigate these effects, achieving performance comparable to that of fully synchronized systems. \textcolor{black}{However, when more than two APs jointly serve a UE, the code rate of DSTBC schemes can limit their SE gains. Hence, we also investigate \acf{DQO}-STBC schemes that achieve full code rate by relaxing the orthogonality constraints.}


\end{abstract}


\section{Introduction} \label{sec:introduction}
\Ac{UC} \acf{CF-mMIMO} systems stand as a pivotal architecture for enabling next-generation wireless networks, including 6G and beyond~\cite{BookCFemil2021,ngo2024ultradense}. Moving beyond traditional cellular structures, \ac{CF-mMIMO} employs a large ensemble of geographically distributed \acp{AP} that collaboratively serve multiple \acp{UE}. This distributed coordination approach yields several substantial advantages over conventional deployments. Foremost among these benefits is the provision of a significantly more \ac{SE} across the entire service area, effectively mitigating performance bottlenecks often experienced by \acp{UE} situated at cell boundaries. Furthermore, the inherent redundancy provided by the numerous distributed connection paths substantially enhances system resilience against effects such as link blockages and deep shadow fading, offering crucial support for high-mobility users and \ac{URLLC}~\cite{Buzzi_CFmMIMO_magazine_2025}. Achieving scalability and energy efficiency through techniques like local processing and \ac{UC} clustering further underscores the attractiveness of \ac{CF-mMIMO} as a foundational technology.

In order to fully realize the performance potential afforded by the network-wide coordination of \acp{AP}, the \ac{DL} transmission critically requires joint coherent transmission. This necessity mandates that the signals transmitted from all serving APs arrive at the targeted \ac{UE} antenna perfectly aligned in phase, ensuring maximal constructive combination. However, maintaining such stringent phase synchronization is highly demanding in practical implementations. Each AP operates using its own local, independent oscillator to modulate the data-carrying signal, resulting in distinct and uncontrollable phase offsets across the network. These uncompensated phase misalignments directly violate the assumption of coherent transmission, leading to phase distortions in the effective \ac{DL} channel vector. As a result, the desired signals may combine destructively, causing severe degradation in the received signal power and significantly impairing the achievable \ac{DL} \ac{SE}.

\subsection{Related Works and Motivation} \label{subsec:related-works}

Mitigating \ac{AP} phase misalignments in \ac{CF-mMIMO} systems remains a significant challenge. Most current literature focuses on mitigating these effects rather than developing inherently immune transmission schemes. One major research thrust involves over-the-air (OTA) synchronization and \ac{TDD} reciprocity calibration protocols~\cite{Rogalin_Sync_2014, Larsson_PhaseCalibration_2023,Cao_Calibration_2023,Xu_Calibration_2023,Larsson_MassiveSynchrony_2024,Xu_Calibration_2024,Ganesan_BeamSync_2024,KHNgo_PhaseCalibration_2025}. While pioneering, these methods often suffer from high computational complexity at the \ac{CPU}~\cite{Rogalin_Sync_2014} or introduce significant measurement overhead that can degrade spectral efficiency (\ac{SE}) as the number of \acp{AP} grows~\cite{KHNgo_PhaseCalibration_2025}.

Another set of works first analyzed the severe performance degradation caused by asynchronous reception and oscillator phase noise (PN)~\cite{Yan_Async_2019, Li_Async_2021, Fang_PN_2021}. Subsequent proposals for robust precoding schemes~\cite{Qi_Async_2023, AsynchronousCellFree, Wang_RobustPrecoding_2023} attempt to counter these effects. However, these solutions often rely on impractical assumptions like perfect \ac{CSI}~\cite{Qi_Async_2023}, are only effective in extreme mismatch scenarios~\cite{AsynchronousCellFree}, or lose their performance gains under severe phase uncertainties~\cite{Wang_RobustPrecoding_2023}.


Recognizing the difficulty of achieving full coherence, a third approach relaxes synchronization requirements. This includes hybrid \ac{AP} grouping~\cite{Quin_PhaseOffset_2024}, partially coherent (PC) transmission~\cite{PM_Larson, PM_GTEL}, and non-coherent (NC) joint transmission~\cite{PM_NCT, Miretti_CoarseSynchronization_2025}. While effective under certain constraints, these schemes introduce new challenges, such as feedback overhead for \ac{UE}-assisted grouping~\cite{Quin_PhaseOffset_2024}, performance reliance on cluster tuning~\cite{PM_Larson}, or fundamental \ac{SNR} and capacity losses relative to the ideal coherent case~\cite{Miretti_CoarseSynchronization_2025, d2024coherent}.

These strategies all fail to fully mitigate phase misalignments. Residual errors persist, and many methods depend on high-quality \ac{CSI}, which is unreliable under pilot contamination. Motivated by this gap, our previous works~\cite{freitas2025PMCellFree,Freitas_UAV_2025} proposed \ac{DSTBC} techniques~\cite{Larsson_Stoica_2003} as an alternative that inherently removes the need for \ac{AP} phase synchronization. While this approach successfully restored \ac{SE} and \ac{BER} performance, the analysis had practical limitations. Specifically, the \ac{ML} detector used in~\cite{freitas2025PMCellFree,Freitas_UAV_2025} incurred high computational complexity, and the analytical proof of phase cancellation relied on an idealistic assumption of perfect signal detection. \textcolor{black}{Moreover, the reduced code rate of orthogonal DSTBC schemes limited the achievable SE when more than two APs served the UE.}

\vspace*{-2mm}
\subsection{Main Contributions}
\textcolor{black}{This paper extends our studies~\cite{freitas2025PMCellFree,Freitas_UAV_2025} by developing a comprehensive analytical framework for \ac{DSTBC}-based \ac{CF-mMIMO} that overcomes the limitations identified in Section~\ref{subsec:related-works}. We provide a quantitative assessment of \ac{DSTBC} performance, demonstrating its potential to achieve synchronous-like performance in asynchronous environments. We also investigate \ac{DQO}-STBC schemes to mitigate the code-rate limitations of orthogonal DSTBC transmission.}
\begin{itemize}
    \item We develop a \ac{DSTBC} methodology based on \textit{amicable orthogonal designs}, $\{\mathbf{A}_n\}_{n=1}^{n_s}$ and $\{\mathbf{B}_n\}_{n=1}^{n_s}$, that mitigates phase misalignments. We show that this structure decouples the \ac{ML} detection criterion into $n_s$ per-symbol terms. This enables low-complexity, individual symbol detection, overcoming the high-complexity \ac{ML} detection limitation of~\cite{freitas2025PMCellFree,Freitas_UAV_2025} and representing a novel approach for \ac{CF-mMIMO}.
    \item We rigorously prove the phase mitigation capability of \ac{DSTBC} without the ideal assumption of perfect detection used in~\cite{freitas2025PMCellFree}. Our derivation shows the desired signal component depends only on the magnitude squared of the effective \ac{DL} channel, confirming the cancellation of phase misalignment terms under realistic conditions.
    \item \textcolor{black}{We investigate \ac{DQO}-STBC schemes for \ac{CF-mMIMO} systems to mitigate the code-rate reduction of orthogonal DSTBC schemes when four APs serve the UE.}
    \item \textcolor{black}{We derive analytical expressions for the \ac{DL} \ac{SINR} of conventional \ac{CF-mMIMO} under phase misalignments. These expressions explicitly characterize the \ac{SINR} as a function of the average phase misalignment at the \acp{AP}.}
    \item Based on these \ac{SINR} results, we derive closed-form achievable \ac{SE} expressions for conventional systems using \ac{MR} precoding, enabling direct performance evaluation without extensive simulations.
    \item For the proposed \ac{DSTBC} scheme, we derive an approximate \ac{SNR} expression for the single-user scenario to provide a baseline understanding of its resilience.
    \item We derive a comprehensive, approximate \ac{SINR} expression for the general multi-user \ac{DSTBC} \ac{CF-mMIMO} system. This expression accounts for residual noise and multi-user interference from the differential detection process, with closed-form calculations for signal and interference components. It enables performance evaluation without relying on intensive Monte Carlo simulations.
\end{itemize}

\vspace*{-3mm}
\subsection{Organization and Notation}
The paper is organized as follows: Section~\ref{sec:introduction} provides the necessary background and motivation, while Section~\ref{sec:system-model} details the system model. \textcolor{black}{The core contributions, which involve the \ac{DSTBC}-based approaches designed to mitigate phase misalignment effects, are presented in Sections~\ref{sec:dl-transmission} and \ref{sec:DQO_STBC}.} In Section~\ref{sec:results}, we present the numerical results demonstrating the effectiveness of our proposed approach. Lastly, Section~\ref{sec:conclusion} summarizes the findings and conclusions of this study.


\textit{Notation:} Boldface lowercase and uppercase letters denote vectors and matrices, respectively. The superscripts $(\cdot)^\mathrm{T}$ and $(\cdot)^\mathrm{H}$ denote the transpose and the conjugate-transpose operations, respectively. The $N\! \times\! N$ identity matrix is denoted as $\mathbf{I}_{N}$, and the cardinality of the set $\mathcal{A}$ is represented by $|\mathcal{A}|$. The trace and euclidean norm are denoted as $\mathrm{tr}(\, \cdot \,)$ and $\left \| \, \cdot \, \right \|$, and $\lfloor \cdot \rfloor$ is the floor operation. $\mathcal{N}_{\mathbb{C}} (\mu,\sigma^{2})$ stands for a complex Gaussian random variable with mean $\mu$ and variance $\sigma^{2}$. $\mathbf{A}\!=\!\mathrm{diag}(\mathbf{a})$ yields a diagonal matrix with the elements of the vector $\mathbf{a}$ on the diagonal, and $\mathbf{a} \!=\!\mathrm{diag}(\mathbf{A})$ yields a vector given by the diagonal elements of $\mathbf{A}$. The operator $\angle \cdot$ denotes the phase.

\vspace*{-3mm}
\section{System Model}
\label{sec:system-model}

We consider a \ac{CF-mMIMO} system consisting of $K$ single-antenna \acp{UE} and $L$ \acp{AP}, each equipped with $N$ antennas, geographically distributed into the coverage area. The \acp{AP} connect to a \ac{CPU} via fronthaul links, which are assumed to be error-free and able to support the data traffic~\cite{BookCFemil2021}. The system operates on \ac{TDD} mode and assumes that the \ac{UL} and \ac{DL} channels are reciprocal. Hence, channel estimation is performed only in the \ac{UL} direction. The channel $\mathbf{h}_{k, l}\in\mathbb{C}^{N}$ between the \ac{UE} $k$ and \ac{AP} $l$ undergoes an independent correlated Rayleigh fading realization, being defined as $\mathbf{h}_{k, l} \sim \mathcal{N}_\mathbb{C} (\mathbf{0}, \mathbf{R}_{k, l})$, where $\mathbf{R}_{k, l} = \mathbb{E}_{\mathbf{h}} \{\mathbf{h}_{k, l} \mathbf{h}_{k, l}^\mathrm{H}\} \in \mathbb{C}^{N \times N}$ is the statistical covariance matrix\footnote{The statistical covariance matrix captures the large-scale fading effects, namely the spatial channel correlation, path loss, and shadowing \cite{BookCFemil2021}.} of the channel. The operator $\mathbb{E}_{\mathbf{{h}}}$ indicates that the expectations are taken with respect to the channel realizations. The channels of different \acp{AP} are assumed to be uncorrelated, i.e., $\mathbb{E}_{\mathbf{{h}}} \{\mathbf{h}_{k,l} \mathbf{h}_{k, l{'}}^\mathrm{H}\} = \mathbf{0}$,  $\forall l \neq l{'}$, as the \acp{AP} are placed several wavelengths apart to ensure macrodiversity. The covariance matrices are assumed to remain constant over multiple coherence blocks, while the channels $\{\mathbf{h}_{k, l}\}$ vary independently from block to block.

\subsection{Phase Misalignment and UL Training}
\label{SSec:PhaseMisalignment}
The impact of phase misalignment in the \ac{DL} channels of \ac{CF-mMIMO} systems is commonly modeled by introducing an uncompensated phase component into the channel vector between UE $k$ and AP $l$. As a result, the true channel between the \ac{UE} $k$ and \ac{AP} $l$ is given by $\mathbf{g}_{k,l} =  e^{-j \vartheta_l} \mathbf{h}_{k,l}$, where $\vartheta_l$ is the oscillator phase\footnote{\textcolor{black}{In some approaches, the phase offsets introduced by the UEs' local oscillators are incorporated into the true DL channel. However, as commonly done in the literature \cite{PM_Larson, PM_GTEL, PM_NCT, Miretti_CoarseSynchronization_2025, d2024coherent}, we neglect UE-side oscillator phase impairments, assuming that they are compensated by standard synchronization mechanisms, such as phase-locked loops (PLLs) in the receiver RF chain.}} in the transmit chain of \ac{AP} $l$. This phase may also include phase noise effects, and is assumed to stay constant for at least the time needed to send two consecutive DSTBC codewords, i.e., for a few symbol intervals \cite{Larsson_Stoica_2003}. \textcolor{black}{We model the oscillator phases as independent Gaussian random variables, i.e., $\vartheta_l \sim \mathcal{N}(0, \sigma_{\mathrm{osc}}^2)$ for $l = \{1, \ldots, L\}$.}


As we focus on \ac{DL} channels, let us consider that each coherence block contains $\tau_{c}$ samples, with $\tau_{d}$ samples being reserved for \ac{DL} data and $\tau_{p}$ for \ac{UL} pilot signals. During the \ac{UL} training phase, the \acp{UE} send pilot sequences of $\tau_{p}$-length to the \acp{AP} for channel estimation. The pilot signals are assumed to be mutually orthogonal and independent of the number of \acp{UE} $K$ to ensure the scalability of the pilot resources. Let $\mathcal{Q}_k \subset \{1, \ldots, K\}$ denote the subset of the \acp{UE} assigned to the pilot $q_k$, including the \ac{UE} $k$. The received pilot signal at \ac{AP} $l$ can be expressed as \cite{BookCFemil2021}
\begin{equation}
\centering
\mathbf{y}_{q_k, l}^{\textrm{pilot}} = \sum\nolimits_{i\in \mathcal{Q}_{k}} \sqrt{\tau_{p} \eta_{i}}\,e^{-j \tilde{\vartheta}_l}\mathbf{h}_{i, l}+\mathbf{n}_{q_k, l},
\label{Eq:RecSigPil}
\end{equation}
\textcolor{black}{where $\eta_{i}$ is the power with which \ac{UE} $i$ transmits the pilot symbol, $\mathbf{n}_{q_k, l} \sim \mathcal{N}_\mathbb{C}\left(\mathbf{0}_N,\sigma_{\mathrm{ul}}^2 \mathbf{I}_{N}\right)$ is the additive noise at the receiver, and $\tilde{\vartheta}_l$ is the oscillator phase during the UL pilot transmission.} Assuming that the covariance matrix $\mathbf{R}_{k, l}$ is known at the \acp{AP} and \ac{CPU}, the \ac{MMSE} estimate of the channel $\mathbf{h}_{k, l}$ is given by \cite{BookCFemil2021}
\begin{equation}
    \widehat{\mathbf{h}}_{k, l} = \sqrt{\tau_{p} \eta_{k}} \mathbf{R}_{k, l}\mathbf{\Psi}_{q_k, l}^{-1} \mathbf{y}_{q_k, l}^{\textrm{pilot}} ,
    \label{Eq:ChanEst}
\end{equation}
where $\mathbf{\Psi}_{q_k, l} = \mathbb{E}_{\mathbf{{h}}} \{(\mathbf{y}_{q_k, l}^{\textrm{pilot}}) (\mathbf{y}_{q_k, l}^{\textrm{pilot}})^\mathrm{H}\}$ represents the covariance matrix of the received pilot signal $\mathbf{y}_{q_k, l}^{\textrm{pilot}}$, being calculated as 
\begin{align}
    \mathbf{\Psi}_{q_k, l}  = \sum\nolimits_{i \in \mathcal{Q}_{k}} \eta_{i} \tau_{p} \mathbf{R}_{i, l} +\sigma_{\mathrm{ul}}^{2} \mathbf{I}_{N}.
\end{align}%
\textcolor{black}{Since oscillator phases fluctuate over time, the phase $\tilde{\vartheta}_l$ in \eqref{Eq:RecSigPil} deviates from that affecting the DL transmission, denoted by $\vartheta_l$. 
To facilitate the analytical treatment of DSTBC, we assume a phase drift between the UL and DL phases. While our analysis relies on this assumption, we also consider, in the simulations, a time-varying model in which the phase evolves symbol-by-symbol according to a Wiener (random-walk) process, as described in Subsection~\ref{SSec:Wiener}.}


\subsection{AP Clustering and DL Data Transmission}
In \ac{UC} \ac{CF-mMIMO} systems, each \ac{UE} is served by a subset of \acp{AP}. The association between the generic $k$-th \ac{UE} and the \acp{AP} can be modeled with the vector $\mathbf{a}_{k} \in \mathbb{C}^{L \times 1}$, in which the element $\mathrm{a}_{k, l}$ is $1$ if \ac{AP} $l$ provides service to \ac{UE} $k$, implying that $l \in \mathcal{M}_{k}$, and $\mathrm{a}_{k, l}$ is $0$ otherwise. Here, $\mathcal{M}_k \subset \{1,\ldots,L\}$ denotes the subset of \acp{AP} serving the \ac{UE} $k$. Moreover, each \ac{AP} is assumed to serve a limited number of \acp{UE}, denoted by $K_{\text{max}}$, to prevent its computational load from scaling with the number of \acp{UE} $K$ in the network, with $K_{\text{max}} \leq \tau_p$ \cite{BookCFemil2021, RestrictedProcessingMarx}. Therefore, by letting $\mathcal{K}_l$ denote the subset of \acp{UE} served by \ac{AP} $l$, one can note that $|\mathcal{K}_l| \leq K_{\text{max}}$.

Once the \ac{UE}'s \ac{AP} cluster is defined, the precoding vector $\mathbf{w}_{k,l} \in \mathbb{C}^{N \times 1}$ between the \ac{UE} $k$ and the \ac{AP} $l$ is generated, for all \acp{AP} in $\mathcal{M}_k$. The precoding vectors can be computed locally at each \ac{AP} or centrally at the \ac{CPU}. 
The precoding vector $\mathbf{w}_{k,l}$ also satisfies $\mathbb{E}_{\widehat{\mathbf{h}}} \big\{ \left\| \mathbf{w}_{k,l} \right\|^{2} \big \} = \rho_{k,l}$, where $\rho_{k,l} \leq \rho_{\text{max}}$ is the fraction of power allocated to the \ac{UE} $k$ regarding the \ac{AP} $l$. Here, $\rho_{\text{max}}$ represents the maximum transmit power at each \ac{AP}. 
After generating the precoding vectors, all \acp{AP} in $\mathcal{M}_k$ jointly transmit the same data symbol to \ac{UE} $k$ in \ac{DL} coherent transmissions. Let $\sum_{i=1}^{K} \mathrm{a}_{i, l} \mathbf{w}_{i,l} s^{p}_{i} \in \mathbb{C}^{N \times 1}$ denote the data signal transmitted by \ac{AP} $l$ at discrete time epoch $p = \{1,\dots, \tau_d\}$, where $s^{p}_{i}$ is the complex symbol intended for \ac{UE} $k$. Moreover, let us denote
\begin{align}
    \mathrm{g}_{k,l}^{\mathrm{(ef)}} &= \mathrm{a}_{k, l}  \mathbf{g}_{k,l}^{\mathrm{H}} \mathbf{w}_{k,l} = \mathrm{a}_{k, l} \,e^{j \vartheta_l} \mathbf{h}_{k,l}^{\mathrm{H}} \mathbf{w}_{k,l}\,,\\
    \widetilde{\mathrm{g}}_{i,k,l}^{\,\mathrm{(ef)}} & = \mathrm{a}_{i, l}\,e^{j\vartheta_l} \mathbf{h}_{k,l}^{\mathrm{H}}  \mathbf{w}_{i,l}\,,
\end{align}
as the \textit{effective} \ac{DL} channels of the \ac{UE} $k$, and the interfering \ac{UE} $i$, respectively, with respect to AP $l$, for $k \neq i$. As a result, the received signal at \ac{UE} $k$  can be expressed as
\begin{equation}
    \centering
    {y}^{p}_{k} = \sum_{l=1}^{L} \mathrm{g}_{k,l}^{\mathrm{(ef)}} s^{p}_{k} + \sum_{l=1}^{L} \sum_{i=1, i\neq k}^{K} \widetilde{\mathrm{g}}_{i,k,l}^{\,\mathrm{(ef)}} s^{p}_{i} + n_{k}^p,
    \label{Eq:ReceivedSignalTraditionalCF}
\end{equation}
where $n_{k}^p \sim \mathcal{N}_{\mathbb{C}}\big ( \mathrm{0}, \sigma_{\mathrm{dl}}^{2} \big )$ is the receiver noise. Assuming that the \ac{PSK} modulation\footnote{The use of \ac{PSK} modulation is motivated by the fact that \ac{DSTBC} schemes are typically designed for unitary constellations. Moreover, since this paper focuses on evaluating \ac{DSTBC} in \ac{CF-mMIMO}, \ac{PSK} modulation is a suitable choice. Rectangular constellations will be investigated in future works.}
is used, the detection of $s^{p}_{k}$ given ${y}^{p}_{k}$ 
can be performed using the \ac{ML} criterion $\hat{s}^{p}_{k} = \mathop{\mathrm{arg\,min}}_{s \in \mathcal{S}} \, \big| {\angle y}^{p}_{k}  - \angle s \big|^{2}$.

\subsection{Precoding and Power Allocation Schemes}

\ac{UC} \ac{CF-mMIMO} systems are implemented using either centralized or distributed processing.
In distributed processing, channel estimation and precoding design are performed locally in the \acp{AP}. In contrast, in centralized processing, these tasks are performed in the \ac{CPU}. The centralized processing typically achieves higher \ac{SE} and requires less signaling on the fronthaul links, as the \ac{CPU} has access to the global \ac{CSI}. In contrast, the distributed processing demands less computational complexity.

To compute the precoding vectors in a scalable way, the \ac{UL}/\ac{DL} duality is considered. 
In distributed processing, $\mathbf{w}_{k, l}$ is selected based on its local channel estimates as
\begin{equation}
    \mathbf{w}_{k, l}=\sqrt{\rho_{k, l}} \frac{\overline{\mathbf{w}}_{k, l}}{\sqrt{\mathbb{E}_{\widehat{\mathbf{h}}}\{\left\|\overline{\mathbf{w}}_{k, l}\right\|^{2}\}}}\,,
\label{Eq:distributed:PrecodingEquation}
\end{equation}
where $\overline{\mathbf{w}}_{k, l}$ defines the direction of the precoding vector.
In centralized processing, the collective precoding vector $\mathbf{w}_{k} \in \mathbb{C}^{NL \times 1}$ is designed based on global channel estimates as
\begin{equation}
\mathbf{w}_{k}\!=\!\big [\mathrm{a}_{k,1}\mathbf{w}_{k, 1}^{\mathrm{T}},...,\mathrm{a}_{k, L}\mathbf{w}_{k, L}^{\mathrm{T}}  \big ]^{\mathrm{T}}\!=\!%
\sqrt{\rho_{k}} \frac{\overline{\mathbf{w}}_{k}}{\sqrt{\mathbb{E}_{\widehat{\mathbf{h}}}\{\left\|\overline{\mathbf{w}}_{k}\right\|^{2}\}}},
\label{Eq:centralized:PrecodingEquation}
\end{equation}
where $\overline{\mathbf{w}}_{k} \in \mathbb{C}^{NL \times 1}$ defines the direction of the collective precoding vector. Moreover, $\rho_{k}$ is the power globally allocated to the \ac{UE} $k$, and $\mathbb{E}_{\widehat{\mathbf{h}}}\{ \left \|\mathbf{w}_{k} \right \|^{2}\} = \rho_{k}$. 
We adopt the \ac{P-MMSE} for the centralized processing, and employ the \ac{MR} and \ac{LP-MMSE} precoding for the distributed processing. These schemes are chosen due to their scalability properties~\cite{BookCFemil2021}.
The \ac{MR} scheme is implemented by setting $\overline{\mathbf{w}}_{k, l} = \widehat{\mathbf{h}}_{k, l}$ in \eqref{Eq:distributed:PrecodingEquation}. This precoding technique has low computational complexity, maximizes the received signal power at the \ac{UE} using only the local \ac{CSI}, but does not mitigate the multi-user interference. Whereas, the \ac{LP-MMSE} precoding scheme suppresses interference only from the \acp{UE} served by each \ac{AP}, relying solely on locally available channel estimates. While it is more complex than the \ac{MR} scheme, it can yield better \ac{SE} performance. The \ac{LP-MMSE} precoding vector, from \ac{AP} $l$ to \ac{UE} $k \in \mathcal{K}_l$, is given by \cite{BookCFemil2021}
\begin{equation}
    \overline{\mathbf{w}}_{k, l}^{\textrm{LP-MMSE}} \!=\! \eta_k\Bigg(\!\sum\limits_{i \in \mathcal{K}_l} \eta_i\left(\widehat{\mathbf{h}}_{i, l} \widehat{\mathbf{h}}_{i, l}^{\mathrm{H}}\!+\!\mathbf{U}_{i, l}\right)\!+\sigma_{\mathrm{ul}}^2 \mathbf{I}_N\!\Bigg)^{\!\!\!-1} {\! \! \widehat{\mathbf{h}}_{k, l}},
\end{equation}
where $\mathbf{U}_{i, l} = \mathbb{E}_{\mathbf{\widetilde{h}}}\{ \mathbf{\widetilde{h}}_{i, l} \mathbf{\widetilde{h}}_{i, l}^{\mathrm{H}} \}$ is the correlation matrix of the estimation error $\mathbf{\widetilde{h}}_{i, l} = \mathbf{h}_{i, l} - \mathbf{\widehat{h}}_{i, l}$. The \ac{P-MMSE} precoding scheme is used in centralized processing, and unlike \ac{MMSE} does not consider all \acp{UE}' channel estimates for suppressing interference, but only the subset of \acp{UE} that most interfere with \ac{UE} $k$, which is defined as $\mathcal{P}_k=\left\{i: \mathbf{a}_k \mathbf{a}_i^{\mathrm{H}} \neq \mathbf{0}_{L\times L}\right\}$. The \ac{P-MMSE} precoding vector is given by
\begin{equation}
    \centering
    \!\overline{\mathbf{w}}_{k}^{\textrm{P-MMSE}} \!=\!\eta_k\left(\sum\nolimits_{i \in \mathcal{P}_k} \eta_i \widehat{\mathbf{h}}_i \widehat{\mathbf{h}}_i^{\mathrm{H}} \!+\! \mathbf{Z}_{\mathcal{P}_k} \!+\! \sigma_{\mathrm{ul}}^2 \mathbf{I}_N\right)^{\!\!-1} {\!} \widehat{\mathbf{h}}_k,
\end{equation}
where $\widehat{\mathbf{h}}_i = \big [\mathrm{a}_{i, 1}\widehat{\mathbf{h}}_{i, 1}^{\mathrm{T}},...,\mathrm{a}_{i, L}\widehat{\mathbf{h}}_{i, L}^{\mathrm{T}} \big ]^{\mathrm{T}} \in \mathbb{C}^{NL \times 1}$ is the collective estimated channel vector and $
\mathbf{Z}_{\mathcal{P}_k}=\sum_{i \in \mathcal{P}_k} \eta_i \mathbf{U}_i$, with $\mathbf{U}_i = \mathrm{diag} \big(\mathrm{a}_{i, 1} \mathbf{U}_{i, 1},..., \mathrm{a}_{i, L} \mathbf{U}_{i, L}\big)$.

Two scalable power allocation strategies are considered in this work.  
For the centralized processing, we adopt the fractional power allocation,\footnote{Various power allocation strategies have been proposed in the literature~\cite{BookCFemil2021,PrecodingAndPowerOptimization,ScalabilityAspects}. However, optimal power allocation is out of the scope of this paper.} as it preserves the ability to mitigate the multi-user interference. The power coefficient $\rho_k$ is computed as \cite{BookCFemil2021, Chen2023}
\begin{equation}
\rho_k=\rho_{\max} \frac{\big(\sum_{l \in \mathcal{M}_k} \beta_{k, l}^{\varsigma}\big)^\kappa \varpi_k^{-\zeta}}{\max _{\ell \in \mathcal{M}_k} \sum_{i \in \mathcal{K}_\ell}\big(\sum_{l \in \mathcal{M}_i} \beta_{i, l}^{\varsigma}\big)^\kappa \varpi_i^{1-\zeta}},
\end{equation}
where $\beta_{k, l} = \mathrm{tr}\{\mathbf{R}_{k, l}\}/N$ denotes the large-scale fading coefficients of the \ac{UE} $k$ to the \ac{AP} $l$.
The term $\varpi_k$ represents the largest fraction of $\rho_k$ that can be allocated to a single \ac{AP} serving the \ac{UE} $k$, i.e., $\varpi_k = \max_{\ell \in \mathcal{M}_k} \mathbb{E}_{\widehat{\mathbf{h}}} \{ \left\| \bar{\mathbf{w}}_{k, \ell} \right\|^2 \}$. Moreover, $\zeta \in[-1,1]$, $\kappa \in[0,1]$, and $\varsigma$ are design parameters.
For the distributed processing, the power is allocated proportionally to the large-scale fading gains of each \ac{UE}, as proposed in~\cite{ScalabilityAspects}. Accordingly, $\rho_{k, l},~k \in \mathcal{K}_{l}$, is given by

\begin{equation}
\centering
\rho_{k, l} = \rho_{\max} \sqrt{\beta_{k, l}} \Big/ \Big(\sum\nolimits_{i \in \mathcal{K}_{l} }\sqrt{\beta_{i, l}} \Big) \,.
\label{Eq:PowerDistributed}
\end{equation}
\subsection{Spectral Efficiency}
\textcolor{black}{In \ac{UC} \ac{CF-mMIMO} systems, although existing capacity bounds and SE expressions implicitly account for phase misalignment effects, they do not explicitly characterize the SE as a direct function of the average AP phase misalignment \cite{MassiveMIMOPhases, cellFreeRician_FaseShifts, HardwareImpairments2020, UplinkOscillators2026}. This is because oscillator phases vary independently across the APs and fluctuate over time, making the analytical characterization of SE degradation more involved. To provide additional insight into this effect, this paper incorporates the statistical characterization of oscillator phase variations into the SE analysis, thereby enabling a clearer visualization of how phase misalignments affect the SINR and SE expressions. In particular, the expected value of $e^{j\vartheta_l}$ is calculated as}
\begin{equation}
    \centering
    \mathbb{E}_{\vartheta} \big\{ e^{j \vartheta_l} \big\} = e^{-\sigma_{\mathrm{osc}}^2/2},
    \label{Eq:AverageTheta}
\end{equation}
\noindent where the operator $\mathbb{E}_{\vartheta}$ indicates that the expectation is taken with respect to $\vartheta_l,~\forall l$. This result allows us to characterize the impact of phase misalignment on the \ac{DL} \ac{SE} of \ac{UC} \ac{CF-mMIMO} systems using two standard capacity bounds, which we refer to as achievable \acp{SE}. These capacity bounds will serve as the upper and lower performance limits of a \ac{UC} \ac{CF-mMIMO} system in our simulations. These capacity bounds are derived based on the signal model in \eqref{Eq:ReceivedSignalTraditionalCF}.

\begin{proposition} \label{prop:SE:upperBound}
Assuming that the \ac{MMSE} channel estimates are available at the \acp{AP}, or at the \ac{CPU}, and that the channel is perfectly known at the receiver, an upper bound \!\!\footnote{In practice, the \ac{UE} does not have access to perfect \ac{CSI}, which may render the term achievable \ac{SE} somewhat misleading under the aforementioned assumptions, particularly because the \ac{UE} also lacks access to the phase misalignment of the \acp{AP}. Nevertheless, we use the term achievable \ac{SE} to assess the full potential of a \ac{UC} \ac{CF-mMIMO} network under ideal conditions. In such a scenario, the \ac{SE} is indeed achievable, allowing us to compute an upper bound that serves as an optimistic performance benchmark for conventional \ac{UC} \ac{CF-mMIMO} systems.} on the achievable \ac{DL} \ac{SE} for \ac{UE}~$k$ is given by
\begin{equation}
\centering
    \mathrm{SE}^{(\mathrm{dl, 1})}_{k} = \frac{\tau_d}{\tau_c} \mathbb{E}_{\mathbf{h},\widehat{\mathbf{h}}} \left\{ \log_{2} \left ( 1 + \mathrm{SINR}^{(\mathrm{dl, 1})}_{k} \right ) \right\},
    \label{Eq:SE_inst}
\end{equation}
where the expectations $\mathbb{E}_{\mathbf{h},\widehat{\mathbf{h}}}$ are computed with respect to the true and estimated channel realizations. The term $\tau_d /\tau_c$ is the fraction of samples per coherence block dedicated to transmitting the \ac{DL} data. Furthermore, $\mathrm{SINR}^{(\mathrm{dl, 1})}_{k}$ represents the \ac{DL} instantaneous effective \ac{SINR}, which is given by
\begin{equation}
\mathrm{SINR}^{(\mathrm{dl, 1})}_{k} \!=\! \frac{\mathbb{E}_{\vartheta} \bigg\{ \Big |\sum\nolimits_{l=1}^{L} e^{j \vartheta_l} {\mathrm{h}}_{k, l}^{\mathrm{(ef)}} \Big|^{2}\bigg\}}{ \sum\limits_{{i=1,\, i\neq k}}^{K} \mathbb{E}_{\vartheta} \bigg \{ \Big| \sum\nolimits_{l=1}^{L} e^{j \vartheta_l}\widetilde{\mathrm{h}}_{i, k, l}^{\mathrm{(ef)}} \Big|^{2}\bigg \} + \sigma_{\mathrm{dl}}^{2}},
\label{Eq:instantaneousSINR_1}
\end{equation}
and where ${\mathrm{h}}_{k, l}^{\mathrm{(ef)}} = \mathrm{a}_{k, l}\mathbf{h}_{k, l}^{\mathrm{H}} \mathbf{w}_{k, l}$, and $\widetilde{\mathrm{h}}_{i, k, l}^{\mathrm{(ef)}} = \mathrm{a}_{i, l}\mathbf{h}_{k, l}^{\mathrm{H}} \mathbf{w}_{i, l}$.
\end{proposition}
\begin{IEEEproof}
    This result readily follows from utilizing \cite[Lemma 3.5]{BookCFemil2021} on the signal model in \eqref{Eq:ReceivedSignalTraditionalCF} while taking the statistical expectation with respect to $\{\vartheta_l\}$.
\end{IEEEproof}
To write~\eqref{Eq:instantaneousSINR_1} as a function of the average value of the oscillator phases at each \ac{AP}, we reformulate the numerator, respectively, the expectation in the denominator of \eqref{Eq:instantaneousSINR_1} as 
\begin{align}
    &\sum\nolimits_{l=1}^{L} \sum\nolimits_{l{'}=1}^{L} \mathbb{E}_{\vartheta}\Big\{e^{j (\vartheta_l - \vartheta_{l{'}})} \Big\} {\mathrm{h}}_{k, l}^{\mathrm{(ef)}} {\mathrm{h}}_{k, l{'}}^{(\textrm{ef})*}\,, \\
    &\sum\nolimits_{l=1}^{L} \sum\nolimits_{l{'}=1}^{L} \mathbb{E}_{\vartheta}\Big\{e^{j (\vartheta_l - \vartheta_{l{'}})} \Big\} \widetilde{\mathrm{h}}_{i, k, l}^{\mathrm{(ef)}} \widetilde{\mathrm{h}}_{i, k, l{'}}^{(\textrm{ef})*}\,.
\end{align}
\textcolor{black}{Note that $\mathbb{E}_{\vartheta}\{e^{j (\vartheta_l - \vartheta_{l{'}})} \} = 1$, if $l = l{'}$, and $\mathbb{E}_{\vartheta}\{e^{j (\vartheta_l - \vartheta_{l{'}})} \} = e^{-\sigma_{\mathrm{osc}}^2}$, if $l \ne l{'}$. Hence, $\mathrm{SINR}^{(\mathrm{dl, 1})}_{k}$ can be rewritten as}
\begin{align}
     \frac{\tilde{\nu} \big|\sum_{l = 1}^L {\mathrm{h}}_{k, l}^{\mathrm{(ef)}} \big|^2 + (1-\tilde{\nu}) \sum_{l = 1}^L \big| {\mathrm{h}}_{k, l}^{\mathrm{(ef)}} \big|^2}{\!\sum\limits_{\substack{i=1\\i\neq k}}^{K} \left[\tilde{\nu}\left| \sum\limits_{l=1}^L \widetilde{\mathrm{h}}_{i, k, l}^{\mathrm{(ef)}} \right|^2 \!\!+\! (1\!-\!\tilde{\nu}) \sum\limits_{l=1}^L \big| \widetilde{\mathrm{h}}_{i, k, l}^{\mathrm{(ef)}} \big|^2\right] \!+\! \sigma_{\mathrm{dl}}^{2}},
    \label{Eq:instantaneousSINR_2}
\end{align}
\textcolor{black}{where $\tilde{\nu} = e^{-\sigma_{\mathrm{osc}}^2}$.}
%
The impact of the phase misalignment on the \ac{DL} \ac{SE} can be easily inferred from \eqref{Eq:instantaneousSINR_2} due to the decay behavior of $\tilde{\nu}$. \textcolor{black}{For instance, when $\sigma_{\mathrm{osc}}^2 = 0$, the resulting \ac{SINR} corresponds to that of a fully coherent system, whereas for $\sigma_{\mathrm{osc}}^2 = \pi$, it approaches the \ac{SINR} of a fully non-coherent system.} Moreover, deriving the \ac{SE} in \eqref{Eq:SE_inst} in closed form is analytically not tractable due to the expectation operator outside the logarithm. However, it can be evaluated numerically via Monte Carlo simulations.


\begin{proposition}
When the \ac{MMSE} channel estimates are available at the \acp{AP}, or at the \ac{CPU}, and the receiver has knowledge of the average effective channel, a lower bound on the achievable \ac{DL} \ac{SE} for \ac{UE} $k$ can be computed as
\begin{equation}
\centering
    \mathrm{SE}^{(\mathrm{dl, 2})}_{k} = \frac{\tau_d}{\tau_c}  \log_{2} \left ( 1 + \mathrm{SINR}^{(\mathrm{dl, 2})}_{k} \right ) ,
    \label{Eq:SE_hardening}
\end{equation}
\begin{figure*}[tb!]
\begin{equation}
    \mathrm{SINR}^{(\mathrm{dl, 2})}_{k} = \frac{\big| \sum_{l=1}^L \mathbb{E}_{\mathbf{h}, \vartheta}\big\{e^{j \vartheta_l} {\mathrm{h}}_{k, l}^{\mathrm{(ef)}} \big\} \big|^{2}}{\sum_{i=1}^{K} \mathbb{E}_{\mathbf{h}, \vartheta} \big\{ \big|\sum_{l=1}^L e^{j \vartheta_l} \widetilde{\mathrm{h}}_{i, k, l}^{\mathrm{(ef)}} \big|^2 \big\} - \big| \sum_{l=1}^L \mathbb{E}_{\mathbf{h}, \vartheta}\big\{e^{j \vartheta_l} {\mathrm{h}}_{k, l}^{\mathrm{(ef)}} \big\} \big|^{2} +\sigma_{\mathrm{dl}}^{2}}
    \label{Eq:SINR_hardening}
\end{equation}
\vspace{-0.8em}
\hrulefill
\end{figure*}
with $\mathrm{SINR}^{(\mathrm{dl, 2})}_{k}$ being computed in \eqref{Eq:SINR_hardening} shown at the top of the next page, where the operator $\mathbb{E}_{\mathbf{h},\vartheta}$ indicates that the expectations are taken over both the channel realizations and the phase misalignments.
\end{proposition}
\begin{IEEEproof}
    This result readily follows from utilizing \cite[Corollary 6.3]{BookCFemil2021} on the signal model in \eqref{Eq:ReceivedSignalTraditionalCF} while taking the statistical expectation with respect to $\{\vartheta_l\}$ as well.
\end{IEEEproof}

Equation \eqref{Eq:SINR_hardening} is also known as the \textit{hardening bound}, which is a typical capacity lower bound utilized in massive MIMO theory that is valid for any choice of precoding vector. To characterize the impacts of phase misalignment in \eqref{Eq:SINR_hardening}, we rely on the fact that $e^{j\vartheta_l}$
is statistically independent of the channel. Then, by following a similar procedure to that used in the derivation of \eqref{Eq:instantaneousSINR_2}, \eqref{Eq:SINR_hardening} can be reformulated as in \eqref{Eq:SINR_hardening_2} shown at the top of the next page.

\begin{figure*}[tb!]
\begin{equation}
    \mathrm{SINR}^{(\mathrm{dl, 2})}_{k} = \frac{\tilde{\nu}\, \big| \sum\nolimits_{l=1}^L \mathbb{E}_{\mathbf{h}}\big\{{\mathrm{h}}_{k, l}^{\mathrm{(ef)}} \big\} \big|^{2}}{\sum\nolimits_{i=1}^{K} \mathbb{E}_{\mathbf{h}}\big\{ \tilde{\nu} \,\big|\sum\nolimits_{l=1}^L \widetilde{\mathrm{h}}_{i, k, l}^{\mathrm{(ef)}} \big|^2 + (1-\tilde{\nu}) \sum\nolimits_{l=1}^L \big|\widetilde{\mathrm{h}}_{i, k, l}^{\mathrm{(ef)}} \big|^2\big\} - \tilde{\nu} \,\big| \sum\nolimits_{l=1}^L \mathbb{E}_{\mathbf{h}}\big\{{\mathrm{h}}_{k, l}^{\mathrm{(ef)}} \big\} \big|^{2} +\sigma_{\mathrm{dl}}^{2}}
    \label{Eq:SINR_hardening_2}
\end{equation}
\vspace{-0.8em}
\hrulefill
\end{figure*}
\begin{corollary}
By using the \ac{MR} precoding scheme with $\overline{\mathbf{w}}_{k, l} = \mathbf{\widehat{h}}_{k, l}$ in \eqref{Eq:distributed:PrecodingEquation}, $\mathbb{E}_{\mathbf{h}} \big\{{\mathrm{h}}_{k, l}^{\mathrm{(ef)}} \big\} $ in \eqref{Eq:SINR_hardening_2} becomes \cite{BookCFemil2021}
\begin{equation}
\mathbb{E}_{\mathbf{h}} \big\{{\mathrm{h}}_{k, l}^{\mathrm{(ef)}} \big\} \!=\! \sqrt{ \eta_k \tau_p \rho_{k, l} \operatorname{tr}(\mathrm{a}_{k, l} \boldsymbol{\Theta}_{k,l})},
\label{Eq:Eq:DSk_HB_2}
\end{equation}
with $\boldsymbol{\Theta}_{k,l} \!=\! \mathbf{R}_{k, l} \mathbf{\Psi}_{q_k, l}^{-1} \mathbf{R}_{k, l}$. $\mathbb{E}_{\mathbf{h}}\!\big\{\big| \!\sum_{l=1}^L \widetilde{\mathrm{h}}_{i, k, l}^{\mathrm{(ef)}} \big|^2 \big \}$ is given by
\begin{align}
&\mathbb{E}_{\mathbf{h}}\Big\{\Big|\sum\nolimits_{l=1}^L \widetilde{\mathrm{h}}_{i, k, l}^{\mathrm{(ef)}} \Big|^2 \Big \} \!=\! \sum\nolimits_{l=1}^L \rho_{i, l} \frac{\operatorname{tr}(\mathrm{a}_{i, l} \boldsymbol{\Theta}_{i,l} \mathbf{R}_{k, l})}{\operatorname{tr}(\boldsymbol{\Theta}_{i,l})} \nonumber\\
&\quad\quad +\!\Bigg|\sum\limits_{l=1}^L \sqrt{\frac{\eta_k \tau_p \rho_{i, l}}{\operatorname{tr}(\boldsymbol{\Theta}_{i,l})}} \operatorname{tr}(\mathrm{a}_{i, l} \widetilde{\boldsymbol{\Theta}}_{i,k,l})\Bigg|^2 \mathbb{I}_{i, k},
\label{Eq:IS_kil_HB_2}
\end{align}
where $\widetilde{\boldsymbol{\Theta}}_{i,k,l} \!=\! \mathbf{R}_{i, l} \mathbf{\Psi}_{q_i, l}^{-1} \mathbf{R}_{k, l}$ and the binary indicator $\mathbb{I}_{i, k} \in\{0,1\}$ specifies the \acp{UE} sharing the same pilot signal, with $\mathbb{I}_{i, k} = 1$ corresponding to $i \in \mathcal{Q}_k$, and $\mathbb{I}_{i, k} = 0$ otherwise. Besides, the expectations $\mathbb{E}_{\mathbf{h}}\big\{\big|\widetilde{\mathrm{h}}_{i, k, l}^{\mathrm{(ef)}} \big|^2 \big\}$ are obtained as
\begin{align}
&\mathbb{E}_{\mathbf{h}}\big\{ \big| \widetilde{\mathrm{h}}_{i, k, l}^{\mathrm{(ef)}} \big|^2 \big\} = \rho_{i, l} \frac{\operatorname{tr}(\mathrm{a}_{i, l} \boldsymbol{\Theta}_{i,l} \mathbf{R}_{k, l})}{\operatorname{tr}(\boldsymbol{\Theta}_{i,l})} \nonumber \\
&\quad\quad+\! 
\Bigg|\sqrt{\frac{\eta_k \tau_p \rho_{i, l}}{\operatorname{tr}(\boldsymbol{\Theta}_{i,l})}} \operatorname{tr}(\mathrm{a}_{i, l} \widetilde{\boldsymbol{\Theta}}_{i,k,l})\Bigg|^2 \mathbb{I}_{i, k}.
\label{Eq:IS_kil_HB_3}
\end{align}
\end{corollary}
\begin{IEEEproof}
The closed-form expressions for the expectations associated with $\mathbb{E}_{\mathbf{h}}\big\{\sum_{l=1}^L \big|\widetilde{\mathrm{h}}_{i, k, l}^{\mathrm{(ef)}} \big|^2 \big\}$ can be computed by leveraging the properties of jointly Gaussian vectors [\citenum{BookCFemil2021}, Lemma~B.5]. That is, given two circularly-symmetric complex normals vectors $\widetilde{\mathbf{x}}$ and $\widetilde{\mathbf{y}}$, $\mathbb{E} \big\{ \left|\widetilde{\mathbf{x}}^{\mathrm{H}} \widetilde{\mathbf{y}} \right|^{2} \big\}$ is calculated as
\begingroup\makeatletter\def\f@size{10}\check@mathfonts
\begin{equation}
    \centering
     \operatorname{tr} \big( \mathbb{E} \big\{ \widetilde{\mathbf{x}}\widetilde{\mathbf{x}}^{\mathrm{H}} \big\} \mathbb{E} \big\{ \widetilde{\mathbf{y}}\widetilde{\mathbf{y}}^{\mathrm{H}} \big\} \big) + \big| \operatorname{tr} \big( \mathbb{E} \big\{ \widetilde{\mathbf{y}}\widetilde{\mathbf{x}}^{\mathrm{H}} \big\} \big) \big|^{2}
    \label{Eq:covariance_secondTerm}.
\end{equation}
\endgroup
By setting $\overline{\mathbf{w}}_{i, l} = \mathbf{\widehat{h}}_{i, l}$ in \eqref{Eq:distributed:PrecodingEquation}, one can compute
\begin{equation}
    \centering
    \widetilde{\mathrm{h}}_{i, k, l}^{\mathrm{(ef)}} = \mathrm{a}_{i, l} \sqrt{\rho_{i, l}} \mathbf{h}_{k, l}^{\mathrm{H}} \mathbf{\widehat{h}}_{i, l}\Big/\sqrt{\mathbb{E}_{\mathbf{h}}\big\{\big\|\mathbf{\widehat{h}}_{i, l}\big\|^{2}\big\}}.
\end{equation}
\noindent Let us define the vectors $\widetilde{\mathbf{x}}$ and $\widetilde{\mathbf{y}}$ as $\widetilde{\mathbf{x}} = \mathrm{a}_{i, l} \sqrt{\rho_{i, l}} \mathbf{h}_{k, l}^{\mathrm{H}}$ and $\widetilde{\mathbf{y}} = \mathbf{\widehat{h}}_{i, l}\Big/\sqrt{\mathbb{E}_{\mathbf{h}}\big\{\big\|\mathbf{\widehat{h}}_{i, l}\big\|^{2}\big\}}$. By applying $\widetilde{\mathbf{x}}$ and $\widetilde{\mathbf{y}}$ in \eqref{Eq:covariance_secondTerm}, we obtain $\mathbb{E}_{\mathbf{h}}\big\{\big|\widetilde{\mathrm{h}}_{i, k, l}^{\mathrm{(ef)}}\big|^2 \big\}$ as in \eqref{Eq:IS_kil_HB_3}.
\end{IEEEproof}

\section{DL Data Transmission Resilient to Phase Misalignment Effects}
\label{sec:dl-transmission}

The concept of \ac{DSTBC} was introduced in MIMO systems as an extension of STBC schemes. Its aim is to enhance transmission diversity by leveraging the multiple antennas at the transmitter and potentially at the receiver. More specifically, \ac{DSTBC} techniques serve as the differential counterpart of space-time block coding, with the key distinction that they do not require \ac{CSI} knowledge at the receiver. More precisely, in our proposed scheme, the channel estimates are used at the network side to compute the precoding vectors $\mathbf{w}_{k,l}$ for all \acp{AP} serving the \ac{UE} $k$ aimed at suppressing multiuser interference; however, the detection scheme does not need such knowledge. 

\subsection{Differential Transmission} 
Differential transmission in \ac{DSTBC} methods involves splitting the data symbols for the \ac{UE} into several space-time codewords. The receiver decodes the data by analyzing two consecutive received codewords, also called received signal blocks. To explain this process: consider the \ac{UE} $k$ which is supported by $L_k$ \acp{AP}, where $L_k = |\mathcal{M}_k|$. Let $P$ denote the number of symbols periods that a space-time codeword encompasses, and let $\mathcal{S}_{k} = \{ s^{1}_{k}, s^{2}_{k}, \ldots, s^{N_{sym}}_{k} \}$ be the set of $N_{sym}$ complex information symbols selected from a unitary constellation for the \ac{UE} $k$. This set, $\mathcal{S}_{k}$, is divided by the \ac{CPU} into smaller groups, each containing $n_s$ complex symbols. The number of these subsets in $\mathcal{S}_{k}$ is represented by $G$, which is calculated as $G = \lfloor \tau_d/P \rfloor$. Furthermore, the $t$-th subset of $\mathcal{S}_{k}$ is defined as $\mathcal{S}^{t}_{k} = \{ s^{(t-1)n_s +1}_{k}, \ldots, s^{t n_s}_{k} \}$, with the cardinality of $\mathcal{S}^{t}_{k}$ being equal to $|\mathcal{S}^{t}_{k}| = n_s$.
Then, each subset $\mathcal{S}^{t}_{k}$ is mapped by the \ac{CPU} onto a code matrix $\mathbf{X}^{t}_{k} \in \mathbb{C}^{L_k \times P}$, where $t$ also denotes the index of the code matrix. To facilitate detection, $\mathbf{X}^{t}_{k}$ is typically designed as a unitary matrix, i.e., $\mathbf{X}^{t}_{k} (\mathbf{X}^{t}_{k})^\mathrm{H} = \mathbf{I}_{L_k}$. In addition, these matrices are often designed to be orthogonal, meaning that the inner product between distinct rows of $\mathbf{X}^{t}_{k}$ is zero. The orthogonality property further contributes to reducing the detection complexity at the receiver. Nonetheless, orthogonal STBC matrices possess only a semi-unitary structure. In particular, the product $\mathbf{X}^{t}_{k} (\mathbf{X}^{t}_{k})^\mathrm{H}$ results in $\mathbf{X}^{t}_{k} (\mathbf{X}^{t}_{k})^\mathrm{H} = \sum_{n=1}^{n_s} \big|s^{n}_{k}\big|^2 \mathbf{I}_{L_k}$. To transform $\mathbf{X}^{t}_{k}$ into a unitary matrix, $\mathbf{X}^{t}_{k}$ it is typically scaled by a factor of $1/\sqrt{n_s}$, which yields to
\begin{equation}
    \centering
    \mathbf{X}^{t}_{k} (\mathbf{X}^{t}_{k})^\mathrm{H} = \frac{1}{n_s}\sum\nolimits_{n=1}^{n_s} \big|s^{n}_{k}\big|^2 \mathbf{I}_{L_k} = \mathbf{I}_{L_k},
\end{equation}
\noindent provided that $s^{n}_{k}$ is taken from a unitary constellation, i.e., $|s^{n}_{k}| = 1$. For simplicity, this paper also focuses on square code matrices, i.e., by assuming $P = L_k$. 

\begin{example*}[Orthogonal Code Matrices]
A well-established orthogonal code matrix in the literature is the Alamouti matrix, which encodes two complex symbols ($n_s = 2$) over two symbol periods ($P = 2$), resulting in a code rate of $R = n_s/P = 1$. For instance, when $t = 1$, the Alamouti matrix is expressed as
\begin{equation}
\mathbf{X}^{t}_{k} =
\frac{1}{\sqrt{2}}
\begin{bmatrix}
s^{1}_{k} & \phantom{-}(s^{2}_{k})^{*} \\
s^{2}_{k} & -(s^{1}_{k})^{*}
\end{bmatrix}.
\label{Eq:Alamouti}
\end{equation}
One can note that $s^{1}_{k}$ and $s^{2}_{k}$ are the two consecutive elements of $\mathcal{S}^{t}_{k}$ for $t = 1$. For $t > 1$, \eqref{Eq:Alamouti} would involve the symbols from the corresponding subset $\mathcal{S}^{t}_{k}$. For instance, when $t = 2$, the elements of \eqref{Eq:Alamouti} correspond to $s^{3}_{k}$ and $s^{4}_{k}$. For configurations with a higher number of \acp{AP} serving the \ac{UE}, other orthogonal matrices exist; as an example, if $L_k = 4$, the following orthogonal code matrix can be employed
\begin{equation}
\mathbf{X}^{t}_{k} = 
\frac{1}{\sqrt{3}}
\begin{bmatrix}
\phantom{\:}s^{1}_{k} & \phantom{\::}0 & s^{2}_{k} & -s^{3}_{k} \\ 
\phantom{\,,}0 & \phantom{\,,}s^{1}_{k} & \phantom{\::}(s^{3}_{k})^{*} & \phantom{\:::}(s^{2}_{k})^{*} \\
-(s^{2}_{k})^{*} & -s^{3}_{k} & \phantom{\::}(s^{1}_{k})^{*} & \phantom{\::}0 \\
\phantom{\:::}(s^{3}_{k})^{*} & -s^{2}_{k} & \phantom{\:}0 & \phantom{\:::}(s^{1}_{k})^{*}
\end{bmatrix},
\label{Eq:codeMatrix_Nt_4}
\end{equation}
\noindent where \eqref{Eq:codeMatrix_Nt_4} contains the elements of $\mathcal{S}^{t}_{k}$ for $t = 1$. 
This code matrix sends $n_s = 3$ symbols over $P = 4$ symbol periods, thus having the code rate of $R = 3/4$.    
\end{example*}
\begin{figure}[!t]
   \centering\hspace{-20mm}
   \resizebox{.85\columnwidth}{!}{\input{scenario}}
  \caption{Illustration of how the information matrix $\mathbf{C}^{t}_{k}$ is row-wise split among the \acp{AP} serving \ac{UE}~$k$. The \ac{CPU} assigns each row of matrix $\mathbf{C}^{t}_{k} \in \mathbb{C}^{L_k \times L_k}$ to an \ac{AP} serving the \ac{UE} $k$ via the mapping $m(l,k)$, with $l \in \mathcal{M}_k$. In this illustration, a matrix of dimension $\mathbf{C}^{t}_{k} \in \mathbb{C}^{4 \times 4}$ is considered. Thus, $\mathbf{C}^{t}_{k}$ is composed of four rows, each comprising a duration time of four symbols. To exemplify the meaning of the mapping $m(l,k)$, note that, since \ac{AP} $7$ is assigned the fourth row of $\mathbf{C}^{t}_{k}$, it follows that $m(7,k) = 4$.}
  \vspace{-5mm}
  \label{Fig:splitC}
\end{figure}
Once defined the code matrices, the \ac{CPU} encodes $\mathbf{X}^{t}_{k}$ differentially by forming the matrix $\mathbf{C}^{t}_{k} \in \mathbb{C}^{L_k \times L_k}$, given by
\begin{equation}
    \centering
    \mathbf{C}^{t}_{k} = \mathbf{C}^{t-1}_{k} \mathbf{X}^{t}_{k}.
    \label{Eq:EncodingMatrix}
\end{equation}
$\mathbf{C}^{t}_{k}$ conveys the information signal to be transmitted to the \ac{UE} $k$, with $\mathbf{C}^{0}_{k} = \mathbf{I}_{L_k}$. Given that $\mathbf{X}^{t}_{k}$ is a unitary matrix, the matrix $\mathbf{C}^{t}_{k}$ is also unitary, i.e., $\mathbf{C}^{t}_{k} (\mathbf{C}^{t}_{k})^\mathrm{H} = \mathbf{I}_{L_k}$. However, note that \eqref{Eq:EncodingMatrix} cannot be directly transmitted to the \ac{UE} $k$. Recall that the \ac{UE} $k$ is served by $L_k$ \acp{AP}, whose indices may range from $1$ to $L$. Meanwhile, the row indices of $\mathbf{C}^{t}_{k}$ span from $1$ to $L_k$, which may not directly correspond to the indices of the \acp{AP} serving the \ac{UE} $k$, as $L_{k}$ is typically smaller than $L$. Hence, before transmitting $\mathbf{C}^{t}_{k}$ to \ac{UE} $k$, the \ac{CPU} distributes the rows of $\mathbf{C}^{t}_{k}$ among the \acp{AP} serving the \ac{UE} $k$, i.e., for all \ac{AP} $l$ in $\mathcal{M}_k$. The row of $\mathbf{C}^{t}_{k}$ assigned to \ac{AP} $l$ is \cite{freitas2025PMCellFree}
\begin{equation}
    \centering
    \mathbf{c}^{t}_{k,l} = \big[ \mathbf{C}^{t-1}_{k} \big]_{m(l,k),:} \mathbf{X}^{t}_{k},
    \label{Eq:EncodingMatrixCellFree}
\end{equation}
where $l\!\in\!\mathcal{M}_{k}$ and $m(l,k)\!\in\!\left\{1, ..., L_{k} \right\}$ is a mapping function that assigns a specific row of the matrix $\mathbf{C}^{t}_{k}$ to \ac{AP} $l$ (see Fig.~\ref{Fig:splitC} for a graphical example). 
Let $\sum_{k=1}^{K} \mathrm{a}_{k, l}\mathbf{w}_{k,l} \mathbf{c}^{t}_{k,l} \in \mathbb{C}^{N \times L_{k}}$ represent the signal sent by \ac{AP} $l$. The received signal block can be expressed as 
\begin{equation}
    \centering
 \mathbf{y}^{t}_{k} = \sum\nolimits_{l=1}^{L} \mathrm{g}_{k,l}^{\mathrm{(ef)}} \big[ \mathbf{C}^{t-1}_{k} \big]_{m(l,k),:} \mathbf{X}^{t}_{k} +  \widetilde{\mathbf{n}}_{k}^{t},
 \label{Eq:ReceivedSignalCellFreeSimplified}
\end{equation}
where $\mathrm{g}_{k,l}^{\mathrm{(ef)}}\big[ \mathbf{C}^{t-1}_{k} \big]_{m(l,k),:}$ is unknown at the \ac{UE} $k$. Moreover, $\widetilde{\mathbf{n}}_{k}^{t}$ denotes the combined effects of interference and receiver noise, being given by
\begin{equation}
    \centering
     \widetilde{\mathbf{n}}_{k}^{t} = \sum_{i=1, i \neq k}^{K} \sum_{l=1}^{L} \widetilde{\mathrm{g}}_{i,k,l}^{\,\mathrm{(ef)}} \big[ \mathbf{C}^{t-1}_{i} \big]_{m(l,i),:} \mathbf{X}^{t}_{i} + \mathbf{n}^{t}_{k}\,,
    \label{Eq:Noise_t}
\end{equation}
where 
$\mathbf{n}^{t}_{k} \sim \mathcal{N}_{\mathbb{C}} ( \mathbf{0}, \sigma_{\mathrm{dl}}^{2} \mathbf{I}_{L_k
} )$ represents the receiver noise.

\subsection{Differential Detection}
The code matrix $\mathbf{X}^{t}_{k}$ can be decoded by using two consecutive received blocks, i.e., $\mathbf{y}^{t}_{k}$ and $\mathbf{y}^{t-1}_{k}$, where $\mathbf{y}^{t-1}_{k}$ is given by $\mathbf{y}^{t-1}_{k} = \sum_{l=1}^{L} \mathrm{g}_{k,l}^{\mathrm{(ef)}} \big[ \mathbf{C}^{t-1}_{k} \big]_{m(l,k),:} + \widetilde{\mathbf{n}}^{t-1}_{k}$, and with $\widetilde{\mathbf{n}}^{t-1}_{k} = \sum_{i=1, i \neq k}^{K} \sum_{l=1}^{L} \widetilde{\mathrm{g}}_{i,k,l}^{\,\mathrm{(ef)}} \big[ \mathbf{C}^{t-1}_{i} \big]_{m(l,i),:} + \mathbf{n}^{t-1}_{k}$. The \ac{ML} detection of $\mathbf{X}^{t}_{k}$ corresponds to compute \cite{Larsson_Stoica_2003}
\begin{equation}
    \centering
    \hat{\mathbf{X}}^{t}_{k} = \mathop{\mathrm{arg\,min}}_{\mathbf{X} \in \mathcal{X}_{N}} \, -2\mathrm{Re} \left \{ \mathrm{tr} \left \{  \mathbf{X} \, \left (\mathbf{y}^{t}_{k} \right)^{\mathrm{H}} \mathbf{y}^{t-1}_{k} \right \} \right \} + \mathrm{const},
    \label{Eq:DifferentialDecodingTraceRealMin}
\end{equation}
or equivalently $\hat{\mathbf{X}}^{t}_{k} \!=\! \mathop{\mathrm{arg\,max}}_{\mathbf{X} \in \mathcal{X}_{N}} \, \mathrm{Re} \big \{ \mathrm{tr} \big \{  \mathbf{X} \, \big (\mathbf{y}^{t}_{k} \big)^{\mathrm{H}} \mathbf{y}^{t-1}_{k} \big \} \big \}$, where $\mathcal{X}_{N}$ represents the set of all possible code matrices for a given unitary constellation $\mathcal{S}$.
The detection of $\hat{\mathbf{X}}^{t}_{k}$ in \eqref{Eq:DifferentialDecodingTraceRealMin} can be computationally intensive as it involves an exhaustive search over all possible codewords $\mathbf{X} \in \mathcal{X}_{N}$. Nonetheless, the detection complexity can be significantly reduced if the codewords belonging to $\mathcal{X}_{N}$ are designed as orthogonal code matrices, as previously mentioned. This is because orthogonal code matrices allow the \ac{ML} detection of the symbols in $\mathbf{X}^{t}_{k}$ to be decoupled. Consequently, each symbol can be detected individually rather than jointly detecting the entire code matrix. To facilitate data detection, a generic codeword $\mathbf{X}$ is redefined as
\begin{equation}
    \centering
    \mathbf{X} = \frac{1}{\sqrt{n_s}}\sum\nolimits_{n=1}^{n_s} \bar{s}^{n} \mathbf{A}_n + j\tilde{s}^{n} \mathbf{B}_n,
    \label{Eq:DecomposedX}
\end{equation}
\noindent where the complex symbol $s^{n} \in \mathcal{S}$ is composed of its real and imaginary parts $\bar{s}^{n}$, $\tilde{s}^{n}$, such that $s^{n} = \bar{s}^{n} + j\tilde{s}^{n}$. Moreover, the matrices $\mathbf{A}_n \in \mathbb{Z}^{L_{k} \times L_{k}}$ and $\mathbf{B}_n  \in \mathbb{Z}^{L_{k} \times L_{k}}$ are referred to as \textit{amicable orthogonal designs}, which consist of a set of predefined matrices that satisfy the following conditions
\begin{align}
    \mathbf{A}_n \mathbf{A}_n^{\mathrm{H}} &= \mathbf{I}_{L_{k}},          & \mathbf{B}_n \mathbf{B}_n^{\mathrm{H}} &= \mathbf{I}_{L_{k}}, \nonumber \\
    \mathbf{A}_n \mathbf{A}_{n{'}}^{\mathrm{H}} &= -\mathbf{A}_{n{'}} \mathbf{A}_n^{\mathrm{H}}, & \mathbf{B}_n \mathbf{B}_{n{'}}^{\mathrm{H}} &= -\mathbf{B}_{n{'}} \mathbf{B}_n^{\mathrm{H}}, \quad n \neq n{'}, \nonumber \\
    \mathbf{A}_n \mathbf{B}_{n{'}}^{\mathrm{H}} &= \mathbf{B}_{n{'}} \mathbf{A}_n^{\mathrm{H}}.
    \label{Eq:amicable_conditions}
\end{align}
\noindent with $n$ and $n{'}$ being the indices that identify different matrices. The number of distinct matrices in sets $\{\mathbf{A}_n\}_{n=1}^{n_s}$ and $\{\mathbf{B}_n\}_{n=1}^{n_s}$ 
is $n_s$ for each set. For instance, for the Alamouti code ($n_s = 2$), these sets are defined as
\begin{equation}
\begin{aligned}
\mathbf{A}_1 &= \begin{bmatrix} 1 & \phantom{\:::}0 \\ 0 & -1 \end{bmatrix}, & \mathbf{B}_1 &= \begin{bmatrix} 1 & \phantom{\:::}0 \\ 0 & \phantom{\:::}1 \end{bmatrix}, \\
\mathbf{A}_2 &= \begin{bmatrix} 0 & \phantom{\:::}1 \\ 1 & \phantom{\:::}0 \end{bmatrix}, & \mathbf{B}_2 &= \begin{bmatrix} 0 & -1 \\ 1 & \phantom{\:::}0 \end{bmatrix}.
\end{aligned}
\end{equation}
For other code matrices, like the one in \eqref{Eq:codeMatrix_Nt_4}, there are additional unique sets of amicable orthogonal designs.\footnote{We omit details in this regard due to space constraints. The methodology for generating amicable orthogonal designs can be found in \cite{Larsson_Stoica_2003}.}

Once the set of amicable orthogonal designs is defined, we can substitute \eqref{Eq:DecomposedX} into \eqref{Eq:DifferentialDecodingTraceRealMin}. Then, we obtain a \ac{ML} criterion that can be minimized with respect to each symbol $s^{n}$ belonging to the unitary constellation $ \mathcal{S}$. Let $s^{\widetilde{m}(n,t)}_{k} \subset \mathcal{S}^{t}_{k}$ denote the $n$-th complex symbol among the $n_s$ symbols transmitted within the code matrix $\mathbf{X}^{t}_{k}$, for $n = 1, \ldots, n_s$. Specifically, $s^{\widetilde{m}(n,t)}_{k}$ is one of the symbols belonging to the subset $\mathcal{S}^{t}_{k}$, defined as $\mathcal{S}^{t}_{k} = { s^{(t-1)n_s + 1}_{k}, \ldots, s^{t n_s}_{k} }$. Here, $\widetilde{m}(n,t)$ is a mapping function that associates the value of $n$ to the corresponding position of $s^{\widetilde{m}(n,t)}_{k}$ in $\mathcal{S}^{t}_{k}$. For instance, $n=1$ indicates that $s^{\widetilde{m}(n,t)}_{k}$ corresponds to $s^{(t-1)n_s + 1}_{k}$, whereas $n=n_s$ refers to $s^{t n_s}_{k}$. Let us define
\begin{equation}
    \centering
    \mathbf{Y}^{t}_{k} = \left (\mathbf{y}^{t}_{k} \right)^{\mathrm{H}} \mathbf{y}^{t-1}_{k}\; \in \mathbb{C}^{L_{k} \times L_{k}}.
    \label{Eq:Y_ML_k}
\end{equation}
The \ac{ML} criterion used to detect $\hat{s}^{\widetilde{m}(n,t)}_{k}$ after substituting \eqref{Eq:DecomposedX} into \eqref{Eq:DifferentialDecodingTraceRealMin} is given by
\begin{equation}
\begin{aligned}
 \sum_{n=1}^{n_s} \big( -\mathrm{Re} \big \{ \mathrm{tr} \big \{ \mathbf{A}_n \, \mathbf{Y}^{t}_{k} \big \} \big \} \bar{s}^{n} + \mathrm{Im} \big \{ \mathrm{tr} \big \{ \mathbf{B}_n \, \mathbf{Y}^{t}_{k} \big \} \big \} \tilde{s}^{n} \big).
\end{aligned}
\label{Eq:ML_amicable}
\end{equation}
\noindent It can be noted that the \ac{ML} metric in \eqref{Eq:DifferentialDecodingTraceRealMin} decouples into a sum of $n_s$ terms in \eqref{Eq:ML_amicable}, each involving exactly one complex symbol. Consequently, the detection of $\hat{s}^{\widetilde{m}(n,t)}_{k}$ is decoupled from the detection of $\hat{s}^{\widetilde{m}(n{'},t)}_{k}$, for $n \neq n{'}$. Therefore, the complex symbol $\hat{s}^{\widetilde{m}(n,t)}_{k}$ can also be detected by
\begin{equation}
\begin{aligned}
 \mathop{\mathrm{arg\,min}}_{s^{n} \in \mathcal{S}} -\mathrm{Re} \big \{ \mathrm{tr} \big \{ \mathbf{A}_n \, \mathbf{Y}^{t}_{k} \big \} \big \} \bar{s}^{n} + \mathrm{Im} \big \{ \mathrm{tr} \big \{ \mathbf{B}_n \, \mathbf{Y}^{t}_{k} \big \} \big \} \tilde{s}^{n},
\end{aligned}
\label{Eq:ML_amicable2}
\end{equation}
\noindent where the detected symbol $\hat{s}^{\widetilde{m}(n,t)}_{k}$ is composed of its real and imaginary parts, $\bar{\hat{s}}^{\widetilde{m}(n,t)}_{k}$ and $\tilde{\hat{s}}^{\widetilde{m}(n,t)}_{k}$, respectively. Thus, the product $\mathrm{Re} \big \{ \mathrm{tr} \big \{ \mathbf{A}_n \, \mathbf{Y}^{t}_{k} \big \} \big \} \bar{s}^{n} $ is used to detect $\bar{\hat{s}}^{\widetilde{m}(n,t)}_{k}$, while $\mathrm{Im} \big \{ \mathrm{tr} \big \{ \mathbf{B}_n \, \mathbf{Y}^{t}_{k} \big \} \big \} \tilde{s}^{n}$ is utilized to detect $\tilde{\hat{s}}^{\widetilde{m}(n,t)}_{k}$. 

\subsection{Resilience to Phase Misalignment Effects and SNR}

Let us consider the decision statistic related to $\bar{s}^{n}$ in \eqref{Eq:ML_amicable2}, and let $\mathbf{DS}^{t}_{k} = \sum_{l=1}^{L} \mathrm{g}_{k,l}^{\mathrm{(ef)}} \big[ \mathbf{C}^{t}_{k} \big]_{m(l,k),:} \in \mathbb{C}^{1 \times L_k}$ denote the desired signal of \ac{UE} $k$ for the received signal block $\mathbf{y}^{t}_{k}$, where the last equality, i.e., $\big[ \mathbf{C}^{t}_{k} \big]_{m(l,k),:} = \big[ \mathbf{C}^{t-1}_{k} \big]_{m(l,k),:} \mathbf{X}^{t}_{k}$, comes from \eqref{Eq:EncodingMatrixCellFree}. 
Then, by considering the single-user case, we can write $\mathbf{y}^{t}_{k}$ and $\mathbf{y}^{t-1}$ as $\mathbf{y}^{t}_{k} = \mathbf{DS}^{t}_{k} + \mathbf{n}_{k}^{t}$ and $\mathbf{y}^{t-1}_{k} = \mathbf{DS}^{t-1}_{k} + \mathbf{n}_{k}^{t-1}$.
Recall from \eqref{Eq:Y_ML_k} that $\mathbf{Y}^{t}_{k} = \left (\mathbf{y}^{t}_{k} \right)^{\mathrm{H}} \mathbf{y}^{t-1}_{k}$. By letting $\mathrm{ML}^{t}_{k, n}$ denote $\mathrm{tr} \big \{ \mathbf{A}_n \, \mathbf{Y}^{t}_{k} \big \}$ 
in \eqref{Eq:ML_amicable2}, we obtain an expression consisting of four distinct terms, i.e.,
\begin{equation}
    \centering
    \mathrm{ML}^{t}_{k, n} = \mathrm{ML}^{t}_{k, n, 1} + \mathrm{ML}^{t}_{k, n, 2} + \mathrm{ML}^{t}_{k, n, 3} + \mathrm{ML}^{t}_{k, n, 4},
    \label{Eq:ML_k_t}
\end{equation}
where $\mathrm{ML}^{t}_{k, n, 1} \!=\! \mathbf{A}_n \big( \mathbf{DS}^{t}_{k} \big)^{\mathrm{H}} \mathbf{DS}^{t-1}_{k}$. Thus,
\begin{equation}
      \!\!\mathrm{ML}^{t}_{k, n, 1} \!=\!\! \sum_{l = 1}^{L} \sum_{l{'} = 1}^{L} \! \mathrm{tr} \big \{\mathrm{g}_{k,l}^{{\mathrm{(ef)}}\mathrm{H}} \mathrm{g}_{k,l{'}}^{({\rm ef})} \mathbf{A}_n \big( \mathbf{X}^{t}_{k} \big)^{\mathrm{H}}\mathbf{C}^{t-1, t-1}_{k, m, m{'}, l, l{'}} \big \},\!
    \label{Eq:ML_k_t_1}
\end{equation}
\noindent with $\mathbf{C}^{t-1, t-1}_{k, m, m{'}, l, l{'}} = \big[ \mathbf{C}^{t-1}_{k} \big]^{\mathrm{H}}_{m(l,k),:} \big[ \mathbf{C}^{t-1}_{k} \big]_{m{'}(l{'},k),:}$. The term $\mathrm{ML}^{t}_{k, n, 2} = \mathbf{A}_n \big( \mathbf{n}_{k}^{t} \big)^{\mathrm{H}} \mathbf{DS}^{t-1}_{k}$ is obtained as
\begin{equation}
    \mathrm{ML}^{t}_{k, n, 2} = \sum\nolimits_{l = 1}^{L}\mathrm{tr} \big \{\mathrm{g}_{k,l}^{({\rm ef})} \mathbf{A}_n \big( \mathbf{n}^{t}_{k} \big)^{\mathrm{H}} \big[\mathbf{C}^{t-1}_{k} \big]_{m(l,k),:} \big \},
    \label{ML_k_2}
\end{equation}
\noindent whereas $\mathrm{ML}^{t}_{k, n, 3} = \mathbf{A}_n \big( \mathbf{DS}^{t}_{k} \big)^{\mathrm{H}} \mathbf{n}_{k}^{t-1}$ is given by
\begin{equation}
    \mathrm{ML}^{t}_{k, n, 3} = \sum\limits_{l = 1}^{L}\mathrm{tr} \big \{\mathrm{g}_{k,l}^{{\mathrm{(ef)}}\mathrm{H}} \mathbf{A}_n \big[ \mathbf{C}^{t}_{k} \big]^{\mathrm{H}}_{m(l,k),:} \mathbf{n}^{t-1}_{k}\big \},
    \label{Eq:ML_k_3_t}
\end{equation}
\noindent with $\big[ \mathbf{C}^{t}_{k} \big]_{m(l,k),:} = \big[ \mathbf{C}^{t-1}_{k} \big]_{m(l,k),:} \mathbf{X}^{t}_{k}$, which can also be regarded as $\mathbf{c}^{t}_{k,l}$ in \eqref{Eq:EncodingMatrixCellFree}. The term $\mathrm{ML}^{t}_{k, n, 4}$ is calculated as $\mathrm{ML}^{t}_{k, n, 4} = \mathrm{tr} \big\{ \mathbf{A}_n \left( \mathbf{n}^{t}_{k} \right)^{\mathrm{H}} \mathbf{n}^{t-1}_{k}  \big\}$. However, $\mathrm{ML}^{t}_{k, n, 4}$ can be neglected if the noise variance $\sigma_{\mathrm{dl}}^{2}$ is small enough.

To evaluate the \ac{SNR} and assess the robustness of the proposed \ac{CF-mMIMO} system using \ac{DSTBC} schemes against phase misalignments, we begin by solving $\mathrm{ML}^{t}_{k, n, 1}$. Hence, the code matrix $\big(\mathbf{X}^{t}_{k} \big)^{\mathrm{H}}$ in \eqref{Eq:ML_k_t_1} is defined as
\begin{equation}
    \centering
    \!\left(\mathbf{X}^{t}_{k} \right)^{\mathrm{H}} \!=\! \frac{1}{\sqrt{n_s}}\sum_{n{'}=1}^{n_s} \bar{s}^{\widetilde{m}(n{'},t)}_{k} \mathbf{A}_{n{'}}^{\mathrm{H}} - j\tilde{s}^{\widetilde{m}(n{'},t)}_{k} \mathbf{B}_{n{'}}^{\mathrm{H}}.
    \label{Eq:DecomposedX_t_k}
\end{equation}
In the following, we split the computation of \eqref{Eq:ML_k_t_1} into two steps. First, we evaluate \eqref{Eq:ML_k_t_1} by considering the product $\mathbf{A}_n \, \big( \mathbf{X}^{t}_{k} \big)^{\mathrm{H}}$ without including the imaginary components of \eqref{Eq:DecomposedX_t_k}, i.e., without the term $-j\tilde{s}^{\widetilde{m}(n{'},t)}_{k} \mathbf{B}_{n{'}}^{\mathrm{H}}$. Second, we indicate how to calculate the contributions of the imaginary terms of $\left(\mathbf{X}^{t}_{k} \right)^{\mathrm{H}}$ to $\mathrm{ML}^{t}_{k, n, 1}$ in \eqref{Eq:ML_k_t_1}, as shown in Appendix A. To compute \eqref{Eq:ML_k_t_1} without the imaginary components of \eqref{Eq:DecomposedX_t_k}, we define $\big(\mathbf{\overline{X}}^{t}_{k, n} \big)^{\mathrm{H}} = \mathrm{Re} \big \{ \mathbf{A}_n \, \big( \mathbf{X}^{t}_{k} \big)^{\mathrm{H}} \big \}$, which is given by
\begin{equation}
    \centering
     \big(\mathbf{\overline{X}}^{t}_{k, n} \big)^{\mathrm{H}} = \frac{1}{\sqrt{n_s}}  \big(\bar{s}^{\widetilde{m}(n,t)}_{k} \mathbf{I}_{L_{k}} + \mathbf{E}^{t}_{k, n} \big)\,,
    \label{Eq:DecomposedX_t_k_simplified}
\end{equation}
where $\mathbf{E}^{t}_{k, n}$ is obtained as follows
\begin{equation}
    \centering
    \mathbf{E}^{t}_{k, n} = \sum\limits_{n{'}=1, n{'} \neq n}^{n_s} \bar{s}^{\widetilde{m}(n{'},t)}_{k} \mathbf{A}_n \mathbf{A}_{n{'}}^{\mathrm{H}}.
    \label{Eq:E_j_j_t_k}
\end{equation}
\noindent Given that \eqref{Eq:DecomposedX_t_k_simplified} contains only real-valued terms, then
\begin{equation}
      \!\overline{\mathrm{ML}}^{t}_{k, n, 1} \!=\! \sum_{l = 1}^{L} \sum_{l{'} = 1}^{L} \!\mathrm{tr} \Big \{\mathrm{g}_{k,l}^{{\mathrm{(ef)}}\mathrm{H}} \mathrm{g}_{k,l{'}}^{({\rm ef})} \big(\mathbf{\overline{X}}^{t}_{k, n} \big)^{\mathrm{H}} \mathbf{C}^{t-1, t-1}_{k, m, m{'}, l, l{'}} \Big \}.\!
    \label{Eq:ML_k_t_1_real}
\end{equation}
Then, we split the cross-product terms in \eqref{Eq:ML_k_t_1_real} into two cases: $m(l,k) = m{'}(l{'},k)$ and $m(l,k) \neq m{'}(l{'},k)$. However, since \eqref{Eq:DecomposedX_t_k_simplified} consists of two terms, i.e., $\bar{s}^{\widetilde{m}(n,t)}_{k} \mathbf{I}_{L_{k}}/ \sqrt{n_s}$ and $\mathbf{E}^{t}_{k, n}/\sqrt{n_s}$, each case yields two contributions to $\overline{\mathrm{ML}}^{t}_{k, n, 1}$, resulting in a total of four separate calculations. Let us begin by computing the contribution of $\bar{s}^{\widetilde{m}(n,t)}_{k} \mathbf{I}_{L_{k}}/ \sqrt{n_s}$ to \eqref{Eq:ML_k_t_1_real} for the case $m(l,k) = m{'}(l{'},k)$. We denote this specific calculation as $\overline{\mathrm{ML}}^{t}_{k, n, 1, 1}$. By substituting $\bar{s}^{\widetilde{m}(n,t)}_{k} \mathbf{I}_{L_{k}}/ \sqrt{n_s}$ in \eqref{Eq:ML_k_t_1_real}, the expression for $\overline{\mathrm{ML}}^{t}_{k, n, 1, 1}$ is obtained as
\begin{equation}
     \overline{\mathrm{ML}}^{t}_{k, n, 1, 1} = \frac{\bar{s}^{\widetilde{m}(n,t)}_{k}}{\sqrt{n_s}} \sum_{l = 1}^{L}  \mathrm{g}_{k,l}^{{\mathrm{(ef)}}\,\mathrm{H}} \mathrm{g}_{k,l}^{({\rm ef})}\mathrm{tr} \big \{ \mathbf{C}^{t-1, t-1}_{k,m,m,l,l}\big \}.
     \label{Eq:ML_k_t_sum1}
\end{equation}
\noindent Recall that $\big[ \mathbf{C}^{t-1}_{k} \big]_{m{'}(l{'},k),:}$ and $ \big[ \mathbf{C}^{t-1}_{k} \big]_{m(l,k),:}$ are row vectors. Given that $\mathbf{C}^{t-1, t-1}_{k, m, m{'}, l, l{'}}=\big[ \mathbf{C}^{t-1}_{k} \big]^{\mathrm{H}}_{m(l,k),:} \big[ \mathbf{C}^{t-1}_{k} \big]_{m{'}(l{'},k),:}$, and by applying the cyclic property of the trace, we obtain $\mathrm{tr} \big \{ \mathbf{C}^{t-1, t-1}_{k, m, m{'}, l, l{'}} \big \} = \big[ \mathbf{C}^{t-1}_{k} \big]_{m{'}(l{'},k),:} \big[ \mathbf{C}^{t-1}_{k} \big]^{\mathrm{H}}_{m(l,k),:}$. Besides, since the information signal $\mathbf{C}^{t-1}_{k}$ is assumed to be unitary and orthogonal, the following property holds
\begin{equation}
\centering
\mathrm{tr} \big \{ \mathbf{C}^{t-1, t-1}_{k, m, m{'}, l, l{'}} \big \} = \begin{cases}
1 & \text{ for } m(l,k) = m{'}(l{'},k)\,,\\
0 & \text{ for } m(l,k) \neq m{'}(l{'},k)\,.
\end{cases}
\label{Eq:TracesProperties}
\end{equation}
\noindent Consequently, \eqref{Eq:ML_k_t_sum1} simplifies to
\begin{equation}
     \overline{\mathrm{ML}}^{t}_{k, n, 1, 1} = \frac{1}{\sqrt{n_s}}\bar{s}^{\widetilde{m}(n,t)}_{k}  \sum\nolimits_{l = 1}^{L} \big |\mathrm{g}_{k,l}^{{\mathrm{(ef)}}} \big|^{2}  \Bigg.
     \label{Eq:ML_k_t_sum2}.
\end{equation}
Then, we compute the contribution of $\bar{s}^{\widetilde{m}(n,t)}_{k} \mathbf{I}_{L_{k}}/ \sqrt{n_s}$ to \eqref{Eq:ML_k_t_1_real} in the case where $m(l,k) \neq m{'}(l{'},k)$. Let $\overline{\mathrm{ML}}^{t}_{k, n, 1, 3}$ denote this calculation. Accordingly, $\overline{\mathrm{ML}}^{t}_{k, n, 1, 3}$ is given by
\begin{equation}
      \frac{\bar{s}^{\widetilde{m}(n,t)}_{k}}{\sqrt{n_s}} \sum_{l = 1}^{L} \sum_{l{'} = 1, l{'}\neq l}^{L} \mathrm{g}_{k,l}^{{\mathrm{(ef)}}\,\mathrm{H}} \mathrm{g}_{k,l{'}}^{({\rm ef})} \mathrm{tr} \Big \{\mathbf{C}^{t-1, t-1}_{k, m, m{'}, l, l{'}} \Big \} = 0,
    \label{Eq:ML_k_t_4}
\end{equation}
\noindent as $\mathrm{tr} \big \{\mathbf{C}^{t-1, t-1}_{k, m, m{'}, l, l{'}} \big \} = 0$, for $m(l,k) \neq m{'}(l{'},k)$.

\begin{lemma} \label{lemma:ML}
 The contributions of the remaining components, i.e., $\mathbf{E}^{t}_{k, n}$ and $-j\tilde{s}^{\widetilde{m}(n{'},t)}_{k} \mathbf{B}_{n{'}}^{\mathrm{H}}$, to the calculation of $\mathrm{ML}^{t}_{k, n, 1}$ vanish when applying the real-part operation in \eqref{Eq:ML_amicable2}. Thus,
\begin{equation}
     \mathrm{Re}\{\mathrm{ML}^{t}_{k, n, 1}\} = \overline{\mathrm{ML}}^{t}_{k, n, 1, 1}.
\end{equation}   
\end{lemma}
\begin{IEEEproof}
    The proof readily follows by applying the properties of amicable orthogonal designs; details are omitted due to space limitations.
\end{IEEEproof}

Note that the effects of phase misalignment are mitigated in \eqref{Eq:ML_k_t_sum2} as it depends solely on the absolute value of $\mathrm{g}_{k,l}^{\mathrm{(ef)}}$. In addition, \eqref{Eq:ML_k_t_sum2} reveals that it is possible to mitigate such undesired effects in \ac{CF-mMIMO} systems without assuming perfect detection of the code matrix. This contrasts with \cite{freitas2025PMCellFree}, which relies on perfect detection to demonstrate the effectiveness of \ac{DSTBC} techniques in \ac{CF-mMIMO} systems.

\begin{proposition} \label{prop:SNR}
The \ac{SNR} of \ac{UE} $k$ can be expressed as
\begin{equation}
    \mathrm{SNR}_k \approx \frac{\mathbb{E}_{\mathbf{C}} \big \{ | \mathrm{ML}^{t}_{k, n, 1} |^{2}\big \}}{\mathbb{E}_{\mathbf{C}} \big \{ | \mathrm{ML}^{t}_{k, n, 2} |^{2}\big \} + \mathbb{E}_{\mathbf{C}} \big \{ | \mathrm{ML}^{t}_{k, n, 3} |^{2}\big \}}\,,
\end{equation} 
which reduces to computing
\begin{equation}
    \centering
     \mathrm{SNR}_k \approx \frac{\sum_{l = 1}^{L} \big |\mathrm{g}_{k,l}^{{\mathrm{(ef)}}} \big|^{2}}{2n_s\sigma_{\mathrm{dl}}^{2}},
    \label{Eq:SNR_complex_2} 
\end{equation}
where the operator $\mathbb{E}_{\mathbf{C}}$ indicates that the expectations are taken with respect to the information signals $\mathbf{C}^{t}_{k}$.    
\end{proposition}
\begin{IEEEproof}
    See Appendix~\ref{subsec:proof:proposition:SNR}.
\end{IEEEproof}

\subsection{SINR Expression}

The procedure for deriving the \ac{SINR} closely follows the approach used for the \ac{SNR} derivation. However, to accurately account for multi-user interference, the received signal block $\mathbf{y}^{t}_{k}$ must be modeled according to \eqref{Eq:ReceivedSignalCellFreeSimplified}, rather than assuming only the desired signal and noise components. Accordingly, to compute $\mathbf{Y}^{t}_{k} = \left (\mathbf{y}^{t}_{k} \right)^{\mathrm{H}} \mathbf{y}^{t-1}_{k}$ in \eqref{Eq:Y_ML_k}, $\mathbf{y}^{t-1}_{k}$ is modeled as
\begin{equation}
    \centering
    \mathbf{y}^{t-1}_{k} = \sum\nolimits_{l=1}^{L} \mathrm{g}_{k,l}^{\mathrm{(ef)}} \big[ \mathbf{C}^{t-1}_{k} \big]_{m(l,k),:} +  \widetilde{\mathbf{n}}_{k}^{t-1},
 \label{Eq:ReceivedSignalCellFreeSimplified_minus}
\end{equation}
\noindent where the combined effects of interference and receiver noise due to the received signal block $\mathbf{y}^{t-1}_{k}$ is given by
\begin{equation}
    \centering
     \widetilde{\mathbf{n}}_{k}^{t-1} = \sum\nolimits_{i=1, i \neq k}^{K} \sum\nolimits_{l=1}^{L} \widetilde{\mathrm{g}}_{i,k,l}^{\,\mathrm{(ef)}} \big[ \mathbf{C}^{t-1}_{i} \big]_{m(l,i),:} + \mathbf{n}^{t-1}_{k}
    \label{Eq:Noise_t_minus}.
\end{equation}
Recall that $\mathbf{DS}^{t}_{k} = \sum_{l=1}^{L} \mathrm{g}_{k,l}^{\mathrm{(ef)}} \big[ \mathbf{C}^{t}_{k} \big]_{m(l,k),:} \in \mathbb{C}^{1 \times L_k}$ denote the desired signal of \ac{UE} $k$ for the received signal block $\mathbf{y}^{t}_{k}$, where the last equality, i.e., $\big[ \mathbf{C}^{t}_{k} \big]_{m(l,k),:} = \big[ \mathbf{C}^{t-1}_{k} \big]_{m(l,k),:} \mathbf{X}^{t}_{k}$, comes from \eqref{Eq:EncodingMatrixCellFree}. 
Thus, we can define the received signal blocks $\mathbf{y}^{t}_{k}$ and $\mathbf{y}^{t-1}_{k}$ as $\mathbf{y}^{t}_{k} = \mathbf{DS}^{t}_{k} + \widetilde{\mathbf{n}}_{k}^{t}$ and $\mathbf{y}^{t-1}_{k} = \mathbf{DS}^{t-1}_{k} + \widetilde{\mathbf{n}}_{k}^{t-1}$, with $\widetilde{\mathbf{n}}_{k}^{t}$ being calculated in \eqref{Eq:Noise_t}. Let $\mathrm{MLI}^{t}_{k, n}$ denote the operation $\mathrm{tr} \big \{ \mathbf{A}_n \, \mathbf{Y}^{t}_{k} \big \}$ in \eqref{Eq:ML_amicable}. Then, we obtain
\begin{equation}
    \centering
    \mathrm{MLI}^{t}_{k, n} = \mathrm{MLI}^{t}_{k, n, 1} + \mathrm{MLI}^{t}_{k, n, 2} + \mathrm{MLI}^{t}_{k, n, 3} + \mathrm{MLI}^{t}_{k, n, 4},
    \label{Eq:MLI_k_t}
\end{equation}
\noindent with $\mathrm{MLI}^{t}_{k, n, 1} = \mathrm{tr} \big \{\mathbf{A}_n(\mathbf{DS}^{t}_{k})^{\mathrm{H}}\mathbf{DS}^{t-1}_{k}\big \}$ representing the desired signal components and the remaining terms denoting the cross-products that are obtained as $\mathrm{MLI}^{t}_{k, n, 2} = \mathrm{tr} \big \{\mathbf{A}_n(\mathbf{DS}^{t}_{k})^{\mathrm{H}} \widetilde{\mathbf{n}}^{t-1}_{k} \big \}$, $\mathrm{MLI}^{t}_{k, n, 3} = \mathrm{tr} \big \{\mathbf{A}_n( \widetilde{\mathbf{n}}_{k}^{t})^{\mathrm{H}} \mathbf{DS}^{t-1}_{k}\big \}$, and $\mathrm{MLI}^{t}_{k, n, 4}=\mathrm{tr} \big \{\mathbf{A}_n( \widetilde{\mathbf{n}}_{k}^{t})^{\mathrm{H}} \widetilde{\mathbf{n}}^{t-1}_{k}\big \}$. Since the terms in $\mathrm{MLI}^{t}_{k, n}$ contain different numbers of components, the \ac{SINR} is computed based on their individual power contributions.

\begin{proposition} \label{prop:SINR-closed-form}
In \ac{CF-mMIMO} systems employing \ac{DSTBC} techniques, an expression for the \ac{DL} \ac{SINR} at \ac{UE} $k$ can be given in \eqref{Eq:SINR_inst_DSTBC}, and in \eqref{Eq:SINR_inst_DSTBC_2} in closed form. The last term in the denominator of \eqref{Eq:SINR_inst_DSTBC_2} is computed in closed form as in \eqref{Eq:MLI_k_4_4_mean_1}. The expressions \eqref{Eq:SINR_inst_DSTBC}--\eqref{Eq:MLI_k_4_4_mean_1} are shown on top of the next page.   
\end{proposition}
\begin{IEEEproof}
    See Appendix~\ref{subsec:proof:proposition:SINR-closed-form}.
\end{IEEEproof}
\begin{figure*}[tb!]
\begin{equation}
    \mathrm{SINR}^{(\mathrm{dl, 3})}_{k} \!=\! \frac{\mathbb{E}_{\mathbf{C}} \big \{ | \mathrm{MLI}^{t}_{k, n, 1} |^{2}\big \}}{ \sum\nolimits_{z{'}=1}^{2}\mathbb{E}_{\mathbf{C}} \big \{ | \mathrm{MLI}^{t}_{k, n, 2, z{'}} |^{2}\big \} \!+\! \sum\nolimits_{z{'}=1}^{2} \mathbb{E}_{\mathbf{C}} \big \{ | \mathrm{MLI}^{t}_{k, n, 3, z{'}} |^{2}\big \} \!+\! \sum\nolimits_{z{'}=1}^{4}\mathbb{E}_{\mathbf{C}} \big \{ | \mathrm{MLI}^{t}_{k, n, 4, z{'}} |^{2}\big \}}
    \label{Eq:SINR_inst_DSTBC}
\end{equation}
\vspace{-0.8em}
\hrulefill
\end{figure*}

\begin{figure*}[tb!]
\begin{equation}
    \mathrm{SINR}^{(\mathrm{dl, 3})}_{k} \approx \frac{\Big ( \sum_{l = 1}^{L} \big |\mathrm{g}_{k,l}^{{\mathrm{(ef)}}} \big|^{2}/\sqrt{n_s}\Big)^2}{2 \bigg( \sigma_{\mathrm{dl}}^{2} \sum\limits_{l = 1}^{L}  \big| \mathrm{g}_{k,l}^{({\rm ef})}\big|^{2} + \sigma_{\mathrm{dl}}^{2}\sum\limits_{i=1, i \neq k}^{K} \sum\limits_{l=1}^{L} \big|\widetilde{\mathrm{g}}_{i,k,l}^{\,\mathrm{(ef)}}\big|^{2} + \sum\limits_{i=1, i \neq k}^{K} \sum\limits_{l=1}^{L} \sum\limits_{r=1}^{L} \frac{1}{L_k} \big|\mathrm{g}_{k,l}^{\mathrm{(ef)}}\big|^{2} \big|\widetilde{\mathrm{g}}_{i,k,r}^{\,\mathrm{(ef)}}\big|^{2}\bigg) + \mathbb{E}_{\mathbf{C}} \big \{ | \mathrm{MLI}^{t}_{k, n, 4, 4} |^{2}\big \}}
    \label{Eq:SINR_inst_DSTBC_2}
\end{equation}
\vspace{-0.8em}
\hrulefill
\end{figure*}
\begin{figure*}[tb!]
\begin{equation}
    \mathbb{E}_{\mathbf{C}} \big\{ | \mathrm{MLI}^{t}_{k, n, 4, 4} |^{2} \big\} \approx \sum_{\substack{i = 1 \\ i \neq k}}^{K} \Bigg( \sum_{l = 1}^{L} \frac{\big|\widetilde{\mathrm{g}}_{i,k,l}^{\,\mathrm{(ef)}}\big|^{4}}{n_s}  + \sum_{l = 1}^{L} \sum_{\substack{r = 1 \\ r \neq l}}^{L} \big| \widetilde{\mathrm{g}}_{i,k,l}^{\,\mathrm{(ef)}} \, \widetilde{\mathrm{g}}_{i,k,r}^{\,\mathrm{(ef)}} \big|^{2} \bigg( \frac{1}{L_k} + \widetilde{L}_k \bigg)  \Bigg) + \sum_{\substack{i = 1 \\ i\neq k}}^{K} \sum_{\substack{v = 1 \\ v \neq k, v \neq i}}^{K} \sum_{l = 1}^{L} \sum_{r = 1}^{L} \frac{\big| \widetilde{\mathrm{g}}_{i,k,l}^{\,\mathrm{(ef)}} \, \widetilde{\mathrm{g}}_{v,k,r}^{\mathrm{(ef)}} \big|^{2}}{L_k}
    \label{Eq:MLI_k_4_4_mean_1}
\end{equation}
\vspace{-0.8em}
\hrulefill
\end{figure*}

{\color{black}
\subsection{Complexity Analysis and System Overhead}

The proposed scheme introduces additional computational requirements at both the CPU and the UE due to differential encoding and data detection. Specifically, computing \eqref{Eq:EncodingMatrix} at the CPU requires the multiplication of two $L_k \times L_k$ matrices, resulting in $L_k^3$ complex multiplications. However, since $L_k$ is typically small (e.g., $L_k \leq 4$), this additional computational cost is negligible compared to other signal processing tasks, such as precoding and channel estimation.

At the UE, the main operations associated with differential data detection are: (i) computing the product $(\mathbf{y}^t_{k,j})^{\mathrm{H}} \mathbf{y}^{t-1}_{k,j}$ in \eqref{Eq:Y_ML_k}, requiring $L_k^2$ complex multiplications; (ii) evaluating \eqref{Eq:ML_amicable2}, which involves two $L_k \times L_k$ matrix multiplications per detected symbol, resulting in a total of $2 n_s L_k^3$ additional complex multiplications at the receiver; and (iii) a codebook search with complexity $\mathcal{O}(N_s M_o)$, where $M_o$ is the modulation order. Overall, the additional computational cost remains moderate due to the small size of $L_k$, making the proposed scheme suitable for practical implementations. In addition, the proposed approach eliminates the need for CSI acquisition and hardware calibration at the UE, thereby reducing signaling overhead and implementation complexity.
}

{\color{black}
\section{Differential Quasi-Orthogonal Space-Time Block Coding and Time-Varying Phase Noise} 
\label{sec:DQO_STBC}


DSTBC schemes can mitigate phase misalignments in UC CF-mMIMO systems. However, for more than two serving APs (i.e., $L_k > 2$), their performance is limited by the code rate imposed by the orthogonality requirement of the space-time codewords. In this context, \ac{DQO}-STBC schemes provide an alternative to this limitation. By relaxing the orthogonality constraint, they can enable full code rates for $L_k > 2$. The DQ-OSTBC scheme considered in this work is based on~\cite{DQoSTBC2007}, where it was originally derived for four transmit antennas, and is here adapted to $L_k = 4$ serving APs. These schemes follow a different procedure from that used with orthogonal code matrices for both data encoding and data detection. Specifically, the data symbols are mapped onto QO-STBC code matrices, and both data encoding and detection are adapted to account for the quasi-orthogonality of these matrices. Moreover, for notational convenience, each code matrix is indexed according to the time index of the last symbol it spans. The overall procedure is divided into three stages, which are described next.

\subsection{First Step: Initialization}
During the initialization phase, the serving APs transmit to UE $k$, at time epoch $4\tilde{t}$, an encoded QO code matrix $\mathbf{C}_{k}^{4\tilde{t}} \in \mathbb{C}^{4 \times 4}$ that does not carry any information. The CPU generates its first row as $\big[ \mathbf{C}^{4\tilde{t}}_{k} \big]_{1,:} = [\mathrm{c}_{k,1}^{4\tilde{t}}, \, \mathrm{c}_{k,2}^{4\tilde{t}}, \, \mathrm{c}_{k,3}^{4\tilde{t}}, \, \mathrm{c}_{k,4}^{4\tilde{t}}]$, from which $\mathbf{C}_{k}^{4\tilde{t}}$ is constructed as
\cite{quasiOrthogonaJafarkhanil, DQoSTBC2007}
\begin{equation}
\mathbf{C}_{k}^{4\tilde{t}}
=
\begin{bmatrix}
\mathrm{c}_{k,1}^{4\tilde{t}} & \mathrm{c}_{k,2}^{4\tilde{t}} & \mathrm{c}_{k,3}^{4\tilde{t}} & \mathrm{c}_{k,4}^{4\tilde{t}} \\[0.3em]
- \big(\mathrm{c}_{k,2}^{4\tilde{t}}\big)^{*} & \big(\mathrm{c}_{k,1}^{4\tilde{t}}\big)^{*} & - \big(\mathrm{c}_{k,4}^{4\tilde{t}}\big)^{*} & \big(\mathrm{c}_{k,3}^{4\tilde{t}}\big)^{*} \\[0.3em]
- \big(\mathrm{c}_{k,3}^{4\tilde{t}}\big)^{*} & - \big(\mathrm{c}_{k,4}^{4\tilde{t}}\big)^{*} & \big(\mathrm{c}_{k,1}^{4\tilde{t}}\big)^{*} & \big(\mathrm{c}_{k,2}^{4\tilde{t}}\big)^{*} \\[0.3em]
\mathrm{c}_{k,4}^{4\tilde{t}} & - \mathrm{c}_{k,3}^{4\tilde{t}} & - \mathrm{c}_{k,2}^{4\tilde{t}} & \mathrm{c}_{k,1}^{4\tilde{t}}
\end{bmatrix}.
\end{equation}
\noindent Hence, the received signal at UE $k$ is given by
\begin{equation}
     \mathbf{y}_{k}^{4\tilde{t}} = \sum\nolimits_{l=1}^{L} \mathrm{g}_{k,l}^{\mathrm{(ef)}} \big[ \mathbf{C}^{4\tilde{t}}_{k} \big]_{m(l,k),:}^{\mathrm{T}} +  \widetilde{\mathbf{n}}_{k}^{4\tilde{t}},
     \label{eq:ReceivedSignalVector}
\end{equation}
where all interference and additive noise affecting UE $k$ are grouped into $\widetilde{\mathbf{n}}_{k}^{4\tilde{t}} \sim \mathcal{N}_{\mathbb{C}} ( \mathbf{0}, \sigma_{\mathrm{dl}}^{2} \mathbf{I}_{4} ) $. From $\mathbf{y}_{k}^{4\tilde{t}} \in \mathbb{C}^{1 \times 4}$, the UE generates the matrix $\mathbf{Y}_{k}^{4\tilde{t}} \in \mathbb{C}^{4 \times 4}$, which will later be used for data detection. Given that $\mathbf{y}_{k}^{4\tilde{t}} = [y_{k}^{4\tilde{t}-3}, \, y_{k}^{4\tilde{t}-2}, \, y_{k}^{4\tilde{t}-1}, \, y_{k}^{4\tilde{t}}]$, $\mathbf{Y}_{k}^{4\tilde{t}}$ is generated at UE $k$ as
\begin{equation}
\mathbf{Y}_{k}^{4\tilde{t}}
=
\begin{bmatrix}
y_{k}^{4\tilde{t}-3} & (y_{k}^{4\tilde{t}-2})^{*}   & (y_{k}^{4\tilde{t}-1})^{*}   & y_{k}^{4\tilde{t}} \\
y_{k}^{4\tilde{t}-2} & - (y_{k}^{4\tilde{t}-3})^{*} & (y_{k}^{4\tilde{t}})^{*}     & - y_{k}^{4\tilde{t}-1} \\
y_{k}^{4\tilde{t}-1} & (y_{k}^{4\tilde{t}})^{*}     & - (y_{k}^{4\tilde{t}-3})^{*} & - y_{k}^{4\tilde{t}-2} \\
y_{k}^{4\tilde{t}}   & - (y_{k}^{4\tilde{t}-1})^{*} & - (y_{k}^{4\tilde{t}-2})^{*} & y_{k}^{4\tilde{t}-3}
\end{bmatrix}.
\label{eq:Y_k_4t}
\end{equation}

\subsection{Second Step: Differential encoding}
Let $\mathcal{\widetilde{S}}_k^{4\tilde{t}+4} = \{s_{k,1}^{4\tilde{t}+4} \,,\, s_{k,2}^{4\tilde{t}+4}  \,,\, \tilde{s}_{k,3}^{4\tilde{t}+4}  \,,\, \tilde{s}_{k,4}^{4\tilde{t}+4}\}$ denote the set of symbols intended for UE $k$ at time epoch $4\tilde{t}+4$. The first distinction regarding orthogonal space-time code matrices lies in the selection of symbols. Specifically, the data symbols to be transmitted to the UE $k$ are selected pairwise from distinct constellations. That is, $s_{k,1}^{4\tilde{t}+4}$ and $s_{k,2}^{4\tilde{t}+4}$ are obtained from constellation $\mathcal{S}_1$, whereas $\tilde{s}_{k,3}^{4\tilde{t}+4}$ and $\tilde{s}_{k,4}^{4\tilde{t}+4}$ are drawn from constellation $\mathcal{S}_2$. Furthermore, the constellation $\mathcal{S}_1$ is rotated by an angle\footnote{\textcolor{black}{This design choice is motivated by the code's quasi-orthogonal structure. In particular, if a single constellation were used, certain symbol combinations (in quite rare circumstances) could lead to degenerate cases where the encoded signals become zero, which would propagate through the differential encoding process and severely degrade performance \cite{DQoSTBC2007}.}} $\phi$ with respect to $\mathcal{S}_2$, which is calculated as $\phi = \pi/M_o$ for \ac{PSK} modulation.

Once $\mathcal{\widetilde{S}}_k^{4\tilde{t}+4}$ is defined, the CPU generates the symbol vector $\mathbf{s}_{k}^{4\tilde{t}+4} = [s_{k,1}^{4\tilde{t}+4}, \, s_{k,2}^{4\tilde{t}+4}, \, \tilde{s}_{k,3}^{4\tilde{t}+4}, \, \tilde{s}_{k,4}^{4\tilde{t}+4}]$. In the following, $\mathbf{s}_{k}^{4\tilde{t}+4}$ is differentially encoded as
\begin{equation}
    [\mathbf{C}_{k}^{4\tilde{t}+4}]_{1,:} = \mathbf{s}_{k}^{4\tilde{t}+4} \; \mathbf{C}_{k}^{4\tilde{t}}/\widetilde{C}_{k}^{4\tilde{t}},
    \label{eq:EncodingDQ_OSTBC}
\end{equation}
where the normalization by $\widetilde{C}_{k}^{4\tilde{t}} = \sqrt{[\mathbf{C}_{k}^{4\tilde{t}}]_{1,:} \, [\mathbf{C}_{k}^{4\tilde{t}}]_{1,:}^{\mathrm{H}}}$ is required to avoid large peak power variations in the transmitted signals. Moreover, the symbols in $\mathcal{\widetilde{S}}_k^{4\tilde{t}+4}$ must be normalized to ensure constant average transmit power, which can be achieved by properly scaling those drawn from the constellations $\mathcal{S}_1$ and $\mathcal{S}_2$. The remaining rows of $\mathbf{C}_{k}^{4\tilde{t}+4}$ are then generated at the CPU following the same structure as $\mathbf{C}_{k}^{4\tilde{t}}$. 

\begin{figure*}[tb!]
\begin{equation}
\Big[\hat{s}_{k,1}^{4\tilde{t}+4}, \hat{\tilde{s}}_{k,4}^{4\tilde{t}+4} \Big]
= \arg \min_{\substack{s_{1} \in \mathcal{S}_1 \\ \tilde{s}_{4} \in \mathcal{S}_2}}
\Big\{ \Big| [\mathbf{Y}_k^{4\tilde{t}+4}]_{1,:} [\mathbf{Y}_k^{4\tilde{t}}]_{1,:}^{\mathrm{H}} - \big( Z s_{1} + Q \tilde{s}_{4} \big) \Big|^2
+ \Big| [\mathbf{Y}_k^{4\tilde{t}+4}]_{1,:} [\mathbf{Y}_k^{4\tilde{t}}]_{4,:}^{\mathrm{H}} - \big( Z \tilde{s}_{4} + Q s_{1} \big) \Big|^2 \Big\}
\label{eq:dataDetecion1}
\end{equation}

\vspace{0.3em}

\begin{equation}
\Big[\hat{s}_{k,2}^{4\tilde{t}+4}, \hat{\tilde{s}}_{k,3}^{4\tilde{t}+4} \Big]
= \arg \min_{\substack{s_{2} \in \mathcal{S}_1 \\ \tilde{s}_{3} \in \mathcal{S}_2}}
\Big\{ \Big| [\mathbf{Y}_k^{4\tilde{t}+4}]_{1,:} [\mathbf{Y}_k^{4\tilde{t}}]_{2,:}^{\mathrm{H}} - \big( Z s_{2} - Q \tilde{s}_{3} \big) \Big|^2
+ \Big| [\mathbf{Y}_k^{4\tilde{t}+4}]_{1,:} [\mathbf{Y}_k^{4\tilde{t}}]_{3,:}^{\mathrm{H}} - \big( Z \tilde{s}_{3} - Q s_{2} \big) \Big|^2 \Big\}
\label{eq:dataDetecion2}
\end{equation}

\vspace{-0.6em}
\hrulefill
\end{figure*}

\subsection{Third Step: Data Detection}
At time epoch $4\tilde{t} + 4$, the UE $k$ receives 
\begin{equation}
     \mathbf{y}_{k}^{4\tilde{t}+4} = \sum\nolimits_{l=1}^{L} \mathrm{g}_{k,l}^{\mathrm{(ef)}} \big[ \mathbf{C}^{4\tilde{t}+4}_{k} \big]_{m(l,k),:}^{\mathrm{T}} +  \widetilde{\mathbf{n}}_{k}^{4\tilde{t}+4},
     \label{eq:ReceivedSignalVector2}
\end{equation}
from which $\mathbf{Y}_{k}^{4\tilde{t}+4}$ is constructed similarly to $\mathbf{Y}_{k}^{4\tilde{t}}$. The transmitted symbols are then detected using the pairwise least-squares detector in \eqref{eq:dataDetecion1} and \eqref{eq:dataDetecion2}, shown at the top of the next page, where $Z$ and $Q$ denote the channel and amplitude power estimates, respectively. Both are computed as~\cite{DQoSTBC2007}
\begin{align}
Z & \approx [\mathbf{Y}_k^{4\tilde{t}}]_{1,:} \, [\mathbf{Y}_k^{4\tilde{t}}]_{1,:}^{\mathrm{H}} \\
Q & \approx [\mathbf{Y}_k^{4\tilde{t}}]_{4,:} \, [\mathbf{Y}_k^{4\tilde{t}}]_{1,:}^{\mathrm{H}}\;.
\end{align}

\subsection{Pros and Cons of DQ-OSTBC schemes}
\label{ssec:ProsAndCons}


Although DQ-OSTBC schemes achieve full code rates, they also introduce some trade-offs. Starting from \eqref{eq:EncodingDQ_OSTBC}, note that $[\mathbf{C}_{k}^{4\tilde{t}+4}]_{1,:} (\mathbf{C}_{k}^{4\tilde{t}})^{\mathrm{H}} = \mathbf{s}_{k}^{4\tilde{t}+4} \, \mathbf{C}_{k}^{4\tilde{t}} (\mathbf{C}_{k}^{4\tilde{t}})^{\mathrm{H}} / \widetilde{C}_{k}^{4\tilde{t}}$. In orthogonal space-time code matrices, the product $\mathbf{C}_{k}^{4\tilde{t}} (\mathbf{C}_{k}^{4\tilde{t}})^{\mathrm{H}}$ yields a diagonal matrix, eliminating interference from the previously transmitted codeword during data detection. In contrast, for DQ-OSTBC we obtain
\begin{equation}
[\mathbf{C}_{k}^{4\tilde{t}+4}]_{1,:} (\mathbf{C}_{k}^{4\tilde{t}})^{\mathrm{H}} =\begin{aligned}
 \frac{\mathbf{s}_{k}^{4\tilde{t}+4}}{\widetilde{C}_{k}^{4\tilde{t}}}
\begin{bmatrix}
V_1 & 0 & 0 & V_2 \\
0 & V_1 & V_2 & 0 \\
0 & -V_2 & V_1 & 0 \\
V_2 & 0 & 0 & V_1
\end{bmatrix},
\end{aligned}
\label{eq:DQ-OSTBC_interference}
\end{equation}
where $V_2 = 2 \, \mathrm{Re} \big( [\mathbf{C}_{k}^{4\tilde{t}}]_{1,1} [\mathbf{C}_{k}^{4\tilde{t}}]_{1,4}^{*} - [\mathbf{C}_{k}^{4\tilde{t}}]_{1,2} [\mathbf{C}_{k}^{4\tilde{t}}]_{1,3}^{*} \big)$ and $V_1 = \sum_{i=1}^{4} \big| [\mathbf{C}_{k}^{4\tilde{t}}]_{1,i} \big|^2$, with $V_2$ indicating residual interference from symbols transmitted in the previous codeword. Regarding computational complexity, the detector in \eqref{eq:dataDetecion1} and \eqref{eq:dataDetecion2} requires: (i) four row-vector products, requiring $4L_k^2$ complex multiplications; and (ii) a codebook search with complexity $\mathcal{O}(2M^{2})$, which is higher than in the orthogonal case, whose complexity is approximately $\mathcal{O}(N_s M_o)$. Therefore, achieving full code rate comes at the cost of increased receiver complexity and residual interference during data detection\footnote{\textcolor{black}{The DQO-STBC schemes presented here can also be extended to non-unit-modulus constellations, such as QAM. However, for a fair comparison with the classical STBC schemes considered in this paper, their evaluation under QAM modulation is left for future work.}}.

}



\section{Numerical Results} 
\label{sec:results}

We consider a \ac{CF-mMIMO} network comprising $L$ \acp{AP}, each equipped with $N$ antennas, and $K$ single-antenna \acp{UE}. The $L$ \acp{AP} are distributed over a coverage area of $0.5\; \mathrm{km} \times 0.5 \; \mathrm{km}$ following a \ac{HCPP}\footnote{In this approach, the minimum allowed distance between any two \acp{AP} is $d_{\text{min}} = \sqrt{A/L}$, where $A$ is the coverage area. Initially, \acp{AP} are randomly placed using a homogeneous Poisson point process with density $1/d_{\text{min}}$. \acp{AP} too close to others are iteratively replaced until the distance constraint is met.}, whereas the $K$ \acp{UE} are uniformly distributed within the same region. We focus on \ac{DL} channels and assume that $\tau_{c} = 200$, $\tau_{d} = 190$, and $\tau_{p} = 10$. Moreover, the wrap-around technique is employed to provide a better balance regarding the amount of interference affecting each \ac{AP}. The total transmission powers of \acp{UE} and \acp{AP} are $p_{k} = 100\,\mathrm{mW}$, $\rho_{\text{max}} = 200\,\mathrm{mW}$. For computing the centralized power coefficients $\{\rho_k\}$, we consider the popular fractional power allocation strategy with parameters $\varsigma = 0.2$, $\zeta = -0.5$ and $\kappa = 0.5$ \cite{BookCFemil2021, Chen2023}.

The \ac{SE} is computed using the \ac{BER} for schemes that do not admit an expression for the \ac{SE}, such as \ac{DSTBC}. For this purpose, the  \ac{BER} is calculated via Monte-Carlo simulations over different channel realizations and \ac{AP}/\ac{UE} locations, referred to as setups. In each setup, the received symbols are detected using the ML criterion and then mapped to bits via Gray coding. In the following, the \ac{BER} is calculated for each \ac{UE} over all channel realizations as \cite{freitas2025PMCellFree}
\begin{equation}
    \centering
    \widetilde{\mathrm{SE}}_k = P_f \log_2 \left( M_o \right) \left(1-\mathbb{E}_{\mathbf{h}}\left\{ \mathrm{BER}_{k} \right \} \right)\,,
    \label{Eq:SE_BER}
\end{equation}
\noindent \textcolor{black}{where $P_f$ represents the pre-log factor, i.e., the fraction of samples in each coherence block dedicated for DL data transmission. The value of $P_f$ depends on the transmission scheme and is defined as $(\tau_d - 2\tau_g)/\tau_c$ for conventional \ac{CF-mMIMO} and $(G-1)n_s/\tau_c$ for \ac{CF-mMIMO} with \ac{DSTBC} and \ac{DQO}-STBC. Here, $\tau_g$ denotes the guard interval associated with the resources required for phase calibration in conventional \ac{CF-mMIMO} with coherent DL transmission \cite{KHNgo_PhaseCalibration_2025}.} The term $\mathbb{E}_{\mathbf{h}}\left\{ \mathrm{BER}_{k} \right \}$ is the average \ac{BER} across all channel realizations, and $M_o=8$ is the modulation order. 
\begin{table}[t!]
	\centering
	\caption{Main simulation settings.}
	\label{tab:UMiParameters}
	\begin{tabular}{cc}
    \toprule
		\textbf{Parameter} & \textbf{Value} \\ \midrule
		Shadow fading standard deviation, $\sigma_\mathrm{S F}$ & $4 \mathrm{~dB}$ \\ 
		\ac{AP}/\ac{UE} antenna height, $h_\mathrm{AP}, h_\mathrm{UE}$ & $11.65     \mathrm{~m}, 1.65 \mathrm{~m}$ \\
		RX \ac{NF} & $8 \mathrm{~dB}$ \\
		Carrier frequency, bandwidth ($B$) & $3.5 \mathrm{GHz}, 20 \mathrm{MHz}$ \\
		\acp{ASD} & $\sigma_\varphi = \sigma_\theta = 15^{\circ}$\\
		Antenna spacing & 1/2 wavelength distance\\
    \bottomrule
	\end{tabular}
    \vspace{-5mm}
\end{table}

The \ac{AP} clustering scheme that associates the \ac{UE} with the $L_k$ \ac{AP} presenting the strongest channel in its vicinity is utilized \cite{RestrictedProcessingMarx}. In this scheme, the \ac{UE} requests a connection to the $L_k$ \acp{AP} with the greatest channel gains in its surroundings. Then, the \acp{AP} accept the connection request from \ac{UE} $k$ if it has availability for a new connection, i.e., $| \mathcal{K}_l | < \tau_p$ or if the channel gain from \ac{UE} $k$ exceeds that of the currently served \ac{UE} with the weakest channel gain at the \ac{AP}. 

The 3GPP \ac{UMi} path loss model is adopted for modeling the propagation channel~\cite{3GPPSPEC}.
It is considered that the shadowing terms of an \ac{AP} to different \acp{UE} are correlated. Moreover, the computation of the matrices $\mathbf{R}_{kl}$ follows the \textit{local scattering spatial correlation model} \cite{BookCFemil2021}. Table\,\ref{tab:UMiParameters} exhibits the parameters used in the \ac{UMi} and $\mathbf{R}_{kl}$ models \cite{BookCFemil2021,3GPP5GNR}.

\subsection{Impacts of Phase Misalignment in \ac{CF-mMIMO} systems}

\begin{figure}[t]
    \centering
    \begin{subfigure}{.225\textwidth}
    \centering
    \includegraphics[width=\textwidth]{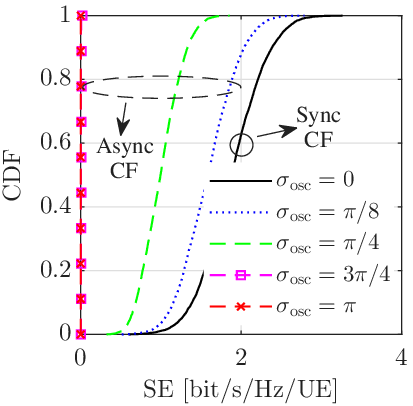}
    \caption{\ac{MR} - Lower Bound}
    \label{Fig:CDF_var_alpha_MR_L_k_8_HB}
    \end{subfigure}
    \begin{subfigure}{.225\textwidth}
    \centering
    \vspace{0.02cm}
    \includegraphics[width=\textwidth]{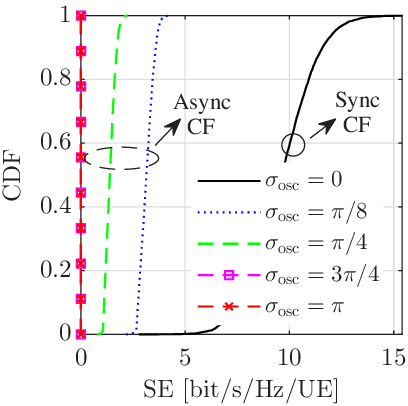}
    \caption{\ac{P-MMSE} - Lower Bound}
    \label{Fig:CDF_var_alpha_P_MMSE_L_k_8_HB}
    \end{subfigure}
    \begin{subfigure}{.225\textwidth}
    \centering
    \includegraphics[width=\textwidth]{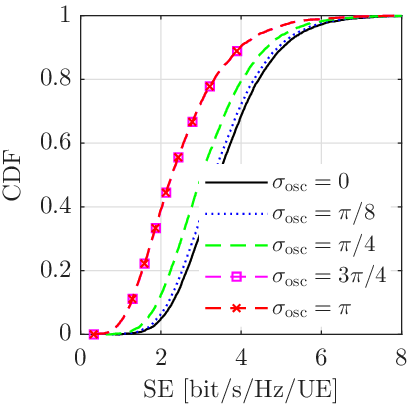} 
    \caption{\ac{MR} - Upper Bound}
    \label{Fig:CDF_var_alpha_MR_L_k_8_Inst}
    \end{subfigure}
    \begin{subfigure}{.225\textwidth}
    \centering
    \includegraphics[width=\textwidth]{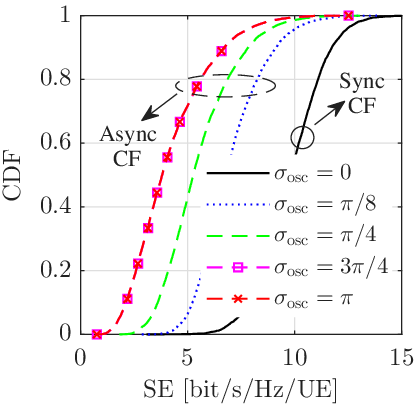} 
    \caption{\ac{P-MMSE} - Upper Bound}
    \label{Fig:CDF_var_alpha_P_MMSE_L_k_8_Inst}
    \end{subfigure}
    \caption{Comparison of the per-UE \ac{DL} \ac{SE} of \ac{CF-mMIMO} systems in the presence of phase misalignment effects by varying the value of $\sigma_{\mathrm{osc}}$ in \eqref{Eq:instantaneousSINR_2} and \eqref{Eq:SINR_hardening_2} from 0 to $\pi$. Eqs. \eqref{Eq:instantaneousSINR_2} and \eqref{Eq:SINR_hardening_2} refer to the upper and lower bounds, respectively. Parameters settings: $L = 40$, $K = 20$, $N = 4$ and $L_k = 8$.}
    \label{Fig:CDFofSE_conventionalCF}
    \vspace{-5mm}
\end{figure}

Fig.\,\ref{Fig:CDFofSE_conventionalCF} shows the \acp{CDF} of the \ac{SE} for \ac{UC} \ac{CF-mMIMO} systems with (Async CF) and without (Sync CF) phase misalignment effects. \textcolor{black}{The phase misalignment level (the standard deviation $\sigma_{\mathrm{osc}}$ of $\vartheta_l$) is varied, and both centralized and distributed processing are evaluated using the lower and upper capacity bounds derived in \eqref{Eq:instantaneousSINR_2} and \eqref{Eq:SINR_hardening_2}, with each \ac{UE} being served by $L_k = 8$ \acp{AP}.}

In Figs.\,\ref{Fig:CDF_var_alpha_MR_L_k_8_HB} and \ref{Fig:CDF_var_alpha_P_MMSE_L_k_8_HB}, the \ac{SE} is computed using the capacity lower bound in \eqref{Eq:SINR_hardening_2}. In these ones, it can be observed that phase misalignment effects can severely degrade the \ac{SE} of \ac{UC} \ac{CF-mMIMO} systems, particularly when $\sigma_{\mathrm{osc}} \geq \pi/4$.
One can also note that the centralized processing is more susceptible to these effects, as it relies on accurate channel estimates of many \acp{UE} to suppress interference effectively. For instance, when $\sigma_{\mathrm{osc}}$ increases from $0$ to $\pi/4$, the \ac{SE} of the 95\% likely \acp{UE} drops by approximately 43\% and 84\% for the \ac{MR} and \ac{P-MMSE} schemes, respectively.
\begin{figure}[t]
    \centering
    \begin{subfigure}{.225\textwidth}
    \centering
    \includegraphics[width=\textwidth]{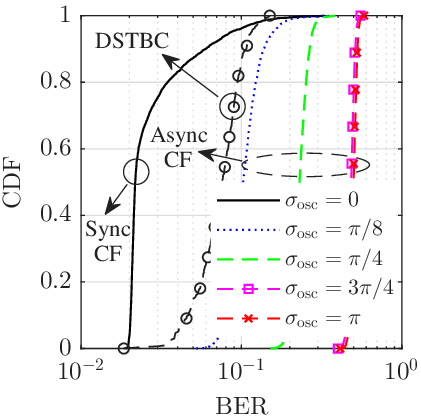} \caption{BER}
    \label{fig:alpha_DSTBC_BER}
    \end{subfigure}
    \begin{subfigure}{.225\textwidth}
    \centering
    \vspace{0.02cm}
    \includegraphics[width=\textwidth]{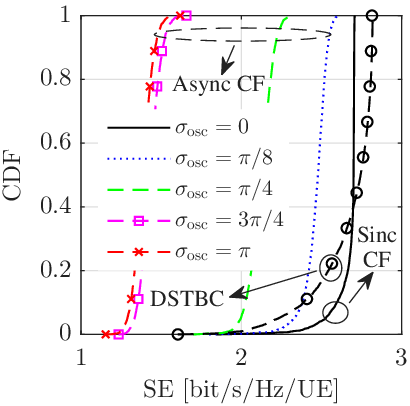} \caption{SE}
    \label{Fig:SE_alpha_DSTBC_BER}
    \end{subfigure}
     \caption{CDF of (a) \ac{BER} (b) per-UE \ac{SE} for conventional and \ac{DSTBC}-based \ac{CF-mMIMO} systems under varying phase misalignment levels. Parameters: $L = 100$, $K = 20$, $L_k = 2$, $N = 4$. Precoding scheme: \ac{P-MMSE}.}
    \label{Fig:CDF_DSTBC_BER}
    \vspace*{-3mm}
\end{figure}

The degradation in \ac{SE} is also observable when it is computed using the capacity upper bound in \eqref{Eq:instantaneousSINR_2}. However, the impact is less pronounced as depicted in Figs.\,\ref{Fig:CDF_var_alpha_MR_L_k_8_Inst} and \ref{Fig:CDF_var_alpha_P_MMSE_L_k_8_Inst}. This is because the \ac{SINR} in \eqref{Eq:instantaneousSINR_2} is derived from \eqref{Eq:instantaneousSINR_1}, where the expectation is taken after the squared magnitude operation, which is analogous to multiplying a term by its complex conjugate. This procedure is similar to that performed by \ac{DSTBC} schemes in \eqref{Eq:Y_ML_k}, which generates better resilience to the effects of phase misalignment. This robustness can be further understood by analyzing the numerator of \eqref{Eq:instantaneousSINR_2}, where the term $(1-\tilde{\nu}) \sum_{l = 1}^L \big| {\mathrm{h}}_{k, l}^{\mathrm{(ef)}} \big|^2$ remains significant even when $\sigma_{\mathrm{osc}} = \pi$ increases, since $\tilde{\nu} = e^{-\sigma_{\mathrm{osc}}^2}$. In contrast, the numerator of \eqref{Eq:SINR_hardening_2}, given by $\tilde{\nu} \big| \sum_{l=1}^L \mathbb{E}_{\mathbf{h}}\big\{ {\mathrm{h}}_{k, l}^{\mathrm{(ef)}} \big\} \big|^{2}$, directly scales with $\tilde{\nu}$ and thus decreases as $\sigma_{\mathrm{osc}}$ increases. On the other hand, it is important to note that \eqref{Eq:instantaneousSINR_2} is derived under idealized assumptions, such as the \ac{UE} having access to the perfect \ac{CSI}. In other words, although it exhibits reduced susceptibility to phase misalignment effects, this benefit relies on assumptions that may not hold in practical scenarios.

\subsection{Performance of \ac{DSTBC} schemes in \ac{CF-mMIMO}}

Fig.\,\ref{Fig:CDF_DSTBC_BER} compares the \ac{SE} achieved by the proposed solution with that of conventional \ac{UC} \ac{CF-mMIMO} systems. To this end, each \ac{UE} is served by $L_k = 2$ \acp{AP}, and the \ac{SE} is computed according to \eqref{Eq:SE_BER}. One can note that the proposed approach outperforms \ac{UC} \ac{CF-mMIMO} systems affected by phase misalignment and approaches the performance of perfectly synchronized systems. The proposed approach can even perform slightly better than synchronized \ac{CF-mMIMO} systems (for approximately 45\% of the \acp{UE}), as the product in \eqref{Eq:Y_ML_k} tends to reinforce the desired signal. Therefore, this effect becomes particularly evident when \acp{UE} experience strong channels and low levels of interference. \textcolor{black}{Moreover, the $2\tau_g$ samples required for phase synchronization introduce an additional pre-log penalty in conventional UC CF-mMIMO, further reducing its achievable SE.}
\begin{figure}[t]
    \centering
    \begin{subfigure}{.225\textwidth}
    \centering
    \includegraphics[width=\textwidth]{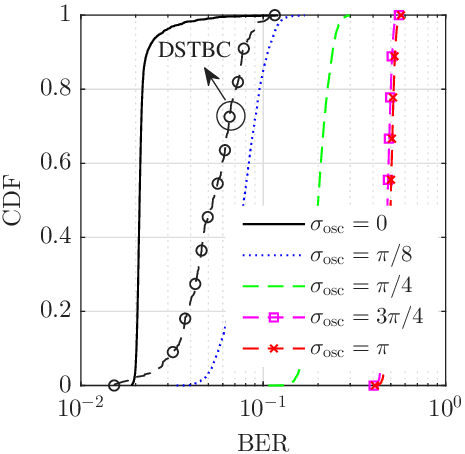} \caption{BER}
    \label{Fig:alpha_DSTBC_BER_L_k_4}
    \end{subfigure}    
    \begin{subfigure}{.223\textwidth}
    \centering
    \vspace{0.02cm}
    \includegraphics[width=\textwidth]{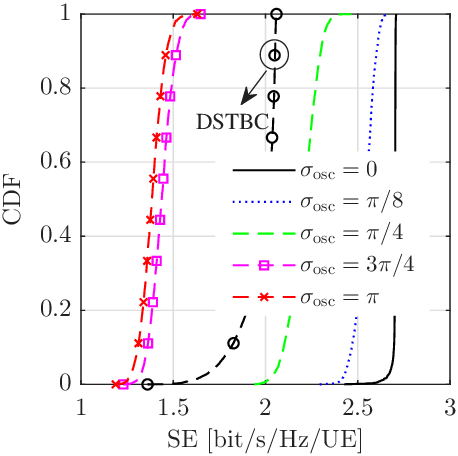} \caption{SE}
    \label{Fig:SE_alpha_DSTBC_BER_L_4}
    \end{subfigure}
     \caption{CDF of (a) \ac{BER} (b) per-UE \ac{SE} for conventional and \ac{DSTBC}-based \ac{CF-mMIMO} systems under varying phase misalignment levels. Parameters: $L = 100$, $K = 20$, $L_k = 4$, $N = 4$. Precoding scheme: \ac{P-MMSE}.}
    \label{Fig:SE_alpha_DSTBC_BER_Lk_4}
    \vspace*{-4mm}
\end{figure}

\begin{figure}[t]
	\centering	\includegraphics[width=.38\textwidth,keepaspectratio=true]{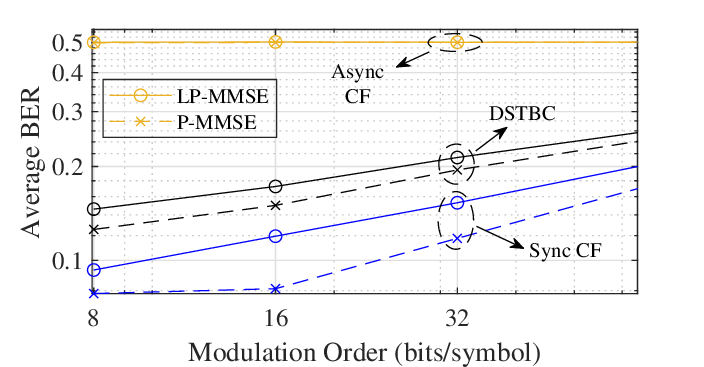}
	\caption{Average \ac{BER} versus the modulation order. Parameters setting: $L = 40$, $K = 20$, $L_k = 2$, $N = 4$, and $\sigma_{\mathrm{osc}} = \pi$. Precoding scheme: \ac{P-MMSE}.}
	\label{Fig:BER_M_Order}
    \vspace*{-2mm}
\end{figure}

However, the gains afforded by the proposed solution tend to decrease for $L_k > 2$, as illustrated in Fig.\,\ref{Fig:SE_alpha_DSTBC_BER_Lk_4}. It is possible to observe that asynchronous \ac{UC} \ac{CF-mMIMO} systems can outperform \ac{DSTBC}-based systems when $\sigma_{\mathrm{osc}} \leq \pi/4$, mainly due to the negative impact of the code rate in \ac{DSTBC} schemes. For instance, when $L_k = 4$, the code rate equals 3/4, thereby reducing the \ac{SE} achieved by the proposed approach by 25\%. In other words, although \ac{DSTBC} schemes offer a feasible approach to mitigate phase misalignment in \ac{CF-mMIMO} systems, the code rate remains a limiting factor in these schemes, particularly for $L_k \geq 4$. On the other hand, the gains provided by the proposed approach are evident for $L_k = 2$ and remain consistent even for higher modulation orders, as depicted in Fig.\,\ref{Fig:BER_M_Order}. Moreover, the proposed approach also presents improvements in terms of \ac{BER} in Fig.\,\ref{Fig:SE_alpha_DSTBC_BER_Lk_4}.

\subsection{\ac{SINR} analysis}

\begin{figure}[t]
    \centering
    \begin{subfigure}{.225\textwidth}
    \centering
    \includegraphics[width=\textwidth]{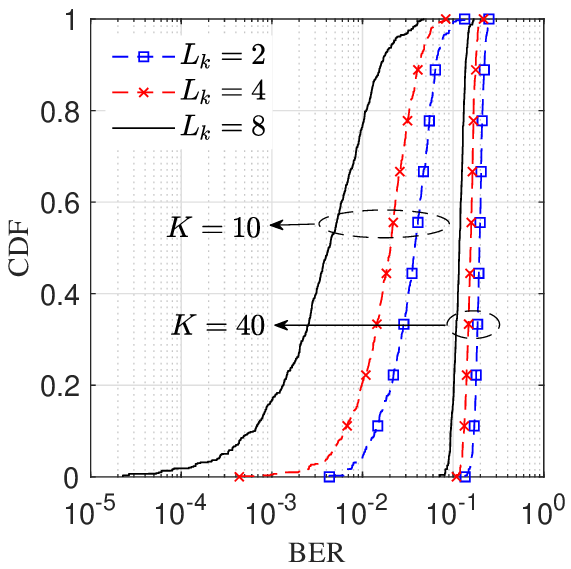} \caption{BER}
    \label{Fig:CDF_BER_LP_MMSE_DSTBC_K}
    \end{subfigure}    
    \begin{subfigure}{.225\textwidth}
    \centering
    \vspace{0.02cm}
    \includegraphics[width=\textwidth]{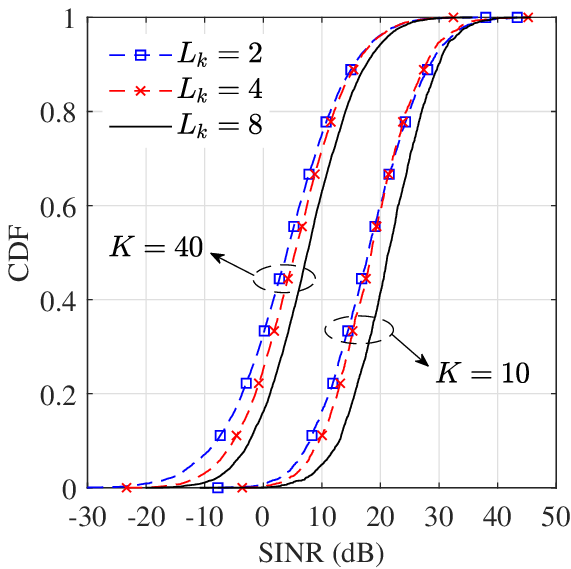} \caption{SINR}
    \label{Fig:CDF_SINR_DSTBC_LP_MMSE}
    \end{subfigure}
     \caption{\ac{BER} (a) and \ac{SE} (b) of \ac{CF-mMIMO} systems using \ac{DSTBC} schemes by varying the number of \acp{AP} serving the \ac{UE}, i.e., $L_{k}$. Parameters setting: $L = 40$, $K = 20$, and $N = 4$. Precoding scheme: \ac{LP-MMSE}.}
    \label{Fig:SINR_DSTBC_K_LP_MMSE}
    \vspace*{-4mm}
\end{figure}

To provide a more detailed performance evaluation of the proposed \ac{DSTBC}-based system, Fig.\,\ref{Fig:SINR_DSTBC_K_LP_MMSE} presents results based on the derived \ac{SINR} expression, supported by \ac{BER} simulations. 
As expected, the \ac{SINR} slightly increases with the number of \acp{AP} $L_k$ serving the \ac{UE}, but the gains are modest. This is because the product $\left (\mathbf{y}^{t}_{k} \right)^{\mathrm{H}} \mathbf{y}^{t-1}_{k}$ in \eqref{Eq:Y_ML_k} can amplify not only the desired signal but also the interference as more \acp{AP} serve the \ac{UE}, as observed in \eqref{Eq:SINR_inst_DSTBC_2}. The same observation can be made with respect to the \ac{BER} as depicted in Fig.\,\ref{Fig:CDF_BER_LP_MMSE_DSTBC_K}.

It is worth noting that we include both the \ac{SINR} and \ac{BER} results in Fig.\,\ref{Fig:SINR_DSTBC_K_LP_MMSE} because the \ac{BER} provides practical insights into error performance and helps validate the analytical \ac{SINR}. However, \ac{BER} evaluation can be computationally intensive and depends on the modulation scheme. In contrast, the derived \ac{SINR} enables a more general and efficient performance assessment of \ac{CF-mMIMO} systems using \ac{DSTBC} schemes.

{\color{black}
\subsection{Impact of Time-Varying Phase Noise}

\label{SSec:Wiener}

In the preceding analysis, to facilitate the analytical treatment of the proposed schemes, we adopted a quasi-static phase model in which the oscillator phases remain approximately constant over two consecutive codewords. Here, we relax this assumption and consider a Wiener (random-walk) phase-noise model, where the oscillator phase at AP $l$ evolves on a symbol-by-symbol basis as
\begin{equation}
\vartheta_l^{p} = \vartheta_l^{p-1} + \Delta_l^{p},
\label{eq:PhaseNoiseTime}
\end{equation}
where $p \in \{1,\ldots,\tau_d\}$ and $\Delta_l^{p} \sim \mathcal{N}(0,\sigma_{\mathrm{osc}}^2)$ is the Wiener phase increment. This model is adopted to evaluate the robustness of DSTBC schemes under time-varying phase noise and the validity of the quasi-static phase assumption for different oscillator phase-noise variances. The remaining signal processing operations, such as UL training and precoding generation, follow Section~\ref{sec:system-model} and are omitted for brevity.

We now evaluate the performance of the proposed approach and conventional UC CF-mMIMO systems assuming symbol-by-symbol oscillator phase variations following the Wiener model described in Subsection~\ref{SSec:Wiener}. Accordingly, Fig.~\ref{Fig:SE_alpha_DSTBC_BER_Lk_wiener} presents the CDF of SE for different AP oscillator phase-noise variances $\sigma_{\mathrm{osc}}^2$. In particular, $\sigma_{\mathrm{osc}}^2 = -50$ dB corresponds to the coherent regime, while larger values of $\sigma_{\mathrm{osc}}^2$ represent increasingly non-coherent scenarios associated with lower oscillator quality. For the proposed DSTBC-based scheme, the oscillator phase-noise variance is fixed at $\sigma_{\mathrm{osc}}^2 = -20$ dB.

\begin{figure}[htb!]
    \centering
    \begin{subfigure}{.225\textwidth}
    \centering
    \includegraphics[width=\textwidth]{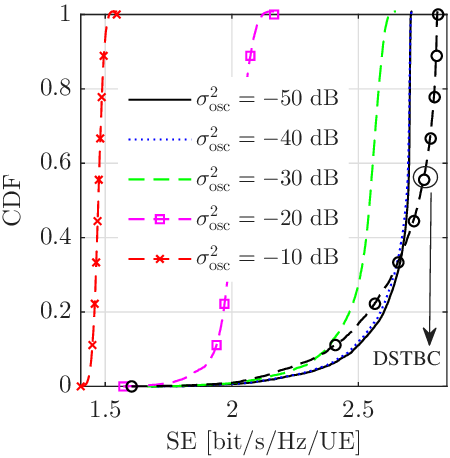} \caption{$L_k = 2$}
    \label{Fig:alpha_DSTBC_BER_L_k_4_wiener}
    \end{subfigure}    
    \begin{subfigure}{.225\textwidth}
    \centering
    \vspace{0.02cm}
    \includegraphics[width=\textwidth]{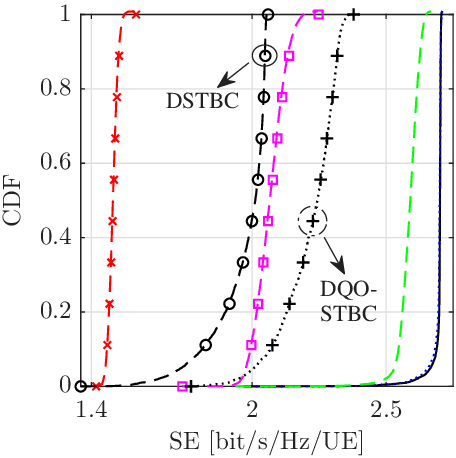} \caption{$L_k = 4$}
    \label{Fig:SE_alpha_DSTBC_BER_L_4_wiener}
    \end{subfigure}
     \caption{\textcolor{black}{CDF of \ac{BER} per-UE \ac{SE} for conventional and differential-based \ac{CF-mMIMO} systems under different levels of AP oscillator phase-noise variance for (a) $L_k = 2$ and (b) $L_k = 4$. Parameters setting: $L = 40$, $K = 20$, $N = 4$, and $\tau_c = 200$. Precoding scheme: \ac{P-MMSE}.}}
    \label{Fig:SE_alpha_DSTBC_BER_Lk_wiener}
    \vspace*{-4mm}
\end{figure}

One can note that the proposed DSTBC-based scheme restores the performance of conventional CF-mMIMO in all non-coherent scenarios and even outperforms coherent CF-mMIMO for the 70\% likely \acp{UE} when $L_k = 2$, as also observed in Fig.\,\ref{Fig:CDF_DSTBC_BER}. Similarly, for $L_k = 4$, the reduced code rate of DSTBC schemes ($R=3/4$) limits their performance, thereby reducing the gains relative to conventional CF-mMIMO. To mitigate this limitation, DQO-STBC schemes can be employed to restore the full code rate when $L_k = 4$. However, the lack of orthogonality introduces residual symbol interference from previously transmitted codewords (as discussed in Subsection \ref{ssec:ProsAndCons}), which limits the gains achieved by DQO-STBC schemes. Overall, the proposed approach remains effective even under time-varying phase noise, since the quasi-static phase assumption is approximately satisfied in DSTBC transmission, since two consecutive DSTBC codewords span only $2L_k$ symbol periods. Consequently, the accumulated phase variation within this interval remains limited.

To support this explanation, Fig.~\ref{Fig:AverageSE_oscillator_quality_tau_c} presents the average SE for both systems under different AP oscillator phase-noise variances and coherence block lengths $\tau_c$. Note that the SE of DSTBC-based schemes remains nearly constant up to $\sigma_{\mathrm{osc}}^2 = -20$ dB, after which it starts to decrease due to the larger phase variations between consecutive symbols. Nevertheless, DSTBC-based transmission still outperforms conventional CF-mMIMO at $\sigma_{\mathrm{osc}}^2 = -10$ dB, demonstrating robustness under distinct phase-noise conditions. Moreover, the SE of conventional CF-mMIMO degrades more rapidly as $\tau_c$ increases due to stronger phase accumulation over longer coherence blocks. In contrast, increasing $\tau_c$ does not degrade the performance of DSTBC-based schemes, since phase variations remain limited over two consecutive codewords. Finally, although not shown in the revised manuscript due to space constraints, our results confirm that DQO-STBC schemes can improve the performance of DSTBC transmission for different values of $\sigma_{\mathrm{osc}}^2$ and are particularly attractive for lower modulation orders, such as 4-PSK, approaching the performance of orthogonal DSTBC schemes when binary PSK (BPSK) modulation is employed.

\begin{figure}[htb!]
    \centering
    \begin{subfigure}{.23\textwidth}
    \centering
    \includegraphics[width=\textwidth]{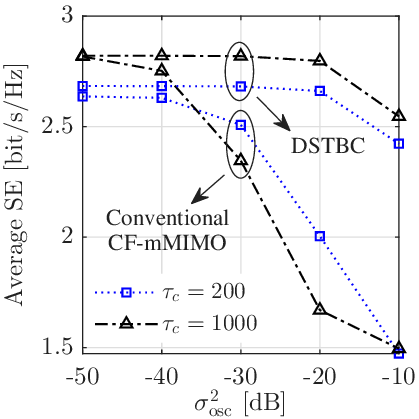} \caption{$L_k = 2$}
    \label{Fig:AverageSE_oscillator_quality_L_k_2_tau_c}
    \end{subfigure}    
    \begin{subfigure}{.23\textwidth}
    \centering
    \vspace{0.02cm}
    \includegraphics[width=\textwidth]{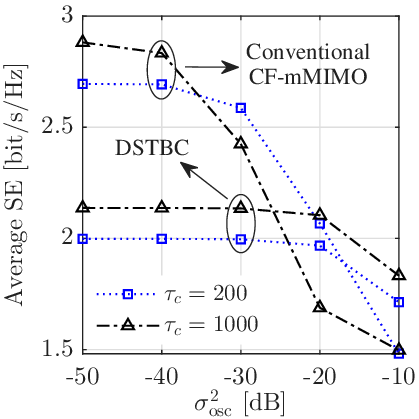} \caption{$L_k = 4$}
    \label{Fig:AverageSE_oscillator_quality_L_k_4_tau_c}
    \end{subfigure}
     \caption{\textcolor{black}{Average \ac{SE} of conventional and \ac{DSTBC}-based \ac{CF-mMIMO} systems versus AP oscillator phase-noise variance for (a) $L_k = 2$ and (b) $L_k = 4$. Here: $L = 40$, $K = 20$, $N = 4$. Precoding scheme: \ac{P-MMSE}.}}
    \label{Fig:AverageSE_oscillator_quality_tau_c}
    \vspace*{-4mm}
\end{figure}

\subsection{Impact of Doppler Shift Effect}

We now evaluate the impact of user mobility on the proposed DSTBC-based transmission. In high-mobility scenarios, the relative movement between UEs and APs introduces Doppler shifts, leading to temporal channel variations and outdated CSI during DL data transmission. To account for this effect, we adopt a first-order auto-regressive channel model commonly used to describe channel aging effects in CF-mMIMO systems. Specifically, the channel between AP $l$ and UE $k$ at time epoch $p$ evolves as~\cite{DopplerShift2021}
\begin{equation}
h_{l,k}^{p}
= \tilde{\rho}_k h_{l,k}^{p-1} + \sqrt{1-\tilde{\rho}_k^2}\, \tilde{z}_{l,k}^{p},
\end{equation}
with $\tilde{\rho}_k = J_0(2\pi f_{D,k}T_s)$ being the temporal correlation coefficient, where $J_0(\cdot)$ denotes the zeroth-order Bessel function of the first kind, $T_s$ is the sampling time, and $f_{D,k} = (v_k f_c)/c$ is the Doppler frequency for a user moving at speed $v_k$, with $f_c$ and $c$ denoting the carrier frequency and the speed of light, respectively. Moreover, $\tilde{z}_{l,k}^{p} \sim \mathcal{CN}(0,\beta_{l,k}\mathbf{I}_N)$ denotes the innovation component, uncorrelated with $h_{l,k}^{p-1}$. Here, $f_c = 3.5$\,GHz, $T_s = 35.68\mu$\,s and $\beta_{l,k}=\mathrm{tr}(\mathbf{R}_{k, l})/N$.

Fig.~\ref{Fig:UEspeed} presents the average SE versus the UE speed. One can note that the performance of both conventional CF-mMIMO and DSTBC-based schemes degrades as the UE speed increases due to stronger temporal channel variations. Nevertheless, the proposed DSTBC-based approach outperforms conventional CF-mMIMO under high mobility, since it does not rely on explicit CSI during data detection and is therefore less sensitive to channel aging effects.

\begin{figure}[htb!]
	\centering	\includegraphics[width=.4\textwidth,keepaspectratio=true]{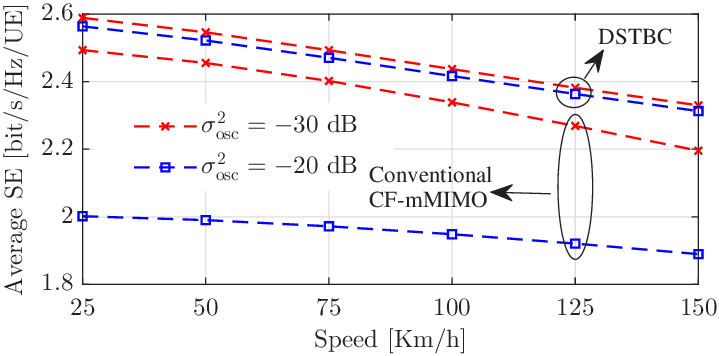}
	\caption{Average SE versus UE speed. Parameters setting: $L = 40$, $K = 20$, $L_k = 2$ and $N = 4$. Precoding scheme: P-MMSE.}
	\label{Fig:UEspeed}
    \vspace*{-2mm}
\end{figure}


}

\section{Conclusions} \label{sec:conclusion}

This paper investigated \ac{DSTBC} techniques as a robust solution for phase misalignments in \ac{UC} \ac{CF-mMIMO} systems. We first quantified the performance degradation in conventional systems by deriving capacity bounds and a closed-form \ac{SE} for \ac{MR} precoding under random phase misalignments. This analysis, covering both centralized and distributed processing, confirmed that misalignments severely degrade \ac{SE}, especially in centralized systems, unless the errors are negligible or perfect \ac{CSI} is available. We then derived \ac{SNR} and \ac{SINR} expressions for the proposed \ac{DSTBC} scheme. Our results demonstrate that \ac{DSTBC} effectively restores performance, approaching the \ac{SE} and \ac{BER} of a perfectly synchronized system without requiring perfect \ac{CSI}. While the inherent \ac{DSTBC} code rate can limit the peak \ac{SE} when many \acp{AP} serve a \ac{UE}, the approach consistently provides significant \ac{BER} gains. \textcolor{black}{Moreover, DQO-STBC schemes were shown to partially mitigate the code-rate limitation.} \textcolor{black}{Future work includes extending the analysis of orthogonal DSTBC to a wider range of modulation schemes, such as quadrature amplitude modulation (QAM), as well as incorporating practical aspects such as fronthaul limitations.}

\appendix

\subsection{Proof of Proposition~\ref{prop:SNR}}
\label{subsec:proof:proposition:SNR}

In order to compute the \ac{SNR} in \eqref{Eq:SNR_complex_2}, we need to compute $\mathbb{E}_{\mathbf{C}} \big \{ | \mathrm{ML}^{t}_{k, n, 2} |^{2}\big \}$ and $\mathbb{E}_{\mathbf{C}} \big \{ | \mathrm{ML}^{t}_{k, n, 3} |^{2}\big \}$.

\subsubsection{Compute $\mathbb{E}_{\mathbf{C}} \big \{ | \mathrm{ML}^{t}_{k, n, 2} |^{2}\big \}$} By applying the cyclic property of the trace in \eqref{ML_k_2}, let us  rewrite $\mathrm{ML}^{t}_{k, n, 2}$ as
\begin{equation}
\mathrm{ML}^{t}_{k, n, 2} = \sum\nolimits_{l = 1}^{L} \big[\mathbf{C}^{t-1}_{k} \big]_{m(l,k),:} \mathrm{g}_{k,l}^{({\rm ef})} \mathbf{A}_n \left( \mathbf{n}^{t}_{k} \right)^{\mathrm{H}}\,.    
\end{equation}
Hence, the term $| \mathrm{ML}^{t}_{k, n, 2} |^{2} \!=\! \mathrm{ML}^{t}_{k, n, 2}( \mathrm{ML}^{t}_{k, n, 2})^{\mathrm{H}}$ becomes
\begin{equation}
    \sum\nolimits_{l = 1}^{L} \sum\nolimits_{l{'} = 1}^{L}  \big[\mathbf{C}^{t-1}_{k} \big]_{m{'}(l{'},k),:} \mathbf{F}^{t}_{k, l, l{'}, n} \big[\mathbf{C}^{t-1}_{k} \big]_{m(l,k),:}^{\mathrm{H}},
\end{equation}
\noindent with $\mathbf{F}^{t}_{k, l, l{'}, n} = \big(\mathrm{g}_{k,l}^{({\rm ef})}\big)^{\mathrm{H}} \mathrm{g}_{k,l{'}}^{({\rm ef})} \mathbf{A}_n \left( \mathbf{n}^{t}_{k} \right)^{\mathrm{H}} \mathbf{n}^{t}_{k} \mathbf{A}_n^{\mathrm{H}}$. 
Given that $| \mathrm{ML}^{t}_{k, n, 2} |^{2}$ is a scalar, it can also be written as
\begin{equation}
    | \mathrm{ML}^{t}_{k, n, 2} |^{2} = \sum_{l = 1}^{L} \sum_{l{'} = 1}^{L} \mathrm{tr} \big \{ \big( \mathbf{C}^{t-1, t-1}_{k, m, m{'}, l, l{'}} \big)^{\mathrm{H}} \mathbf{F}^{t}_{k, l, l{'}, n} \big \},
\end{equation}
\noindent where $\mathbf{C}^{t-1, t-1}_{k, m, m{'}, l, l{'}} = \big[ \mathbf{C}^{t-1}_{k} \big]^{\mathrm{H}}_{m(l,k),:} \big[ \mathbf{C}^{t-1}_{k} \big]_{m{'}(l{'},k),:}$. Since matrices $\mathbf{C}^{t-1, t-1}_{k, m, m{'}, l, l{'}}$ and $\mathbf{F}^{t}_{k, l, l{'}, n}$ are statistically independent, $\mathbb{E}_{\mathbf{C}} \big \{ | \mathrm{ML}^{t}_{k, n, 2} |^{2}\big \}$ is obtained as
\begin{equation}
    \centering
     \sum\nolimits_{l = 1}^{L} \sum\nolimits_{l{'} = 1}^{L} \mathrm{tr} \big \{ \mathbb{E}_{\mathbf{C}} \big \{ \mathbf{C}^{t-1, t-1}_{k, m, m{'}, l, l{'}} \big \} \mathbb{E}_{\mathbf{C}} \big \{ \mathbf{F}^{t}_{k, l, l{'}, n} \big \} \big \}.
    \label{Eq:VarML_k_1_3}
\end{equation}
\noindent Provided that $\mathbf{A}_n \mathbf{A}_n^{\mathrm{H}} = \mathbf{I}_{L_{k}}$, the term $\mathbb{E}_{\mathbf{C}} \big \{ \mathbf{F}^{t}_{k, l, l{'}, n}\big\}$ is given by $ \mathbb{E}_{\mathbf{C}} \big \{ \mathbf{F}^{t}_{k, l, l{'}, n}\big\} = \big(\mathrm{g}_{k,l}^{({\rm ef})}\big)^{\mathrm{H}} \mathrm{g}_{k,l{'}}^{({\rm ef})} \sigma_{\mathrm{dl}}^{2} \mathbf{I}_{L_{k}}$. Then, by computing $\mathrm{tr} \big \{ \mathbf{C}^{t-1, t-1}_{k, m, m{'}, l, l{'}} \big \}$ in \eqref{Eq:TracesProperties}, we have
\begin{equation}
    \centering
     \mathbb{E}_{\mathbf{C}} \big \{ | \mathrm{ML}^{t}_{k, n, 2} |^{2}\big \} = \sigma_{\mathrm{dl}}^{2} \sum\nolimits_{l = 1}^{L}  \big| \mathrm{g}_{k,l}^{({\rm ef})}\big|^{2}.
    \label{Eq:VarML_k_2}
\end{equation}

\subsubsection{Compute $\mathbb{E}_{\mathbf{C}} \big \{ | \mathrm{ML}^{t}_{k, n, 3} |^{2}\big \}$} Let us rewrite $\mathrm{ML}^{t}_{k, n, 3}$ in \eqref{Eq:ML_k_3_t} as $\mathrm{ML}^{t}_{k, n, 3} = \sum_{l = 1}^{L} \mathrm{g}_{k,l}^{{\mathrm{(ef)}}\,\mathrm{H}} \mathbf{n}^{t-1}_{k} \mathbf{A}_n \big[ \mathbf{C}^{t}_{k} \big]^{\mathrm{H}}_{m(l,k),:}$, with $\big[ \mathbf{C}^{t}_{k} \big]_{m(l,k),:} = \big[ \mathbf{C}^{t-1}_{k} \big]_{m(l,k),:} \mathbf{X}^{t}_{k}$. Thus,
\begin{equation}
    \!| \mathrm{ML}^{t}_{k, n, 3} |^{2} \!=\! \sum_{l = 1}^{L} \sum_{l{'} = 1}^{L} \mathrm{tr} \big \{ \big( \mathbf{C}^{t, t}_{k, m, m{'}, l, l{'}} \big)^{\mathrm{H}} \mathbf{F}^{t-1}_{k, m, m{'}, l, l{'}, n} \big \},
\end{equation}
where 
$\mathbf{C}^{t, t}_{k, m, m{'}, l, l{'}} = \big[ \mathbf{C}^{t}_{k} \big]^{\mathrm{H}}_{m(l,k),:} \big[ \mathbf{C}^{t}_{k} \big]_{m{'}(l{'},k),:}$. Hence,
\begin{equation}
    \centering
     \mathbb{E}_{\mathbf{C}} \big \{ | \mathrm{ML}^{t}_{k, n, 3} |^{2}\big \} = \sigma_{\mathrm{dl}}^{2} \sum\nolimits_{l = 1}^{L}  \big| \mathrm{g}_{k,l}^{({\rm ef})}\big|^{2}.
    \label{Eq:VarML_k_5}
\end{equation}

\subsection{Proof of Proposition~\ref{prop:SINR-closed-form}}
\label{subsec:proof:proposition:SINR-closed-form}

To compute the \ac{SINR} in \eqref{Eq:SINR_inst_DSTBC_2}, it is necessary to evaluate the power of each component in the denominator of \eqref{Eq:SINR_inst_DSTBC}.

\subsubsection{Compute $\mathbb{E}_{\mathbf{C}} \big \{ | \mathrm{MLI}^{t}_{k, n, 2, 1} |^{2}\big \}$ and $\mathbb{E}_{\mathbf{C}} \big \{ | \mathrm{MLI}^{t}_{k, n, 2, 2} |^{2}\big \}$} Let the term  $\mathrm{MLI}^{t}_{k, n, 2} = \mathrm{tr} \big \{ \mathbf{A}_n \, (\mathbf{DS}^{t}_{k})^{\mathrm{H}} \widetilde{\mathbf{n}}^{t-1}_{k} \big \}$ be rewritten as $\mathrm{MLI}^{t}_{k, n, 2} = \mathrm{MLI}^{t}_{k, n, 2, 1} + \mathrm{MLI}^{t}_{k, n, 2, 2}$, with
\begin{equation}
    \! \mathrm{MLI}^{t}_{k, n, 2, 1} \!=\! \sum_{i=1, i \neq k}^{K} \sum_{l=1}^{L} \sum_{r=1}^{L} \mathrm{tr} \big\{ \mathrm{g}_{k,l}^{\,\mathrm{(ef)}\mathrm{H}} \widetilde{\mathrm{g}}_{i,k,r}^{\,\mathrm{(ef)}}
    \mathbf{A}_n \mathbf{C}^{t, t-1}_{k, i, m, \mu, l, r} \big \},
    \label{Eq:MLI_k_2_1}
\end{equation}
\noindent where $\mathbf{C}^{t, t-1}_{k, i, m, \mu, l, r} = \big[ \mathbf{C}^{t}_{k} \big]^{\mathrm{H}}_{m(l,k),:} \big[ \mathbf{C}^{t-1}_{i} \big]_{\mu(r,i),:}$. In addition, $\big[ \mathbf{C}^{t}_{k} \big]_{m(l,k),:} = \big[ \mathbf{C}^{t-1}_{k} \big]_{m(l,k),:} \mathbf{X}^{t}_{k}$ and $\mathrm{MLI}^{t}_{k, n, 2, 2}$ is calculated as $\mathrm{MLI}^{t}_{k, n, 2, 2} = \mathrm{tr} \big \{ \sum_{l=1}^{L} \mathrm{g}_{k,l}^{\,\mathrm{(ef)} \mathrm{H}}\mathbf{A}_n \big[ \mathbf{C}^{t}_{k} \big]^{\mathrm{H}}_{m(l,k),:} \mathbf{n}^{t-1}_{k} \big \}$. By noting that $\mathrm{MLI}^{t}_{k, n, 2, 2}$ corresponds to $\mathrm{ML}^{t}_{k, n, 3}$ in \eqref{Eq:ML_k_3_t}, $\mathbb{E}_{\mathbf{C}} \big \{ | \mathrm{MLI}^{t}_{k, n, 2, 2} |^{2}\big \}$ can be calculated as
\begin{equation}
    \centering
     \mathbb{E}_{\mathbf{C}} \big \{ | \mathrm{MLI}^{t}_{k, n, 2, 2} |^{2}\big \} = \sigma_{\mathrm{dl}}^{2} \sum\nolimits_{l = 1}^{L}  \big| \mathrm{g}_{k,l}^{({\rm ef})}\big|^{2}.
    \label{Eq:VarMLI_k_2_2}
\end{equation}
For computing $\mathbb{E}_{\mathbf{C}} \big \{ | \mathrm{MLI}^{t}_{k, n, 2, 1} |^{2}\big \}$, the trace in \eqref{Eq:MLI_k_2_1} can be rewritten as  $\mathrm{g}_{k,l}^{\mathrm{(ef)}\mathrm{H}} \widetilde{\mathrm{g}}_{i,k,r}^{\,\mathrm{(ef)}} \big[ \mathbf{C}^{t-1}_{i} \big]_{\mu(r,i),:} \mathbf{A}_n \big[ \mathbf{C}^{t}_{k} \big]^{\mathrm{H}}_{m(l,k),:}$. Hence, $| \mathrm{MLI}^{t}_{k, n, 2, 1} |^{2}$ is given by
\begin{equation}
    \!\sum_{\substack{i=1 \\ i \neq k}}^{K} \sum_{\substack{i{'}=1 \\ i{'} \neq k}}^{K} \sum_{l=1}^{L} \sum_{l{'}=1}^{L} \sum_{r=1}^{L} \sum_{r{'}=1}^{L} U^{t}_{k, i, i{'}, m, m{'}, \mu,\mu{'},l, l{'}, r, r{'}, n},
    \label{Eq:MLI_k_2_1_abs}
\end{equation}
\noindent where $U^{t}_{k, i, i{'}, m, m{'}, \mu,\mu{'},l, l{'}, r, r{'}, n}$ is obtained as
\begin{equation}
\mathrm{tr} \big\{ \mathrm{g}_{k, i, i{'}, l, l{'}, r, r{'}} \mathbf{C}^{t-1, t-1}_{i, i{'}, \mu,\mu{'},r, r{'}} \mathbf{A}_n \mathbf{C}^{t, t}_{k, m, m{'}, l, l{'}} \mathbf{A}_n^{\mathrm{H}} \big \},
\label{Eq:TraceHuge}
\end{equation}
\noindent with $\mathrm{g}_{k, i, i{'}, l, l{'}, r, r{'}} = \mathrm{g}_{k,l}^{\,\mathrm{(ef)}\mathrm{H}} \mathrm{g}_{k,l{'}}^{\mathrm{(ef)}} \widetilde{\mathrm{g}}_{i{'},k,r{'}}^{\,\mathrm{(ef)}\mathrm{H}}\widetilde{\mathrm{g}}_{i,k,r}^{\,\mathrm{(ef)}}$, and the matrix $\mathbf{C}^{t-1, t-1}_{i, i{'}, \mu,\mu{'},r, r{'}}$ being calculated as
\begin{equation}
    \mathbf{C}^{t-1, t-1}_{i, i{'}, \mu,\mu{'},r, r{'}} = \big[ \mathbf{C}^{t-1}_{i{'}} \big]_{\mu{'}(r{'},i{'}),:}^{\mathrm{H}} \big[ \mathbf{C}^{t-1}_{i} \big]_{\mu(r,i),:}.
\end{equation}
Besides, $\mathbf{C}^{t, t}_{k, m, m{'}, l, l{'}} = \big[ \mathbf{C}^{t}_{k} \big]^{\mathrm{H}}_{m(l,k),:} \big[ \mathbf{C}^{t}_{k} \big]_{m{'}(l{'},k),:}$. Hence, one can evaluate $\mathbb{E}_{\mathbf{C}} \big \{U^{t}_{k, i, i{'}, m, m{'}, \mu,\mu{'},l, l{'}, r, r{'}, n} \big \}$ as 
\begin{equation}
\!\mathrm{tr} \big\{ \mathrm{g}_{k, i, i{'}, l, l{'}, r, r{'}} \mathbb{E}_{\mathbf{C}} \big \{\mathbf{C}^{t-1, t-1}_{i, i{'}, \mu,\mu{'},r, r{'}}  \mathbf{A}_n \mathbf{C}^{t, t}_{k, m, m{'}, l, l{'}} \mathbf{A}_n^{\mathrm{H}} \big \} \big\}.
\end{equation}
\noindent Given that the information signals of \acp{UE} $i$, $i'$ and $k$ are independent, the expected values of $\mathbf{C}^{t-1, t-1}_{i, i{'}, \mu,\mu{'},r, r{'}}$ and $\mathbf{C}^{t, t}_{k, m, m{'}, l, l{'}}$ can be obtained separately. Moreover, $\mathbb{E}_{\mathbf{C}} \big\{ \mathbf{C}^{t-1, t-1}_{i, i{'}, \mu,\mu{'},r, r{'}} \} = \mathbf{0}$ for $i \neq i{'}$, as the information signals of different \acp{UE} are statistically uncorrelated. One can also note that since the information signals generated in \eqref{Eq:EncodingMatrix} are composed of zero-mean complex data symbols drawn from a unitary constellation, the following property holds
\begin{equation}
\centering
\mathbb{E}_{\mathbf{C}} \big \{ \mathbf{C}^{t, t}_{k, m, m{'}, l, l{'}} \big \} \!=\! \begin{cases}
\frac{1}{L_k} \mathbf{I}_{L_k} & \text{if } m(l,k) = m{'}(l{'},k)\,,\\
\mathbf{0} & \text{if } m(l,k) \neq m{'}(l{'},k)\,.
\end{cases}
\label{Eq:ExpectationProperties}
\end{equation}
Similarly, it holds that $\mathbb{E}_{\mathbf{C}} \big \{ \mathbf{C}^{t-1, t-1}_{i, i{'}, \mu,\mu{'},r, r{'}} \big \} = (1/L_k) \mathbf{I}_{L_k}$  for $\mu(r,i) = \mu{'}(r{'},i{'})$, and equals $\mathbf{0}$ otherwise. Thus,
\begin{equation}
        \!\mathbb{E}_{\mathbf{C}} \big \{ | \mathrm{MLI}^{t}_{k, n, 2, 1} |^{2}\big \} \!=\! \sum_{\substack{i=1 \\i \neq k}}^{K} \sum_{l=1}^{L} \sum_{r=1}^{L} \frac{1}{L_k} \big|\mathrm{g}_{k,l}^{\mathrm{(ef)}}\big|^{2} \big|\widetilde{\mathrm{g}}_{i,k,r}^{\,\mathrm{(ef)}}\big|^{2}.
    \label{Eq:MLI_k_2_1_mean}
\end{equation}%
\subsubsection{Compute $\mathbb{E}_{\mathbf{C}} \big \{ | \mathrm{MLI}^{t}_{k, n, 3, 1} |^{2}\big \}$ and $\mathbb{E}_{\mathbf{C}} \big \{ | \mathrm{MLI}^{t}_{k, n, 3, 2} |^{2}\big \}$} Let us define $\mathrm{MLI}^{t}_{k, n, 3} - \mathrm{tr} \big \{ \mathbf{A}_n (\widetilde{\mathbf{n}}_{k}^{t})^{\mathrm{H}} \mathbf{DS}^{t-1}_{k} \big \}$ as $\mathrm{MLI}^{t}_{k, n, 3} = \mathrm{MLI}^{t}_{k, n, 3, 1} + \mathrm{MLI}^{t}_{k, n, 3, 2}$, with $\mathrm{MLI}^{t}_{k, n, 3, 1}$ being found as
\begin{equation}
    \sum_{i=1, i \neq k}^{K} \sum_{l=1}^{L} \sum_{r=1}^{L} \mathrm{tr} \big\{ \widetilde{\mathrm{g}}_{i, k, r}^{\,\mathrm{(ef)}\mathrm{H}} \mathrm{g}_{k, l}^{\mathrm{(ef)}}
    \mathbf{A}_n \mathbf{C}^{t, t-1}_{i, k, m, \mu, l, r} \big \}.
    \label{Eq:MLI_k_3_1}
\end{equation}
In \eqref{Eq:MLI_k_3_1}, we have $\mathbf{C}^{t, t-1}_{i, k, m, \mu, l, r} = \big[ \mathbf{C}^{t}_{i} \big]^{\mathrm{H}}_{\mu(r,i),:} \big[ \mathbf{C}^{t-1}_{k} \big]_{m(l,k),:}$, 
and the term $\mathrm{MLI}^{t}_{k, n, 3, 2}$ is calculated as $\mathrm{MLI}^{t}_{k, n, 3, 2} = \mathrm{tr} \big \{ \sum_{l=1}^{L} \mathrm{g}_{k,l}^{\mathrm{(ef)}} \mathbf{A}_n \big( \mathbf{n}^{t}_{k}\big)^{\mathrm{H}} \big[ \mathbf{C}^{t-1}_{k} \big]_{m(l,k),:} \big \}$. Moreover, note that $\mathrm{MLI}^{t}_{k, n, 3, 2}$ corresponds to $\mathrm{ML}^{t}_{k, n, 2}$ in \eqref{ML_k_2}. Therefore, $\mathbb{E}_{\mathbf{C}} \big \{ | \mathrm{MLI}^{t}_{k, n, 3, 2} |^{2}\big \}$ is obtained as follows
\begin{equation}
    \centering
     \mathbb{E}_{\mathbf{C}} \big \{ | \mathrm{MLI}^{t}_{k, n, 3, 2} |^{2}\big \} = \sigma_{\mathrm{dl}}^{2} \sum\nolimits_{l = 1}^{L}  \big| \mathrm{g}_{k,l}^{({\rm ef})}\big|^{2}.
    \label{Eq:VarMLI_k_3_2}
\end{equation}
\noindent The computation of $\mathbb{E}_{\mathbf{C}} \big \{ | \mathrm{MLI}^{t}_{k, n, 3, 1} |^{2}\big \}$ can be derived analogously to $\mathbb{E}_{\mathbf{C}} \big \{ | \mathrm{MLI}^{t}_{k, n, 2, 1} |^{2}\big \}$ given that $\mathrm{MLI}^{t}_{k, n, 3, 1}$ and $\mathrm{MLI}^{t}_{k, n, 2, 1}$ share the same structure, as indicated in \eqref{Eq:MLI_k_3_1} and \eqref{Eq:MLI_k_2_1}. Consequently, we have 
\begin{equation}
        \mathbb{E}_{\mathbf{C}} \big \{ | \mathrm{MLI}^{t}_{k, n, 3, 1} |^{2}\big \} = \sum_{\substack{i=1 \\ i \neq k}}^{K} \sum_{l=1}^{L} \sum_{r=1}^{L} \frac{1}{L_k} \big|\mathrm{g}_{k,l}^{\mathrm{(ef)}}\big|^{2} \big|\widetilde{\mathrm{g}}_{i,k,r}^{\,\mathrm{(ef)}}\big|^{2}.
    \label{Eq:MLI_k_4_1_mean}
\end{equation}

\subsubsection{Computing $\sum_{z{'}=1}^{4}\mathbb{E}_{\mathbf{C}} \big \{ | \mathrm{MLI}^{t}_{k, n, 4, z{'}} |^{2}\big \}$} Let the interference signals at \ac{UE} $k$ during the received blocks $\mathbf{y}^{t}_{k}$ and $\mathbf{y}^{t-1}_{k}$ be defined as $\mathbf{IS}^{t}_k = \sum_{i=1, i \neq k}^{K} \sum_{l=1}^{L} \widetilde{\mathrm{g}}_{i,k,l}^{\,\mathrm{(ef)}} \big[ \mathbf{C}^{t-1}_{i} \big]_{m(l,i),:} \mathbf{X}^{t}_{i}$ and $\mathbf{IS}^{t-1}_k = \sum_{i=1, i \neq k}^{K} \sum_{l=1}^{L} \widetilde{\mathrm{g}}_{i,k,l}^{\,\mathrm{(ef)}} \big[ \mathbf{C}^{t-1}_{i} \big]_{m(l,i),:}$. Hence, the combined effects of noise and interference can be rewritten as $\widetilde{\mathbf{n}}_{k}^{t} = \mathbf{IS}^{t}_k + \mathbf{n}^{t }_{k}$ and $\widetilde{\mathbf{n}}_{k}^{t-1} = \mathbf{IS}^{t-1}_k + \mathbf{n}^{t-1}_{k}$. Since the last term of \eqref{Eq:MLI_k_t} is defined as $\mathrm{MLI}^{t}_{k, n, 4} - \mathrm{tr} \big \{ \mathbf{A}_n\,( \widetilde{\mathbf{n}}_{k}^{t})^{\mathrm{H}} \widetilde{\mathbf{n}}^{t-1}_{k} \big \}$, we have
\begin{equation}
    \centering
    \mathrm{MLI}^{t}_{k, n, 4} \!=\! \sum\nolimits_{z{'}=1}^4 \mathrm{MLI}^{t}_{k, n, 4, z{'}} \,,
    \label{Eq:MLI_4_4}
\end{equation}
\noindent 
\noindent where the first term $\mathrm{MLI}^{t}_{k, n, 4, 1} = \mathbf{A}_n (\mathbf{n}^{t }_{k})^{\mathrm{H}} \mathbf{n}^{t-1}_{k}$ can be neglected if the noise variance is small enough. The second term is defined as $\mathrm{MLI}^{t}_{k, n, 4, 2} = \mathbf{A}_n(\mathbf{n}^{t }_{k})^{\mathrm{H}} \, \mathbf{IS}^{t}_k$, i.e.,
\begin{equation}
    \centering
     \sum\nolimits_{i=1, i \neq k}^{K} \sum\nolimits_{l=1}^{L} \mathrm{tr} \big\{ \widetilde{\mathrm{g}}_{i,k,l}^{\,\mathrm{(ef)}} \mathbf{A}_n \, (\mathbf{n}^{t }_{k})^{\mathrm{H}} \big[ \mathbf{C}^{t-1}_{i} \big]_{m(l,i),:} \big\}.
\end{equation}
\noindent Then, by using the cyclic property of the trace and following the same procedure we utilized to derive \eqref{Eq:MLI_k_2_1_abs} and \eqref{Eq:TraceHuge}. The term $| \mathrm{MLI}^{t}_{k, n, 4, 2} |^{2}$ can be written as
\begin{equation}
    \centering
     | \mathrm{MLI}^{t}_{k, n, 4, 2} |^{2} \!=\! \sum_{\substack{i=1 \\ i \neq k}}^{K} \sum_{\substack{i{'}=1 \\ i{'} \neq k}}^{K} \sum_{l=1}^{L} \sum_{l{'}=1}^{L} \widetilde{U}^{t}_{i, i{'}, k, m, m{'},l, l{'}, n}\,,
\end{equation}
\noindent with $\widetilde{U}^{t}_{i, i{'}, k, m, m{'}, l, l{'}, n}$ being given by
\begin{equation}
\mathrm{tr} \big\{ \mathrm{\widetilde{g}}_{i{'}, i, k, l{'}, l} \mathbf{\widetilde{C}}^{t-1, t-1}_{i{'}, i, k, m{'}, m, l{'}, l} \mathbf{A}_n (\mathbf{n}^{t}_{k})^{\mathrm{H}}\mathbf{n}^{t}_{k} \mathbf{A}_n^{\mathrm{H}} \big \},
\label{Eq:TraceHuge2}
\end{equation}
\begin{figure*}[tb!] 
\setcounter{mytempeqncnt}{\value{equation}}
\setcounter{equation}{110} 
\begin{equation}
    \mathbb{E}_{\mathbf{C}} \big\{ | \mathrm{MLI}^{t}_{k, n, 4, 4} |^{2} \big\} \approx \sum_{\substack{i = 1 \\ i \neq k}}^{K} \Bigg( \sum_{l = 1}^{L} \frac{\big|\widetilde{\mathrm{g}}_{i,k,l}^{\,\mathrm{(ef)}}\big|^{4}}{n_s}  + \sum_{l = 1}^{L} \sum_{\substack{r = 1 \\ r \neq l}}^{L} \big| \widetilde{\mathrm{g}}_{i,k,l}^{\,\mathrm{(ef)}} \, \widetilde{\mathrm{g}}_{i,k,r}^{\,\mathrm{(ef)}} \big|^{2} \bigg( \frac{1}{L_k} + \widetilde{L}_k \bigg)  \Bigg) + \sum_{\substack{i = 1 \\ i\neq k}}^{K} \sum_{\substack{v = 1 \\ v \neq k, v \neq i}}^{K} \sum_{l = 1}^{L} \sum_{r = 1}^{L} \frac{\big| \widetilde{\mathrm{g}}_{i,k,l}^{\,\mathrm{(ef)}} \, \widetilde{\mathrm{g}}_{v,k,r}^{\mathrm{(ef)}} \big|^{2}}{L_k}
    \label{Eq:MLI_k_4_4_mean}
\end{equation}
\setcounter{equation}{\value{mytempeqncnt}}
\hrulefill
\end{figure*}
\noindent where $\mathbf{\widetilde{C}}^{t-1, t-1}_{i{'}, i, k, m{'}, m, l{'}, l} = \big[ \mathbf{C}^{t-1}_{i{'}} \big]_{m{'}(l{'},i{'}),:}^{\mathrm{H}} \big[ \mathbf{C}^{t-1}_{i} \big]_{m(l,i),:}$ and $\mathrm{\widetilde{g}}_{i{'}, i, k, l{'}, l} =  \widetilde{\mathrm{g}}_{i{'},k,l{'}}^{\,\mathrm{(ef)}\mathrm{H}}\widetilde{\mathrm{g}}_{i,k,l}^{\,\mathrm{(ef)}}$
which results in
\begin{equation}
        \mathbb{E}_{\mathbf{C}} \big \{ | \mathrm{MLI}^{t}_{k, n, 4, 2} |^{2}\big \} \!=\! \sigma_{\mathrm{dl}}^{2}\sum\nolimits_{i=1, i \neq k}^{K} \sum\nolimits_{l=1}^{L} \big|\widetilde{\mathrm{g}}_{i,k,l}^{\,\mathrm{(ef)}}\big|^{2}.
    \label{Eq:MLI_k_4_2_mean}
\end{equation}
\noindent The computation of $\mathbb{E}_{\mathbf{C}} \big \{ | \mathrm{MLI}^{t}_{k, n, 4, 3} |^{2}\big \}$ follows a similar procedure to that used to obtain $\mathbb{E}_{\mathbf{C}} \big \{ | \mathrm{MLI}^{t}_{k, n, 4, 2} |^{2}\big \}$, since both have a similar structure. That is, $\mathrm{MLI}^{t}_{k, n, 4, 3}$ is defined as $\mathrm{MLI}^{t}_{k, n, 4, 3} = \mathbf{A}_n\big( \mathbf{IS}^{t}_k \big)^{\mathrm{H}}\mathbf{n}^{t-1}_{k}$. Thus,
\begin{equation}
    \mathrm{MLI}^{t}_{k, n, 4, 3} \!=\! \sum_{\substack{i=1 \\ i \neq k}}^{K} \sum_{l=1}^{L} \mathrm{tr} \big\{ \big(\widetilde{\mathrm{g}}_{i,k,l}^{\,\mathrm{(ef)}} \big)^{\mathrm{H}} \mathbf{A}_n \big[ \mathbf{C}^{t}_{i} \big]^{\mathrm{H}}_{m(l,i),:} \mathbf{n}^{t-1}_{k} \big \}.
\end{equation}
Consequently,
\begin{equation}
        \mathbb{E}_{\mathbf{C}} \big \{ | \mathrm{MLI}^{t}_{k, n, 4, 3} |^{2}\big \} = \sigma_{\mathrm{dl}}^{2}\sum\nolimits_{i=1, i \neq k}^{K} \sum\nolimits_{l=1}^{L} \big|\widetilde{\mathrm{g}}_{i,k,l}^{\,\mathrm{(ef)}}\big|^{2}.
    \label{Eq:MLI_k_4_3_mean}
\end{equation}
The calculations related to the last term of \eqref{Eq:MLI_4_4} involve several summations, since $\mathrm{MLI}^{t}_{k, n, 4, 4} = \mathbf{A}_n\big( \mathbf{IS}^{t}_k \big)^{\mathrm{H}} \mathbf{IS}^{t-1}_k$.
Therefore, from this point onward, we will adopt a compact notation for multiple summations to enhance brevity: a single summation operator applied to multiple indices should be understood as representing multiple summations, one for each distinct index involved. Let $\mathcal{K} = \{1, \ldots, K\}$ denote the indices of all \acp{UE} in the network. By doing so, $\mathrm{MLI}^{t}_{k, n, 4, 4}$ can be defined as
\begin{equation}
    \centering
    \sum_{\substack{1 \leq i \leq K \\ 1 \leq v \leq K \\  \{i, v\} \subset \mathcal{K} \setminus \{k\}}} \sum_{\substack{1 \leq l \leq L \\ 1 \leq r \leq L }}  
    \bar{\mathrm{g}}_{i,v,k,l,r}^{\mathrm{(ef)}} \mathrm{\overline{C}}^{t-1, t}_{v, i, k, \mu, m, r, l, n},
    \label{MLI_4_4}
\end{equation}
with $\bar{\mathrm{g}}_{i,v,k,l,r}^{\mathrm{(ef)}} = \big( \widetilde{\mathrm{g}}_{i,k,l}^{\,\mathrm{(ef)}}\big)^{\mathrm{H}} \widetilde{\mathrm{g}}_{v,k,r}^{\mathrm{(ef)}}$ and $\mathrm{\overline{C}}^{t-1, t}_{v, i, k, \mu, m, r, l, n}$ being calculated as $\mathrm{\overline{C}}^{t-1, t}_{v, i, k, \mu, m, r, l, n} = \big[ \mathbf{C}^{t-1}_{v} \big]_{\mu(r,v),:} \mathbf{A}_n \big[ \mathbf{C}^{t}_{i} \big]^{\mathrm{H}}_{m(l,i),:}$. 
After defining \eqref{MLI_4_4}, $|\mathrm{MLI}^{t}_{k, n, 4, 4} |^{2}$ can be expressed as
\begin{equation}
    \centering
    \sum_{\substack{1 \leq i \leq K \\ 1 \leq i{'} \leq K \\ 1 \leq v \leq K \\ 1 \leq v{'} \leq K \\  \{i, i', v, v'\} \subset \mathcal{K} \setminus \{k\}}} \sum_{\substack{1 \leq l \leq L \\ 1 \leq l{'} \leq L \\ 1 \leq r \leq L \\ 1 \leq r{'} \leq L}}  
    \overline{U}^{t}_{i, i{'}, v, v{'}, k, m, m{'}, \mu,\mu{'},l, l{'}, r, r{'}, n},
    \label{MLI_4_6}
\end{equation}
\noindent where $\overline{U}^{t}_{i, i{'}, v, v{'}, k, m, m{'}, \mu,\mu{'},l, l{'}, r, r{'}, n}$ is defined as
\begin{equation}
\mathrm{tr} \big\{ \mathrm{\bar{g}}_{i, i{'}, v, v{'}, k, l, l{'}, r, r{'}} \mathbf{C}^{t-1, t-1}_{v{'}, v, \mu{'}, \mu, r{'}, r} \widetilde{\mathbf{C}}^{t, t}_{i, i{'}, m, m{'}, l, l{'}, n}, \big \},
\label{Eq:TraceHuge4}
\end{equation}
\noindent where $\mathrm{\bar{g}}_{i, i{'}, v, v{'}, k, l, l{'}, r, r{'}} =  \big( \widetilde{\mathrm{g}}_{i,k,l}^{\,\mathrm{(ef)}}\big)^{\mathrm{H}} \widetilde{\mathrm{g}}_{i{'},k,l{'}}^{\mathrm{(ef)}} \big(\widetilde{\mathrm{g}}_{v{'},k,r{'}}^{\mathrm{(ef)}}\big)^{\mathrm{H}} \widetilde{\mathrm{g}}_{v,k,r}^{\mathrm{(ef)}}$ and $\mathbf{C}^{t-1, t-1}_{v{'}, v, \mu{'}, \mu, r{'}, r}$ is given by
\begin{equation}
    \centering
    \mathbf{C}^{t-1, t-1}_{v{'}, v, \mu{'}, \mu, r{'}, r} = \big[ \mathbf{C}^{t-1}_{v{'}} \big]^{\mathrm{H}}_{\mu{'}(r{'},v{'}),:} \big[ \mathbf{C}^{t-1}_{v} \big]_{\mu(r,v),:}\,.
\end{equation}
The term $\widetilde{\mathbf{C}}^{t, t}_{i, i{'}, m, m{'}, l, l{'}, n}$ is obtained as
\begin{equation}
    \centering
    \widetilde{\mathbf{C}}^{t, t}_{i, i{'}, m, m{'}, l, l{'}, n}= \mathbf{A}_n \big[ \mathbf{C}^{t}_{i} \big]^{\mathrm{H}}_{m(l,i),:} \big[ \mathbf{C}^{t}_{i{'}} \big]_{m{'}(l{'},i{'}),:} \mathbf{A}_n^{\mathrm{H}}.
\end{equation}
\noindent One can note that computing $\mathbb{E}_{\mathbf{C}} \big \{ | \mathrm{MLI}^{t}_{k, n, 4, 4} |^{2}\big \}$ is not trivial, since it involves several cross-products and summations. Thus, we break the calculation of $\mathbb{E}_{\mathbf{C}} \big \{ | \mathrm{MLI}^{t}_{k, n, 4, 4} |^{2}\big \}$ into five cases, each corresponding to a different combination of \ac{UE} indices in \eqref{MLI_4_6}. In the first case, we compute $\mathbb{E}_{\mathbf{C}} \big \{ | \mathrm{MLI}^{t}_{k, n, 4, 4} |^{2}\big \}$ assuming an entirely different set of \acp{UE}, i.e., all indices $v$, $v'$, $i$, and $i'$ are distinct. In the second case, we consider a scenario with only one distinct \ac{UE}, e.g., $i \neq i{'}$, with $i' = v = v'$. In these two cases, $\mathbb{E}_{\mathbf{C}} \big \{ | \mathrm{MLI}^{t}_{k, n, 4, 4} |^{2}\big \}$ is approximately zero as the information signals transmitted by different \acp{UE} are composed of zero-mean complex-valued symbols. The same notion applies when $v \neq v'$ and $i \neq i'$. Consequently, only the fourth and fifth cases remain. In the fourth case, we have $i = i' = v = v'$, while in the fifth case, $i = i'$ and $v = v'$, with $i \neq v$. Therefore, for the sole purpose of computing $\mathbb{E}_{\mathbf{C}} \big \{ | \mathrm{MLI}^{t}_{k, n, 4, 4} |^{2}\big \}$, equation \eqref{MLI_4_6} can be rewritten as
\begin{equation}
    \centering
    \sum_{\substack{1 \leq i \leq K \\ 1 \leq v \leq K \\  \{i, v\} \subset \mathcal{K} \setminus \{k\}}} \sum_{\substack{1 \leq l \leq L \\ 1 \leq l{'} \leq L \\ 1 \leq r \leq L \\ 1 \leq r{'} \leq L}}  
    \overline{U}^{t}_{i, v, k, m, m{'}, \mu,\mu{'},l, l{'}, r, r{'}, n}.
    \label{MLI_4_7}
\end{equation}
\noindent The above expression can be further simplified by analyzing the indices of the information signal rows, such as $m(l,i)$ and $\mu{'}(r{'},v{'})$. To perform this simplification, the following results were obtained numerically through Monte-Carlo simulations. The results reveal that for two distinct \acp{UE} (i.e., $i \neq v$), the expected value in \eqref{MLI_4_7} is non-zero only in two specific scenarios: 1) when $m(l,i) = m{'}(l{'},i{'})$ and $\mu(r,v) = \mu{'}(r{'},v{'})$, with $m(l,i) \neq \mu(r,v)$; and 2) when all row indices are equal, i.e., $m(l,i) = m{'}(l{'},i{'}) = \mu(r,v) = \mu{'}(r{'},v{'})$. Similarly, when $i = v$, the same reasoning holds. Nonetheless, an additional case must be considered, which yields a non-zero contribution when $\mu(r,v) = m(l,i)$ and $\mu{'}(r{'},v{'}) = m{'}(l{'},i{'})$. Based on the above discussion, \eqref{MLI_4_7} simplifies to
\begin{equation}
\resizebox{0.98\hsize}{!}{$\displaystyle \!
    \sum_{\substack{1 \leq l \leq L \\ 1 \leq \mu \leq L}}\sum_{\substack{1 \leq i \leq K \\ 1 \leq v \leq K \\ \{i, v \} \subset \mathcal{K} \setminus \{k\}}}   
    \!\mathbb{E}_{\mathbf{C}} \big \{ \overline{U}^{t}_{i, v, k, \mu, m, l, r, n} \big\} \!+\! \sum_{\substack{1 \leq i \leq K \\ i \neq k }}   
    \!\mathbb{E}_{\mathbf{C}} \big \{ \widetilde{U}^{t}_{i, k, \mu, m, l, r, n} \big\}
    \label{MLI_4_8},
    $}
\end{equation}
\noindent with $\overline{U}^{t}_{i, v, k, \mu, m, l, r, n}$ being given by
\begin{equation}
\mathrm{tr} \big\{ \mathrm{\bar{g}}_{i, v, k, r, l} \mathbf{C}^{t-1, t-1}_{v, \mu, r} \widetilde{\mathbf{C}}^{t, t}_{i, m, l, n} \big \},
\label{Eq:TraceHuge5}
\end{equation}
\noindent where $\mathrm{\bar{g}}_{i, v, k, r, l} =  \big(\widetilde{\mathrm{g}}_{i,k,l}^{\,\mathrm{(ef)}}\big)^{\mathrm{H}} \widetilde{\mathrm{g}}_{i,k,l}^{\,\mathrm{(ef)}} \big( \widetilde{\mathrm{g}}_{v,k,r}^{\mathrm{(ef)}}\big)^{\mathrm{H}} \widetilde{\mathrm{g}}_{v,k,r}^{\mathrm{(ef)}}$. Moreover, $\widetilde{\mathbf{C}}^{t, t}_{i, m, l, n}= \mathbf{A}_n \big[ \mathbf{C}^{t}_{i} \big]^{\mathrm{H}}_{m(l,i),:} \big[ \mathbf{C}^{t}_{i} \big]_{m(l,i),:} \mathbf{A}_n^{\mathrm{H}}$, and $\mathbf{C}^{t-1, t-1}_{v, \mu, r} = \big[ \mathbf{C}^{t-1}_{v} \big]^{\mathrm{H}}_{\mu(r,v),:} \big[ \mathbf{C}^{t-1}_{v} \big]_{\mu(r,v),:}$.
The term $\widetilde{U}^{t}_{i, k, \mu, m, l, r, n}$ is obtained in \eqref{MLI_4_8} as
\begin{equation}
\mathrm{tr} \big\{ \mathrm{\widetilde{g}}_{i, k, r, l} \mathbf{C}^{t-1, t-1}_{i, m, \mu, l, r} \widetilde{\mathbf{C}}^{t, t}_{i, \mu, m, r, l, n} \big \},
\label{Eq:TraceHuge6}
\end{equation}
\noindent where we have $\mathrm{\widetilde{g}}_{i, k, r, l} =  \big(\widetilde{\mathrm{g}}_{i,k,l}^{\,\mathrm{(ef)}}\big)^{\mathrm{H}} \widetilde{\mathrm{g}}_{i,k,l}^{\,\mathrm{(ef)}} \big( \widetilde{\mathrm{g}}_{i,k,r}^{\,\mathrm{(ef)}}\big)^{\mathrm{H}} \widetilde{\mathrm{g}}_{i,k,r}^{\,\mathrm{(ef)}}$ and $\widetilde{\mathbf{C}}^{t, t}_{i, \mu, m, r, l, n} = \mathbf{A}_n \big[ \mathbf{C}^{t}_{i} \big]^{\mathrm{H}}_{\mu(r,i),:} \big[ \mathbf{C}^{t}_{i} \big]_{m(l,i),:} \mathbf{A}_n^{\mathrm{H}}$. 
Moreover, we have that $\mathbf{C}^{t-1, t-1}_{i, m, \mu, l, r} = \big[ \mathbf{C}^{t-1}_{i} \big]^{\mathrm{H}}_{m(l,i),:} \big[ \mathbf{C}^{t-1}_{i} \big]_{\mu(r,i),:}$.

To compute $\mathbb{E}_{\mathbf{C}}\{\overline{U}^{t}_{i, v, k, \mu, m, l, r, n}\}$ in \eqref{Eq:TraceHuge5} we have noted that $\mathbb{E}_{\mathbf{C}} \big\{\mathrm{tr} \big\{\mathbf{C}^{t-1, t-1}_{v, \mu, r} \, \widetilde{\mathbf{C}}^{t, t}_{i, m, l, n} \big \} \big\}$ is approximately
\begin{equation}
\centering
\mathbb{E}_{\mathbf{C}} \big\{\mathrm{tr} \big\{\mathbf{C}^{t-1, t-1}_{v, \mu, r} \widetilde{\mathbf{C}}^{t, t}_{i, m, l, n} \big \} \big\} \!\approx\! 
\begin{cases}
1/n_s & \!\text{if } \mu(r,v) \!=\! m(l,i). \\
1/L_k & \!\text{otherwise},\\
\end{cases}
\label{Eq:Lk_approx1}
\end{equation}
\noindent for $v = i$. In contrast, $\mathbb{E}_{\mathbf{C}} \big\{\mathrm{tr} \big\{\mathbf{C}^{t-1, t-1}_{i, m, \mu, l, r} \widetilde{\mathbf{C}}^{t, t}_{i, \mu, m, r, l, n} \big \} \big\}$ in \eqref{Eq:TraceHuge6} is calculated approximately as
\begin{equation}
\centering
\mathbb{E}_{\mathbf{C}} \big\{\mathrm{tr} \big\{\mathbf{C}^{t-1, t-1}_{i, m, \mu, l, r} \widetilde{\mathbf{C}}^{t, t}_{i, \mu, m, r, l, n} \big \} \big\} = \widetilde{L}_k,
\label{Eq:trace_approx2}
\end{equation}
\noindent where $\widetilde{L}_k \approx 0$ for $L_k = 2$, $\widetilde{L}_k \approx 0.112$ for $L_k = 4$, and $\widetilde{L}_k \approx 1/L_k$ for $L_k = 8$.   
Finally, $\mathbb{E}_{\mathbf{C}} \big\{ | \mathrm{MLI}^{t}_{k, n, 4, 4} |^{2} \big\}$ can be computed as in \eqref{Eq:MLI_k_4_4_mean}, which is shown at the top of the page.%

\bibliographystyle{IEEEtran}
\bibliography{IEEEabrv,bib_file}

\begin{IEEEbiography}[{\includegraphics[width=1in,height=1.25in,clip,keepaspectratio]{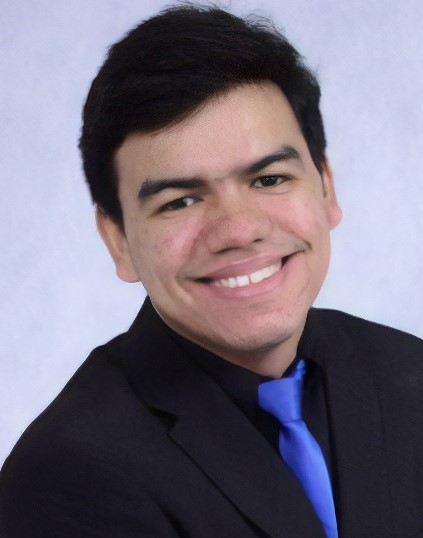}}]{Marx M. M. Freitas} received the B.Sc., M.Sc., and Ph.D. degrees in electrical engineering from the Federal University of Pará (UFPA), Belém, Brazil, in 2018, 2019, and 2024, respectively. From 2022 to 2023, he was a visiting Ph.D. student with the Scuola Superiore Sant'Anna, Pisa, Italy. He is currently a Postdoctoral Fellow with the University of Cassino and Southern Lazio, Cassino, Italy. He has experience in broadband communication systems. His research interests include signal processing, resource allocation, massive MIMO, and mobile transport networks for future wireless communication systems.
\end{IEEEbiography}

\begin{IEEEbiography}
[{\includegraphics[width=1in,height=1.25in,clip,keepaspectratio]{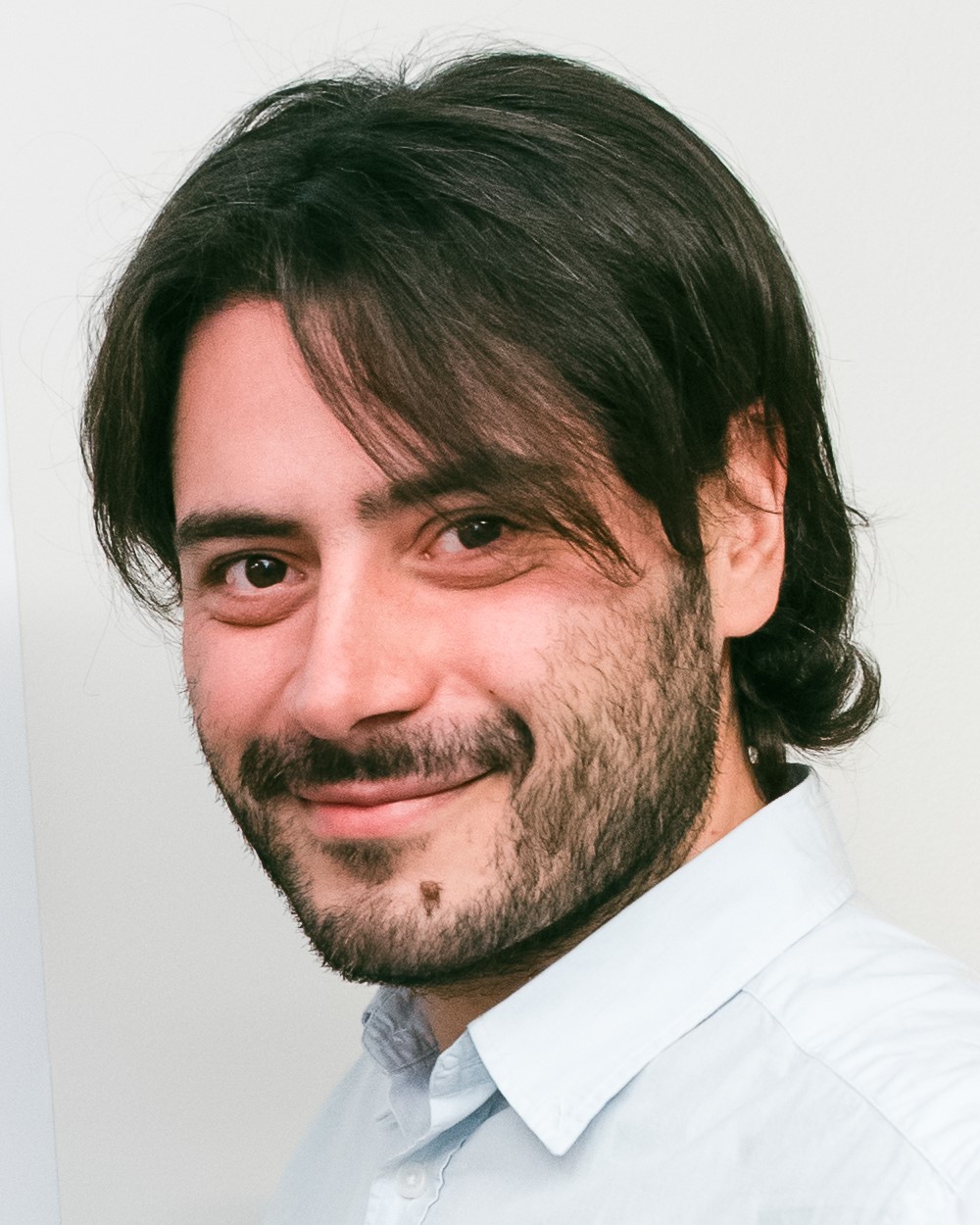}}]{Giovanni Interdonato} (Senior Member, IEEE) 
received the M.Sc. degree in computer and telecommunication systems engineering from the Mediterranea University of Reggio Calabria, Reggio Calabria, Italy, in 2015, 
and the Ph.D. degree in electrical engineering with a specialization in communication systems from Link\"{o}ping University, Link\"{o}ping, Sweden, in 2020. 
From October 2015 to October 2018, he was a Researcher with Ericsson Research, Link\"{o}ping, and a Marie Sk{\l}odowska-Curie Research Fellow in the EU-H2020 ITN project ``5Gwireless''. 
From March 2020 to January 2026, he was an Assistant Professor with the Department of Electrical and Information Engineering (DIEI), University of Cassino and Southern Lazio, Cassino, Italy.
He is currently a Senior Researcher with Ericsson Research, Stockholm, Sweden, where he conducts research in the Radio Systems and Standards area and serves as Ericsson's delegate to the ECC/CEPT, implementing the company's strategy on D2D-IMT radio spectrum policy. 
His main research interests lie in the field of wireless communications and signal processing, with a focus on 6G physical-layer technologies, non-terrestrial networks, radio spectrum and resource management, and communication protocols. He has co-invented about twenty granted patents on massive MIMO and cell-free massive MIMO systems. 

Dr. Interdonato was recognized as an Exemplary Reviewer by \textsc{IEEE Transactions on Communications} in 2021 and as an Exemplary Editor by \textsc{IEEE Communications Letters} in 2022, 2024, and 2025. He received a research grant from the Ericsson Research Foundation in 2019.
He serves as an Associate Editor for \textsc{IEEE Transactions on Wireless Communications}, \textsc{IEEE Communications Letters}, and \textsc{IEEE Open Journal of the Communications Society}.
Dr. Giovanni Interdonato was included in the Stanford University/Elsevier World's Top 2\% Scientists List for 2024.
\end{IEEEbiography}
\vfill

\begin{IEEEbiography}
[{\includegraphics[width=1in,height=1.25in,clip,keepaspectratio]{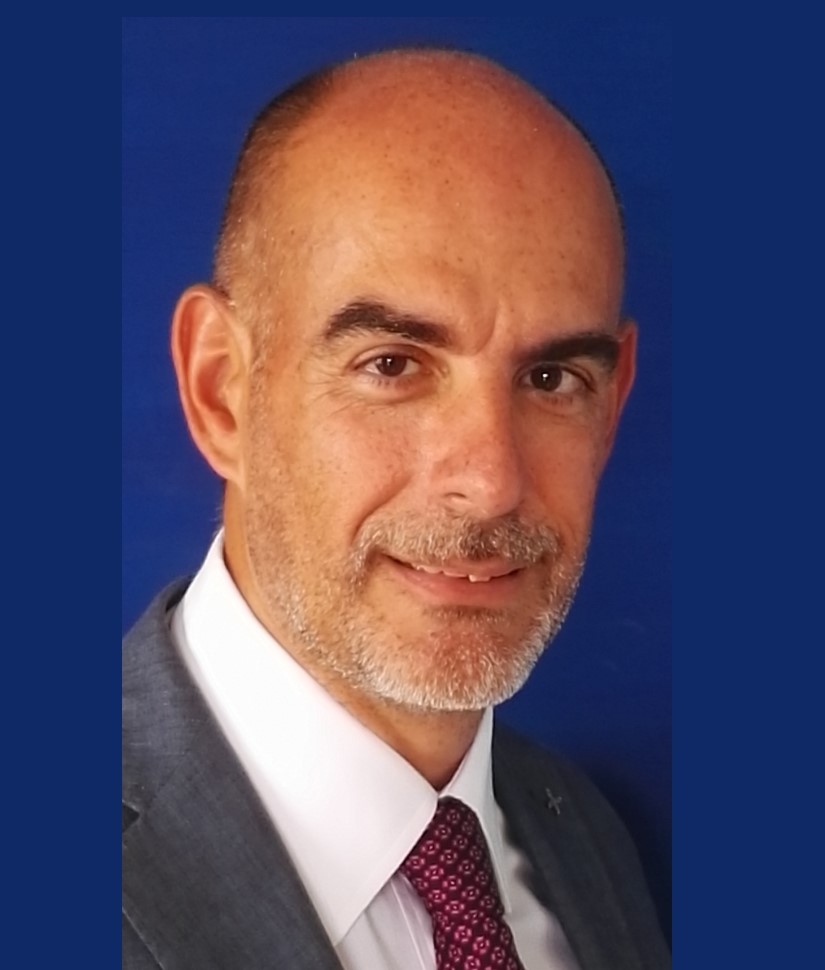}}]{Stefano Buzzi} (Fellow, IEEE) joined the University of Cassino and Lazio Meridionale, Italy, in 2000, first as an Assistant Professor, then as an Associate Professor (since 2002) and, finally, since 2018, as a Full Professor. From 2022 to 2025, he was also affiliated with the Politecnico di Milano, Italy.  He received the M.Sc. degree (summa cum laude) in Electronic Engineering in 1994, and the Ph.D. degree in Electrical and Computer Engineering in 1999, both from the University of Naples ``Federico I''.  He has had short-term research appointments at Princeton University, Princeton (NJ), USA in 1999, 2000, 2001, and 2006. He is a former Associate Editor of IEEE Signal Processing Letters and IEEE Communications Letters and has served as guest editor for five IEEE JSAC special issues. From 2014 to 2020, he was an Editor for the IEEE Transactions on Wireless Communications, and he is currently Editor for the IEEE Transactions on Communications. He also regularly serves as a TPC member for several international conferences.

Dr. Buzzi's research interests are in the broad field of communications and signal processing, with emphasis on wireless communications and beyond-5G systems.  He is currently the General Coordinator of the EU-funded Doctoral Network ISLANDS, on Integrated Sensing and Communications for the Vehicular Environment. He has co-authored 200+ technical peer-reviewed journal and conference papers, and, among these, the highly cited paper ``What will 5G be?'', IEEE JSAC, June 2014.  He was a co-recipient of the 2025 IEEE SPS Best Paper Award for the paper ``Foundations of MIMO radar detection aided by reconfigurable intelligent surfaces,'' IEEE TSP, 2022.  
\end{IEEEbiography}

\end{document}